\documentclass[useAMS,usenatbib]{mn2e}
\usepackage[centerlast]{caption}
\usepackage[mediumspace,mediumqspace,Grey,squaren]{SIunits}
\usepackage[dvips]{graphicx}
\usepackage{afterpage}
\usepackage{amsmath}
\usepackage{amssymb}
\usepackage{mathrsfs}
\usepackage{url}
\usepackage{aas_macros}  
\usepackage{graphicx}
\usepackage{dcolumn}
\usepackage{bm}
\def\be{\begin{equation}}
\def\ee{\end{equation}}
\def\bea{\begin{eqnarray}}
\def\eea{\end{eqnarray}}
\newcommand{\zmin}{\mathrm{z_{min}}}
\newcommand{\zmax}{\mathrm{z_{max}}}

\newcommand{\msun}{M_{\odot}}

\newcommand{\microm}[1]{\unit{#1}{\micro\meter}}

\newcommand{\comment}[1]{}

\newcommand{\De}{\Delta}

\newcommand{\LF}{\left(}
\newcommand{\RF}{\right)}

\newcommand{\Jband}{$J_{125}\;$}
\newcommand{\Hband}{$H_{160}\;$}
\newcommand{\NirJband}{$J_{115}\;$}
\newcommand{\NirHband}{$H_{150}\;$}
\newcommand{\NirKband}{$K_{200}\;$}
\newcommand{\fsmds}{$\mathrm{f_{SMDS}}\;$}
\newcommand{\Msun}{M_{\odot}}
\newcommand{\arsq}{arcmin$^2\;$}
\newcommand{\tento}[1]{$10^{#1}\Msun\;$}
\newcommand{\fdt}{$\mathrm{f_{\De t}}\;$}
\newcommand{\fsurv}{$\mathrm{f_{surv}}\;$}
\newcommand{\zform}{$\mathrm{z_{form}}\;$}

\newcommand{\zstart}{$\mathrm{z_{start}}\;$}
\newcommand{\dNdz}{$\frac{dN}{dz d\theta^2}\;$}
\newcommand{\Mpcyr}{$\rm{Mpc^{-3}yr^{-1}}\;$}
\def\lp{\left(}
\def\rp{\right)}

\begin{document}

\title[Observing Dark Stars with JWST]{Observing Dark Stars with JWST}
\author[Ilie et al. ]{Cosmin Ilie$^{1}$\thanks{E-mail: cilie@umich.edu}, Katherine Freese$^{1,2}$,  Monica Valluri$^{3}$,Ilian T. Iliev$^4$, 
and Paul Shapiro$^{2,5}$\\
$^1$Michigan Center for Theoretical Physics, Physics Department, University of Michigan, Ann Arbor, MI 48109, USA \\
$^2$Texas Cosmology Center, University of Texas, Austin, TX 78712, USA   \\
$^3$ Department of Astronomy, University of Michigan, Ann Arbor, MI 49109, USA \\
$^4$ Department of Physics \& Astronomy, University of Sussex, Brighton BN1 9QH, UK\\
$^5$ Department of Astronomy, University of Texas, Austin, TX 78712, USA
}

\date{\today}

\maketitle

\begin{abstract}

We study the capability of the James Webb Space Telescope (JWST) to
detect Supermassive Dark Stars (SMDS). If the first stars are powered
by dark matter heating in triaxial dark matter haloes, they may grow to
be very large $> 10^6 M_\odot$ and very bright $> 10^9 L_\odot$. These
SMDSs would be visible in deep imaging with JWST and even Hubble Space
Telescope (HST).  Indeed the object detected at $z\sim 10$ in the
recent HST ultra deep field image with the Wide Field Camera 3 (WFC3)
may be a SMDS. We use sensitivity limits from previous HST surveys to
place bounds on the numbers of SMDSs that may be detected in future
JWST imaging surveys.  We showed that SMDS in the mass range
$10^6-10^7 M_\odot$ are bright enough to be detected in all the
wavelength bands of the NIRCam on JWST (but not in the less sensitive
MIRI camera at higher wavelengths). If SMDSs exist at $z \sim$10,
12, and 14, they will be detectable as J-band, H-band, or K-band
dropouts, respectively.  With a total survey area of 150 arcmin$^2$
(assuming a multi-year deep parallel survey with JWST), we find that
typically the number of $10^6 M_\odot$ SMDSs found as H or K-band
dropouts is $\sim 10^5$ \fsmds, where the fraction of early DM haloes
hosting DS is likely to be small, \fsmds$<<1$.  If the SDMS survive down to z=10 where
HST bounds apply, then the observable number of SMDSs as H or K-band dropouts with JWST is $\sim 1-30$.
While individual SMDS are bright enough to be detected by JWST,
standard PopIII stars (without dark matter annihilation) are not, and
would only be detected in first galaxies with total stellar masses of
~$10^6-10^8 M_\odot$.  Differentiating first galaxies at z$>$10 from
SMDSs would be possible with spectroscopy: the SMDS (which are too
cool produce significant nebular emission) will have only absorption lines while the galaxies
are likely to produce emission lines as well. Of particular interest would be the 
HeII emission lines at $\lambda \sim 1.6\mu$  as well as H$\alpha$ lines which would be signatures of early galaxies rather than SMDSs.
The detection of SMDSs with JWST would not only provide
  alternative evidence for WIMPs but would also also provide a
  possible pathway for the formation of massive ($10^4-10^6 M_\odot$)
  seeds for the formation of supermassive black holes that power QSOs
  at $z=6$.

\end{abstract}


\section{Introduction}

The first stars are thought to have formed at $z=10-50$
when the universe was about
200 million years old  in $\sim 10^6 M_\odot$ (mini) halos
consisting of  $85\%$ DM and $15\%$ baryons in the form of H and He
from  big bang nucleosynthesis. Their formation marks the end of the
``dark ages'' of the Universe.   For reviews of the standard
picture of the formation of the first stars see \citet{BarLoe01,Yoshidaetal03,BroLar04,RipAbe05,Brommetal2009}.

\citet{DS2}
first showed that dark matter heating may drastically
alter the picture of formation for these first stars.  
 The canonical example of particle DM is Weakly Interacting Massive
Particles (WIMPs). In many theories WIMPs are their own antiparticles and annihilate with
themselves wherever the DM density is high.   In fact, this annihilation process is exactly
what is responsible in the early Universe for leaving behind the correct relic 
 WIMP abundance today to 
solve the dark matter problem, 24\% of the energy density of the universe.
The same annihilation process would then take place also in the collapsing protostellar
clouds at the centers of minihalos.  At suitably high baryonic density in these clouds, 
the annihilation products get stuck inside the cloud and prevent it from undergoing
further collapse. The annihilation products thermalize with the baryons and provide a
very powerful heat source.   Indeed, the object becomes a "dark star", which, despite its
name, shines very bright.
 The DM -- while only a negligible fraction
of the star's mass -- provides the key power source for the star through
DM heating. Note that the term 'Dark' refers to the power source, not the
content of the star. These first Dark Stars are stars made primarily of hydrogen and helium with a smattering
of dark matter ($<$1\% of the mass consists of DM); yet they shine due
to DM heating.

 Recently there has been much excitement in the
dark matter community about hints of WIMP detections  in a number of experiments:
excess positrons in the PAMELA satellite \citep{PAMELApap,Abdo2010,Adriani2010} may be due to DM annihilation (though alternative astrophysical explanations
are more likely).  Excess  $\gamma$-rays in the  FERMI  satellite \citep{FERMI,Abdo2009,Dobler:2010,FERMI_LAT:2011} may be due to DM annihilation;
and annual modulation \citep{Druiker:1986,Freese:1987} in  direct detection experiments
DAMA \citep{Bernabei:2009} and COGENT \citep{Aalseth:2011}.   The
CRESST experiment \citep{CRESST} also has unexplained events.

The WIMP
annihilation rate is $n_\chi^2 \langle \sigma v \rangle$ where
$n_\chi$ is WIMP density and we take the standard annihilation cross
section
\begin{equation}
\label{eq:sigmav}
\langle \sigma v \rangle = 
3 \times 10^{-26} {\rm cm}^3/{\rm s},
\end{equation}
and WIMP masses in the range 1 GeV-10 TeV.  
WIMP annihilation produces energy at a rate per unit volume 
\begin{equation}
\hat Q_{DM} = n_\chi^2 \langle \sigma v \rangle m_\chi =
\langle \sigma v \rangle \rho_\chi^2/m_\chi ,
\label{eq:Q}
\end{equation}
where $n_\chi$ is the WIMP number density, $m_\chi$ is the WIMP mass,
and $\rho_\chi$ is the WIMP energy density.  The annihilation products
typically are electrons, photons, and neutrinos. The neutrinos escape
the star, while the other annihilation products are trapped in the
dark star, thermalize with the star, and heat it up.  The luminosity
from the DM heating is
\begin{equation}
\label{DMheating}
L_{DM} \sim f_Q \int \hat Q_{DM} dV 
\end{equation}
where $f_Q$ is the fraction of the annihilation energy deposited in
the star (not lost to neutrinos) and $dV$ is the volume element. We
take $f_Q=2/3$ as is typical for WIMPs.

Dark stars are born with masses $\sim 1 \msun$.  They are giant puffy
($\sim 10$ AU), cool (surface temperatures $<10,000$K), yet bright
objects \citep{Freese:2008wh}.  They reside in a large
reservoir ($\sim 10^5 \msun$) of baryons, i.e., $\sim 15$\% of the
total halo mass.  These baryons can start to accrete onto the dark
stars. Dark stars can continue to grow in mass as long as
there is a supply of DM fuel.
We consider two different mechanisms that can continually provide the requisite dark matter fuel,
 allowing them to become supermassive dark stars (SMDS) of mass
 $M_{DS} >10^5 \msun$. : \hfil\break
  1) {\bf Extended Adiabatic Contraction (AC):}   the central dark
  matter density is enhanced due to an increase in the depth of the
  gravitation potential well due to the infall of baryons.  We treat this gravitational
  effect via the Blumenthal method for adiabatic contraction.  While this approach is simple
  to implement, we \citet{DS3} and others \citet{Natarajan:2009,DMfs3}
  have previously shown that it provides
  dark matter densities accurate to within a factor of two, which is perfectly adequate for these 
  studies.  
   In the central cusps of triaxial DM halos DM particles follow a variety of centrophilic orbits \citep[box orbits and chaotic orbits][]{valluri_etal_10} whose population is continuously replenished, allowing DM annihilation to continue much longer than in spherical DM halos.  The period of extended AC can thus last for a very long time (hundreds of
  millions of years or more).  \citet{Freeseetal10} showed that this replenishment of the DM in the central cusp could be used to followed the
 growth of dark stars from their inception at $1
\msun$, till they become supermassive dark stars (SMDS) of mass
 $M_{DS} >10^5 \msun$.   \hfil\break 
2) {\bf Capture:} As a second mechanism for dark matter refueling, we take
the star to be initially powered by the DM from adiabatic
  contraction (AC), but assume the AC phase is  short $\sim
  300,000$ years; once this DM runs out, the star shrinks, its density
  increases, and subsequently the DM is replenished inside the star by
  capture of DM from the surroundings \citep{DMcap,Iocco2008,Sivertsson2010} as it scatters elastically off of nuclei in the star.  In this
  case, the additional particle physics ingredient of WIMP scattering
  is required.  This elastic scattering is the same mechanism that
  direct detection experiments (e.g.  CDMS, XENON, LUX, DAMA, COGENT, COUPP, CRESST) are
  using in their hunt for WIMPs.

 Supermassive dark stars can result from either of these mechanisms
 for DM refueling inside the star. \citet{Umeda:2009} considered  a different
 scenario which also results in SMDSs.  In all of these cases SMDSs can live for a very long
 time, tens to hundreds of million years, or possibly longer (even to
 today). We find that SMDS of mass $M_{DS} > 10^6 \msun$ SDMSs are very bright $> 3
 \times 10^9 L_\odot$ which makes them potentially observable by the James
 Webb Space Telescope (JWST).  
  
    The key ingredient that
 allows dark stars to grow so much larger than ordinary fusion powered
 Population III stars is the fact that dark stars are so much cooler.
 Ordinary Pop III stars have much larger surface temperatures in
 excess of 50,000K. They produce ionizing photons that provide a
 variety of feedback mechanisms that cut off further accretion. \citet{McKeeTan08} have estimated that the resultant Pop III stellar
 masses are $\sim 140\msun$.  The issue of the initial
mass function for Pop~III stars is far from being solved. Recent
simulations \citep[see][]{Clarketal2011,Greifetal2011a,
  Greifetal2011b} indicate that the typical mass of such objects is
much lower that previously thought. Dark stars are very different from
 fusion-powered stars, and their cooler surface temperatures allow
 continued accretion of baryons all the way up to enormous stellar
 masses, $M_{DS} > 10^5\msun$.

 In this paper we discuss detectability of these objects in the upcoming JWST.  In future work we will investigate how well other observations with Herschel, SPITZER, GMT, TMT and other instruments can detect or place bounds on Dark Stars. We restrict our discussion only to SMDS of mass $10^6, 10^7 M_\odot$ (we show that SMDS of $10^5$ and lower are hard to detect). 
Previously  \citet{Zackrisson2010} studied dark stars of even lower masses, and 
concluded that even  $\lesssim$\tento{3} DS could be detected as individual objects
 with JWST  if their fluxes were magnified by gravitational lensing  by a well-placed  foreground cluster.  Since
supermassive dark stars are larger and brighter, they are easier to detect. A preliminary study of detectability with JWST and HST of supermassive Dark Stars was made in \citet{Freeseetal10}
and \citet{Freeseetal10b}. Freese et al.  approximated the spectrum of the SD as a pure blackbody  determined by its temperature and radius and used it to show that  individual SMDS would be detectable with JWST and HST. In this paper we improve our estimate by using
spectra  from the TLUSTY model stellar atmospheres code  for zero-metallicity atmospheres from the work of  \citet{Zackrisson2010b}.

  SMDS formed via Extended AC are easier to detect than those formed with capture.  Those formed
   "with capture" are somewhat hotter (by a factor of few) and have radii
 smaller by a factor of 5-10  for the same stellar mass.  Because they are hotter, their peak wavelength moves out
 of the most sensitive ranges for HST and JWST, and their fluxes in the detectors are lower.

 Once the SMDS run out of DM fuel, they contract and heat up till the core
 reaches $10^8 $K and fusion begins.  Due to their extremely large masses the fusion-powered phase is short and the SMSDs collapse to from massive black holes of mass $10^4-10^6 \msun$.  Again, this
 prediction is different from standard Pop III stars, many of which
 explode as pair-instabilty supernovae \citep{Heger:2002} with
 predicted even/odd element abundance ratios that are not (yet)
 observed in nature.  These massive black holes remnants  could provide the moderately massive "seeds" for the formation of nuclear supermassive black holes accounting for the  existence of
 $10^9~\msun$ BHs \citep{haiman_loeb_01} which are the central engines of the most distant ($z\gtrsim 6$)
 quasars in the Sloan Digital Sky Survey (SDSS) ~\citep{Fan:2001,Fan:2004,Fan:2006}. Indeed direct collapse of very metal-poor, low-angular momentum gas via dynamical instabilities \citep{loeb_rasio_94,begelman_etal_06,lodato_natarajan_06} has been proposed as a way to form massive "seed" black holes of  $10^4-10^6 \msun$ at redshifts of 10-15. These massive seed formation scenarios however, are difficult to  confirm observationally since the BHs form in compact, low luminosity cold gas disks and the BH formation is accompanied by a sudden burst of with a luminosity of $10^9 L_\odot$. In contrast if the "seeds" form from SMDSs, they may well shine  for $10^6-10^7$ years prior to their collapse to a BH, enabling them to be detected by JWST.
 
SMDS  could also make plausible precursors of Intermediate Mass Black Holes;
and account for the  BHs inferred by
extragalactic radio excess seen by the ARCADE experiment \citep{Seiffert:2009}.  In addition, the BH remnants from DS could play a role in
high-redshift gamma ray bursts thought to take place due to accretion
onto early black holes \citep{Narayan:2001}
 
 The possibility that DM
annihilation might have effects on {\it{ today's}} stars was initially
considered in the $'80$s and early $'90$s \citep{krauss,Press:1985,bouquet,salatisi} and has
recently been studied in interesting papers by
\citet{moskalenko,scott1,bertone,scott2,Casanellas:2009,Hooperetal:2010,Scott:2011a}. 

Several authors have explored the repercussions of DM
heating in the first stars including: \citet{DS2, DMcap,Freese:2008wh, DSnl, DMfs1,DMfs2,DMfs3,DMfs4,Schleicher:2009,Gondolo2010,Ripamonti:2010ab,Sivertsson2010,Casanellas:2011,Hirano:2011,Iocco:2011,Ilie:2011,Scott:2011a}.

The effects of DS (and those of the
resultant main sequence stars) on reionization was studied by \citet{Schleicher:2008,Schleicher:2009} and more recently by
 \cite{Scott:2011} as discussed below.

In this paper we follow the approach taken by \citet{Zackrisson2010b,ER:Zackrisson2010b}.  Similar to their work, we use SMDS spectra from the TLUSTY code; compute the formation rate of DSs by counting
DM haloes in N-body simulations; and use HST data to bound the numbers of SMDS that survive to z=10 and therefore
the numbers that may be seen with JWST.
Their study focused on $10^7 M_\odot$ SMDS while we consider lower mass
ones as well.  We go beyond their work by studying SMDS as H and K-band dropouts with JWST, where
JWST can really improve upon all previous data sets.  

The paper is organized as follows. In Section II we describe  the spectra of SMDS obtained by using the TLUSTY
 code.  In Section III we compute formation rate of DSs, by counting DM halos in a   
 N-body simulation of structure formation  at $z>10$  carried out with the Cube P$^3$M code \citep{Ilievetal10}  and  assuming that 
 some fraction \fsmds of these early halos will host DS.
 In Section IV, we examine the detectability of SMDS
 in Hubble Space Telescope (HST).  In fact HST has seen objects out to $z \sim 10$,
 and it is interesting to speculate that HST could already have seen SMDS if they
survive to redshift z=10.  With current  imaging data it is impossible 
 to differentiate  between an early galaxy composed of PopIII stars from a SMDS.  However, the  
 the fact that HST has only seen one
 object at this high redshift  can be used to  set bounds \citep{Zackrisson2010}
 the numbers of dark stars at $z \sim 10$.  
In Section V we show that dark stars may be detected 
in a variety of JWST filters, and in particular may show up as J-band, H-band, or K-band 
dropouts; such a detection would then gives an indication of their redshift.    
In Section VI,  we compare  early galaxies at high redshifts (consisting of PopIII stars with different IMFs) with  SMDS, which will
look very similar with JWST, and start a discussion of ways to differentiate between them. 
In Section VII we conclude and summarize the results of our study.

\section{Dark Star Spectra}\label{DSSpectra}
\begin{figure*}
\begin{center}$
\begin{array}{cc}
\includegraphics[scale=0.50]{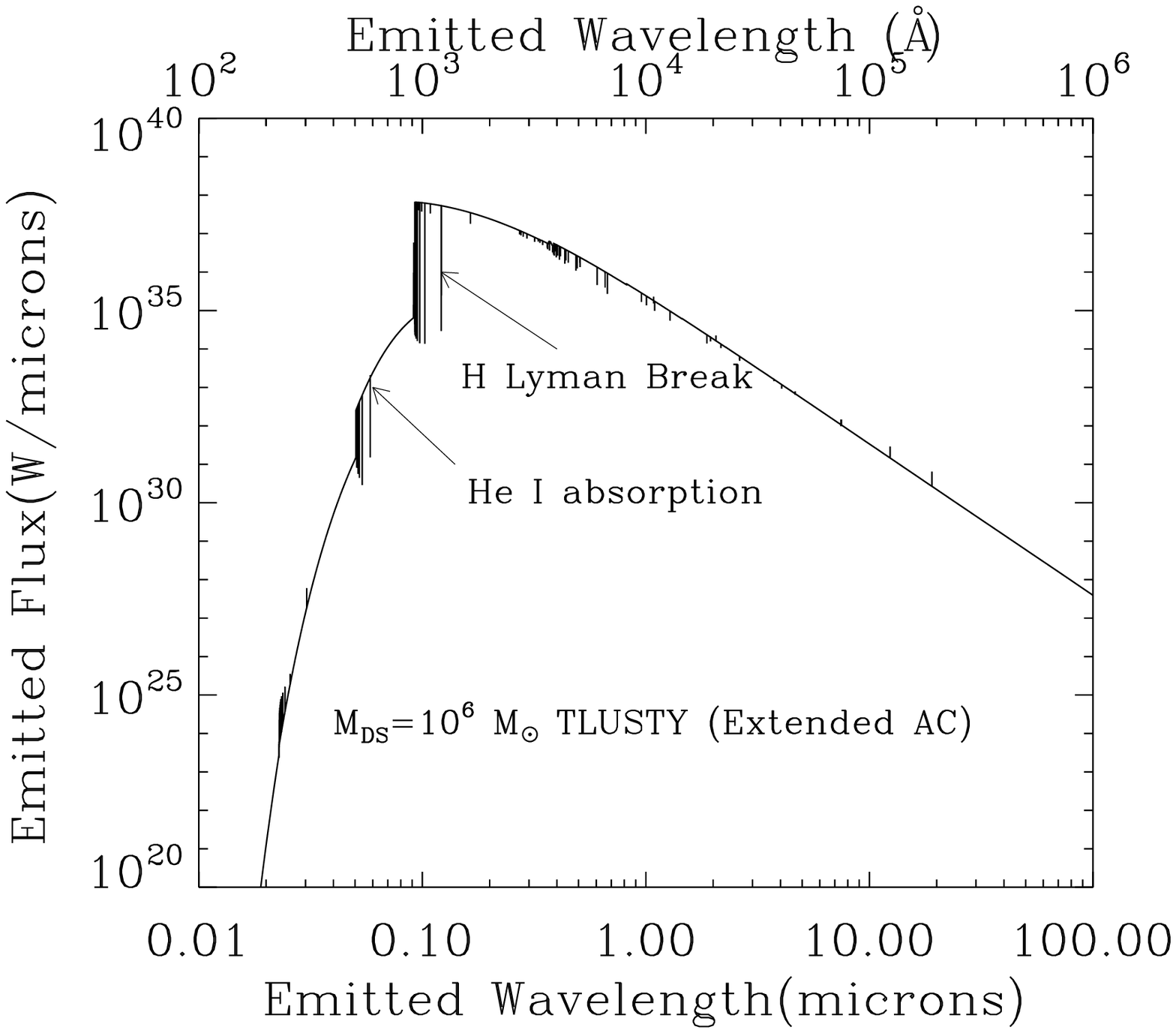}&
\includegraphics[scale=0.50]{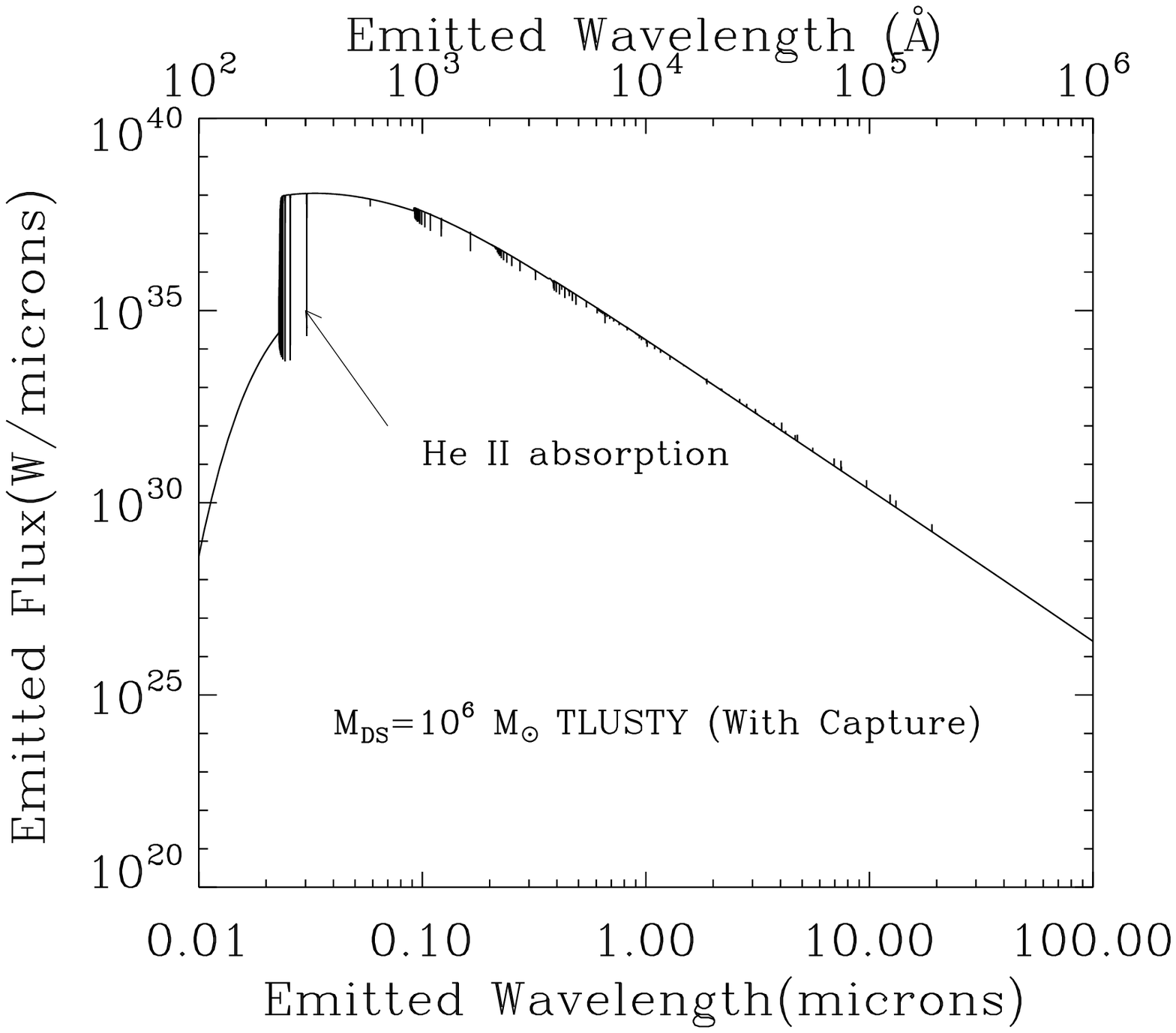} 
\end{array}$
\end{center}
\caption{Expected spectral energy distribution (SEDs) of \tento{6} supermassive dark stars. Left panel: DS with a surface temperature of $1.9\times 10^4$K and formed via the extended adiabatic contraction (AC) only mechanism. Right panel: with a surface temperature of $5.1\times 10^4$  and formed ``with capture''.}
\label{fig:TLUSTYRest}
\end{figure*}

In this section we present spectra of SMDS obtained with the publicly available TLUSTY \citep{TLUSTY} synthetic stellar atmospheres code.  As discussed in \citet{Freeseetal10}, SMDS formed  via captured DM are much hotter than SMDS formed via extended AC.  Also, stars formed via capture undergo a Kelvin-Helmholz contraction phase prior to DM capture,  hence their radii are 5-10 times smaller than those SMDS  of the same mass formed via the extended AC mechanism. Since Dark Stars are composed of primordial hydrogen and helium, no other elements are assumed to be present in the atmosphere and, hence all the observed spectral lines are those of H and He.  However the differences  in the temperature and radii of SMDS formed via these two mechanisms are responsible for the differences in the spectra in the two panels of Figure~ \ref{fig:TLUSTYRest}. The left panel  shows the spectrum for a $10^6 \msun$ dark star with surface temperature $T_{eff} = 1.9\times 10^4$K  which grew via extended AC.   The Lyman edge is seen at roughly 0.1 microns\footnote{Compared to a blackbody of the same temperature,  photons below the Lyman edge have typically been shifted to higher wavelengths (lower energy) by absorption and rescattering. However, the excess  seen at wavelengths just below the Lyman edge is due to photons coming from deeper inside the star (the photosphere is
at roughly an optical depth $\sim 1$, and at this wavelength there is very little absorption).}.
Similarly, Figure~ \ref{fig:TLUSTYRest} (right ) illustrates the spectrum for a $10^6 \msun$  and T$_{eff}=5.1\times 10^4$K  DS which grew via captured DM. The most prominent differences from the left panel are a shift of the peak in the spectrum to lower wavelengths and a steeper UV continuum slope $\beta$ ($f_{\lambda}\propto \lambda^{\beta}$). Despite the fact that the SMDS formed via capture is hotter, its significantly smaller radius  makes it harder to detect  in the near infrared at redshifts of $\sim 10$ and above.

There are significant differences in the spectra the two cases. In the left panel (extended AC),  the lower surface temperature ($\sim 2\times 10^4K$)  implies a significant fraction of neutral H and He remain in the stellar atmosphere resulting in strong absorption lines at wavelengths corresponding to the Lyman series (\microm{0.1216}-\microm{0.0912}). At  shorter wavelengths we notice another break in the spectrum due to neutral helium (HeI) absorption ($\sim$\microm{0.05}-\microm{0.06}). In the right panel (``with capture''), the higher surface temperature (T$_{eff}\sim 5\times 10^4K$) implies that H is ionized hence the  Lyman  absorption lines are weaker. The break in the spectrum in Figure~ \ref{fig:TLUSTYRest} right panel corresponds to absorption by singly ionized helium (HeII) at wavelengths ranging between \microm{0.023} and \microm{0.030}.  In the left panel  HeI lines appear 
at wavelengths $\sim$ [\microm{0.3}-\microm{0.45}], HeII lines at
wavelengths $\sim$\microm{0.46}, and more HeI lines at $\sim$
[\microm{0.47}-\microm{0.7}]. The same lines, with somewhat weaker strength,
are seen in the right panel. In both cases we note a sequence of
absorption lines between $\sim$[\microm{0.8}-\microm{1.0}], which
correspond to HeI absorption.


\section{Dark Star Formation Rate}

The first Dark Stars can form in the early Universe inside minihalos of $\sim 10^6 \Msun$, where 
protostellar clouds collapse via molecular hydrogen cooling until the DM heating sets in.
Later in $10^8 \Msun$ halos, where clouds collapse via atomic line radiative cooling, larger DS can form. To compute the detection rate of SMDS with JWST we need to know the formation rate of $10^6 - 10^8 \Msun$ dark matter halos.  If we assume that a fraction \fsmds
of these halos contain Dark Stars we can use this to compute the formation rate of DSs. We will attempt to set constraints on this fraction by using the fact that a single $z=10$ object  was observed in recent HST ultra deep field observations with the Wide Field Camera 3 \citep[][hereafter HUDF09]{Bouwensetal11}.

A similar study by \citet{Zackrisson2010b, ER:Zackrisson2010b} for the case of \tento{7} SMDS concluded  that the prior null detection of  $z=10$ objects in  first year HUDF09 observation \citep{Bouwens2010Blue},  was sufficient to rule out the detection of \tento{7} SMDS with JWST. However these authors did not consider the effect of the time it takes the SMDS to grow when computing the formation rates for DM halos that could host such objects. This effect is transparent in Table \ref{tb:Rates} in the differences between what we labeled as \zstart (the redshift that should be used when computing the formation rate of DM halos) and \zform (the redshift when the DS reaches its final mass). Consideration of a finite time required for the SMDS to grow (following the formation of its host DM halo)  significantly lowers the bounds predicted from HST,  since to be visible at $z=10$ the more massive DM halos has to have formed at a higher redshifts, where they are rarer. In addition we consider the case of the \tento{6} SMDS, since these objects are likely to be more numerous, are detectable with JWST, and are also subject to bounds from  existing  HST observations.

\begin{figure*}
\begin{center}$
\begin{array}{cc}
\includegraphics[scale=0.50]{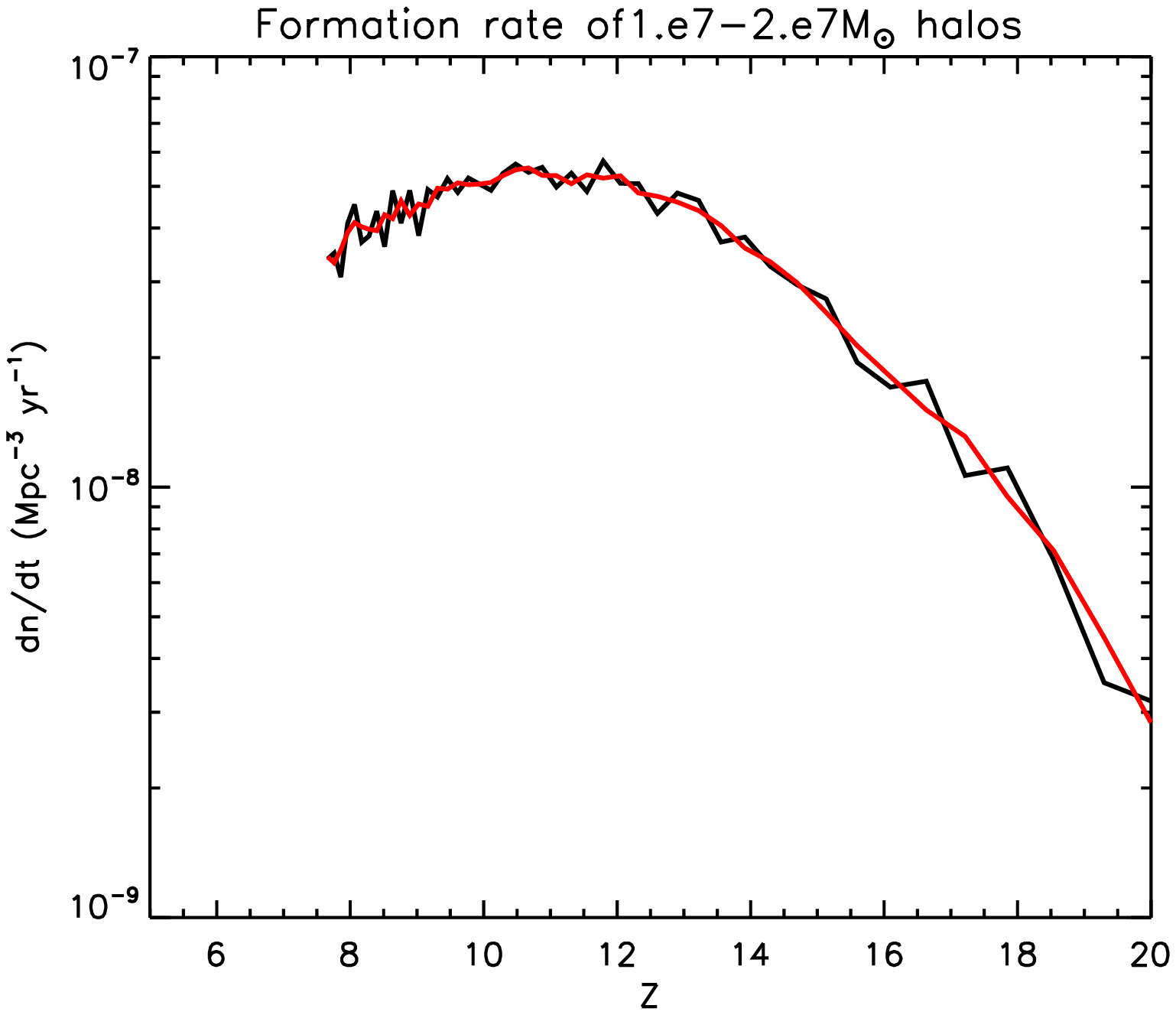} &
\includegraphics[scale=0.50]{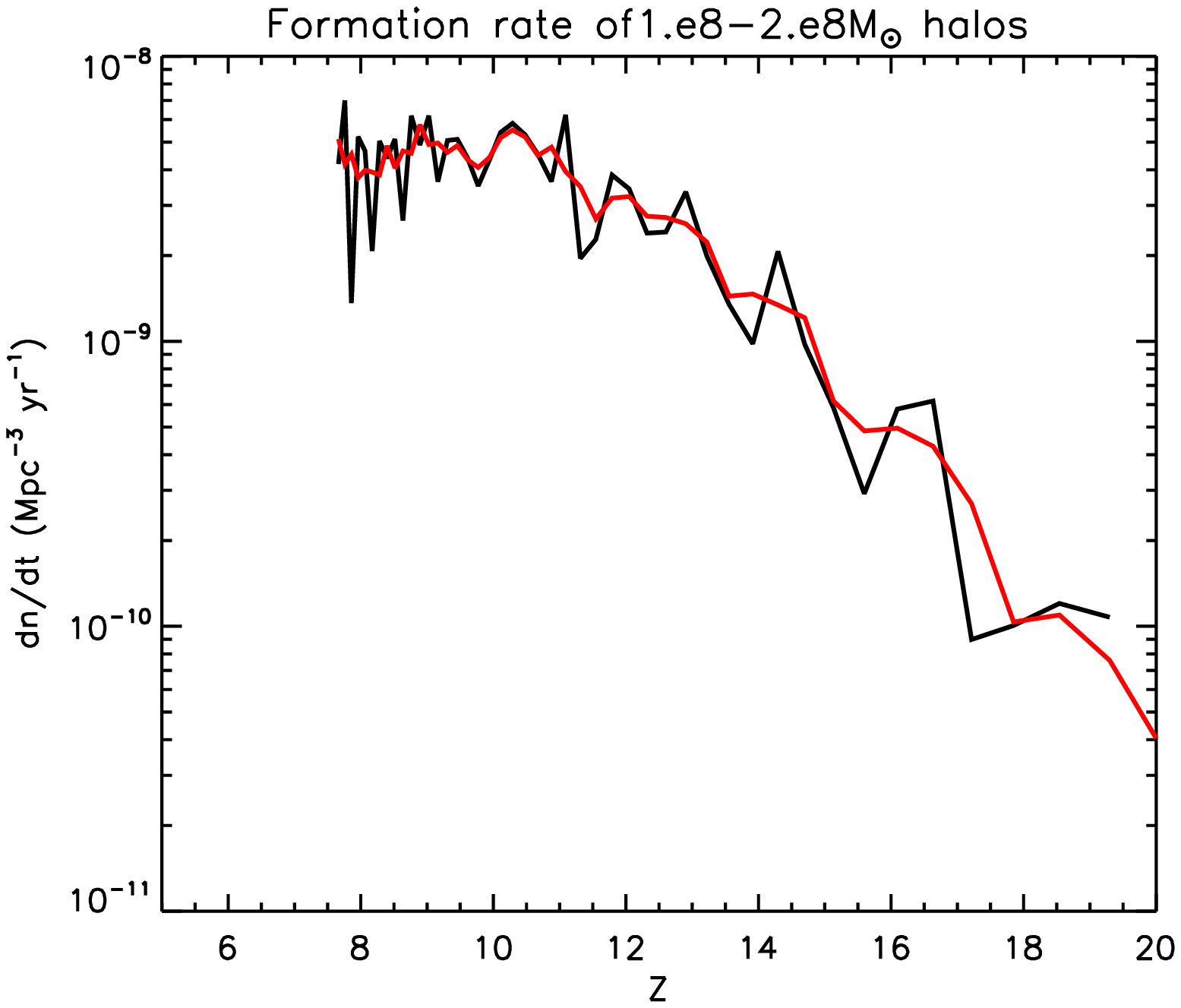}
\end{array}$
\end{center}
\caption{Left: the formation rate of $1-2\times10^7\Msun$ minihalos per comoving Mpc$^3$ and year. These halos are potential hosts for the $10^6\Msun$ SMDS. Right: formation rate for  $1-2\times10^8\Msun$ minihalos in which a $10^7\Msun$ SMDS can form. The black lines correspond to the formation rate computed directly from the N-body simulation and the smoother red lines (obtained by computing a running average)  simply improve visibility of the general trend.}
\label{FRates1}
\end{figure*}

\begin{figure*}
\begin{center}$
\begin{array}{cc}
\includegraphics[scale=0.50]{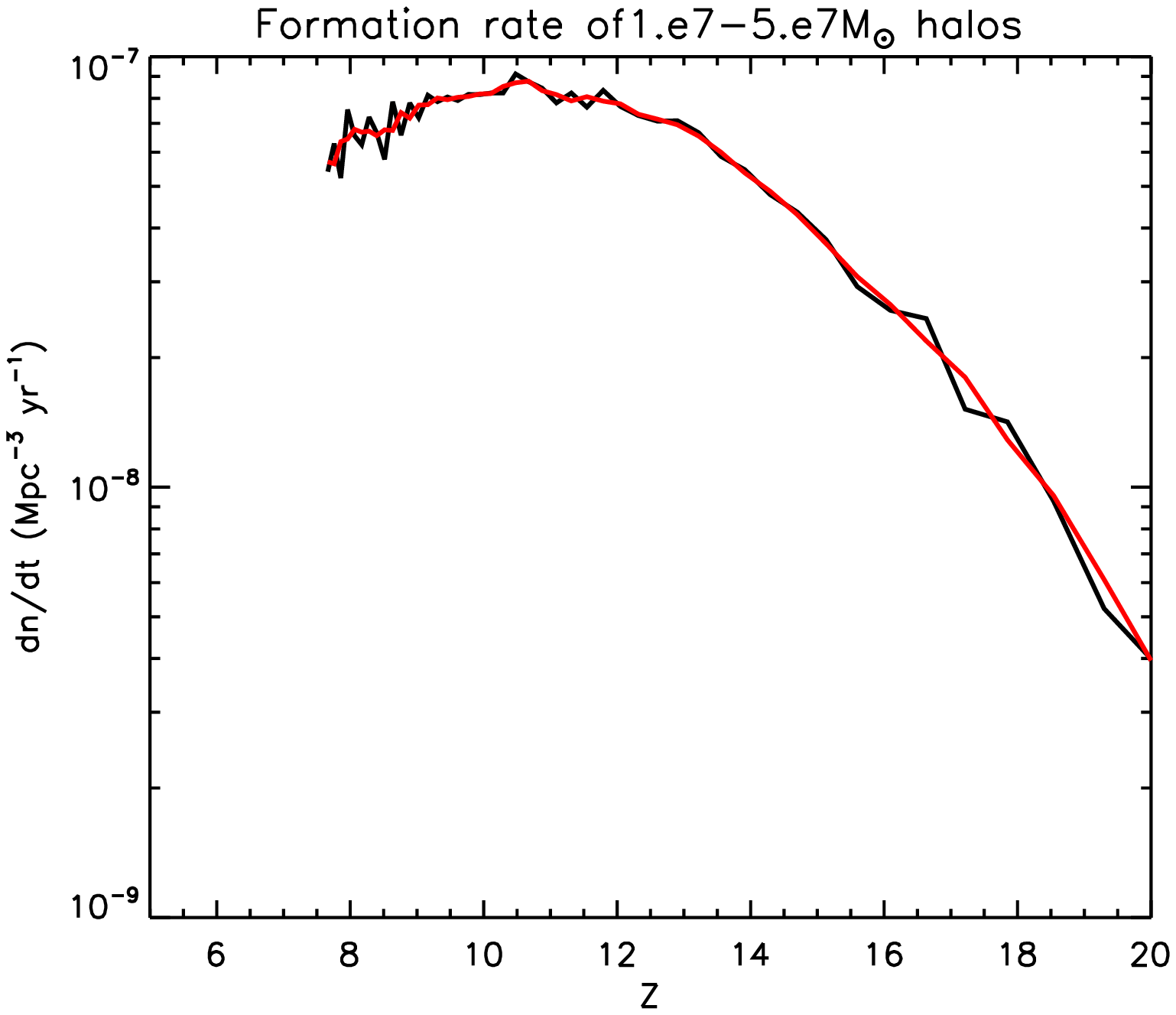} &
\includegraphics[scale=0.50]{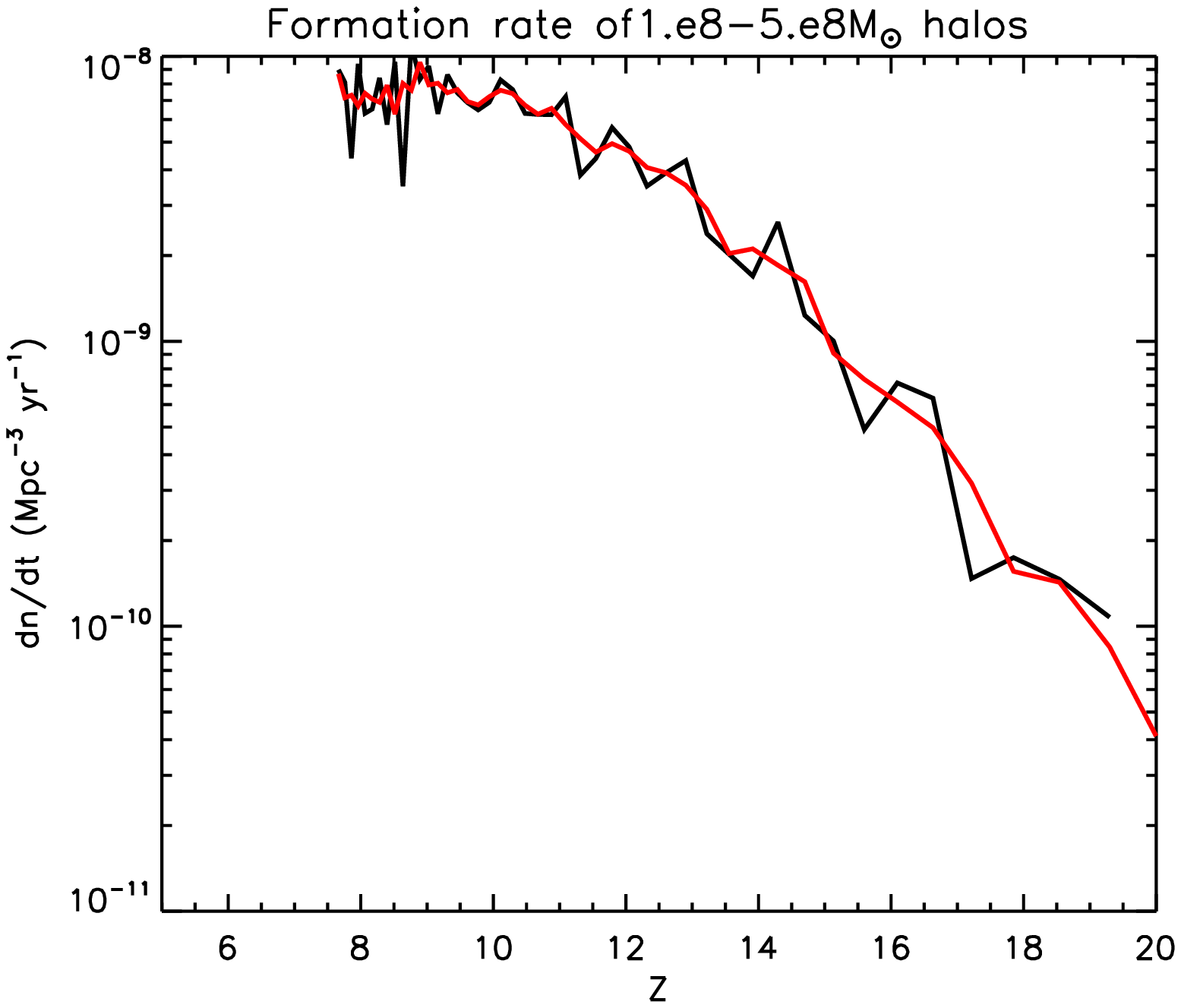}
\end{array}$
\end{center}
\caption{As in Figure~\ref{FRates1}, but with the  larger mass bin width  (see scenario II in text). In the left panel we plot the formation rate of minihalos with a mass in the $1-5\times10^7\Msun$ range, where a DS of $10^6\Msun$ could form. The panel on the right is for halos in the $1-5\times 10^8\Msun$ range, where a DS of $10^7\Msun$ could form.}

\label{FRatesAll}
\end{figure*}

We use  N-Body simulations  of structure formation at high redshifts from \citet{Ilievetal10} carried out with the CubeP$^3$M N-Body code, 
developed from the particle-mesh PM-FAST \citep{Merzetal04}. This high
resolution simulation considers a comoving volume of $6.3 h^{-1}$ Mpc
with  $1728^3$ particles of mass $5.19\times 10^3\Msun$, hence is able
to resolve halos of mass $ \gtrsim 5\times10^5 \msun$.  We compute the
formation rate  ($dn/dt$ as a function of redshift per comoving
Mpc$^3$ per year), of minihalos with masses within different mass
ranges. Figure~\ref{FRates1}  shows the formation rate of halos in
two mass ranges that spans a factor of two in mass ($10^7-2\times 10^7
\msun$ and $10^8-2\times 10^8 \msun$) while  Figure~\ref{FRatesAll} shows the formation rate of halos in two mass ranges that span a factor of five in mass ($10^7-5\times 10^7 \msun$ and $10^8-5\times 10^8 \msun$).  

We computed the formation rate of DM minihalos using two different sets of bin widths, to 
show that the results are relatively insensitive to this issue.
As our canonical case, we computed the formation rate $dn/dt$ of  minihalos per $Mpc^{-3}yr^{-1}$ formed in a bin whose width is a  factor
of two in mass (Figure~ \ref{FRates1}).   $dn/dt$
as a function of redshift is shown for halos in the mass range $(1-2) \times 10^7 \Msun$ (left panel) and
for halos in the mass range $(1-2) \times 10^8 \Msun$ (right panel).
We assume that since the baryonic fraction initially
in the halo is roughly 15\%, we assume that a DS forming in a halo of a given mass can attain at most 10-15\% of the mass of its host halo.  Following Freese et al. 10 we assume that the DS can grow with an accretion rate of  $\sim 1 M_\odot$ to the point
where it consumes a significant fraction of the baryons in the halo.
In other words, we assume that a $10^7\Msun$ SMDS will form in a $(1-2) \times 10^8\Msun$ minihalo.  While this is an unlikely scenario, which involves most of the baryons in the halo being accreted into a single central object, we will see that even with this assumption, detection rates of SMDS with JWST are fairly small. The formation rates in Scenario I are plotted in  Figure~ \ref{FRates1}.

As a check,  we also broadened the range of DM  halo masses in which DS from by allowing halo masses to span a factor of five in mass. Figure~ \ref{FRatesAll} (left panel) shows the formation rate $dn/dt$ as a function of redshift for halos
in the mass range $(1-5) \times 10^7 \Msun$ and the right panel indicates the formation rate of halos in the mass range $(1-5) \times 10^8 \Msun$.
In this scenario the SMDS is $10-50$ times smaller than its host halo, and is more realistic since in this case all the baryons in the halo are not accreted by the DS.

 A comparison of Figures \ref{FRates1} and \ref{FRatesAll} shows that the formation rate of host halos does not vary
significantly between the two scenarios (at most by a factor of 3). Henceforth, in the remainder of this study, we will always take
the halo mass range to span a factor of two in mass.  

We define \zstart to be the (approximate) formation redshift of minihalos capable of hosting DS, 
allowing for an uncertainty of a unit redshift interval; i.e. the minimum redshift of minihalo formation is $\mathrm{z_{min}} =
\mathrm{z_{start}}-1/2$ 
while the maximum redshift is
 $\mathrm{z_{max}} = \mathrm{z_{start}} +1/2$.
 We make a distinction between \zstart the redshift of formation of the DM halo capable of hosting  a DS  (initial $\sim 1 \msun$ 
mini dark stars come into existence very soon after this redshift), and \zform  
the redshift of formation of the SMDS. Between \zstart and \zform  the DS grows by accreting baryons at a rate of 
$10^{-2} - 10^{-1} \msun$/yr  growing over this period to a supermassive size of  $\sim 10^5 - 10^7 \msun$.   This difference between
\zstart and \zform is  crucial to accounting for the differences between the results presented in this paper and previous 
work \citep{Zackrisson2010b} where this additional time required to grow 
supermassive was not allowed for. 

 The formation rate of minihalos  per unit redshift and \arsq is then given by
\be\label{convert}
\frac{dN}{dz d\theta^2} =\frac{dn}{dt}\LF V_c(\zmin)-V_c(\zmax)\RF\frac{C}{4\pi}\Delta t(min;max),
\ee

where, $V_c$ denotes the comoving volume at a given redshift, $C$ is the conversion factor between \arsq and steradians, and $\Delta t(min;max)$ is the cosmic time interval between $\mathrm{z_{min}}$ and $\mathrm{z_{max}}$:
\be\label{time}
\Delta t(min;max)=t_{H}\int_{\zmin}^{\zmax}\frac{1}{(1+z)\LF \Omega_{m}(1+z)^3+\Omega_{\Lambda}\RF^{\frac{1}{2}}} .
\ee  
TheN-body simulations from with the halo formation rates are computed as well as other calculations assume a standard  $\Lambda$CDM Universe in which  $\Omega_m=0.27$ is the cosmic matter density  and $\Omega_\Lambda=0.73$ is the
cosmic dark energy  density or cosmological constant  with parameters from WMAP5 data \citet{WMAP5}.

We consider three possible redshifts \zform by which the DS has accreted enough baryons to become supermassive:
\begin{itemize}
\item Case A:  \zform =10 
\item Case B:  \zform =12  
\nopagebreak
\item Case C:  \zform =15 
\end{itemize}

In principle the accretion rate and the final mass of the SMDS in these three cases will  imply 
three different values of \zstart at which the relevant minihalos formed.
To  simplify the situation we assume a fixed accretion rate of \tento{-1}/yr to determine the values for \zstart (Table \ref{tb:Rates} column 5) and using  corresponding $dn/dt$ values from  Figure~ \ref{FRates1} (Table \ref{tb:Rates} column 6) at $z=$\zstart,  we evaluate \dNdz using Equation(\ref{convert}) (Table \ref{tb:Rates}, column 7). These values of  \dNdz  will be used in sections to follow.

\begin{table*}
  \begin{center}
    \begin{tabular}{ccccccc}
      \hline\hline
      Scenario Name     & Halo Mass Range    & $M_{DS}$     & \zform         & \zstart &        $dn/dt$      & \dNdz        \\
      &  ($\Msun$)               & ($\Msun$)   &                       &          &       (\Mpcyr)      &  arcmin$^{-2}$\\
      \hline
      A               & $(1-2)\times10^8$        & $10^7$                & 10             &13       &    $5\times10^{-9}$  &  235         \\

      B               & $(1-2)\times10^8$       & $10^7$                 &12               & 16      &    $7\times10^{-10}$  &  16         \\

      C               &$(1-2)\times10^8$        & $10^7$                 &15               & 22      &    $1\times10^{-10}$  & 0.77          \\
      \hline
      A               & $(1-2)\times10^7$        & $10^6$                & 10            & 10.7     &    $5\times10^{-8}$   &  4435        \\

      B               & $(1-2)\times10^7$       & $10^6$                  &12              & 12.8     &    $6\times10^{-8}$   &  2965        \\

      C               &$(1-2)\times10^7$        & $10^6$                  &15              & 16        &    $2\times10^{-8}$   &466         \\

      \hline
    \end{tabular} 

    \caption{DM halo formation rates:  $dn/dt$ expressed in \Mpcyr, and \dNdz  as number formed per unit redshift and \arsq for cases considered in the text.
      Top three rows A-C are for a \tento{7} SMDS forming DM halos of mass (1-2) $\times 10^8 \msun$ ;
      bottom three rows are for the \tento{6} SMDS 
      forming in DM halos of mass (1-2) $\times 10^7 \msun$.  We have assumed that
      the DS started accreting baryons with a constant rate of $10^{-1}\Msun/$yr at \zstart and 
      reached its final mass by \zform. }
    \label{tb:Rates} 
  \end{center}
\end{table*}

\section{Supermassive Dark Stars with Hubble Space Telescope}

In this section we examine the observability of dark stars with  existing Hubble Space Telescope (HST) surveys,
speculating that HST may already have seen such objects, if they survive to redshift $z=10$.  We will adopt the standard "dropout technique" pioneered by \citet{steidel_etal_96} and applied recently to J and H band observations of the Hubble Ultra Deep Field (HUDF09) by \citep{Bouwensetal11, Oeschetal11} to detect a candidate galaxy at $z=10$ as a "J-band dropout."  This photometric redshift determination method  requires a 5-sigma detection of an object  in one band but a non-detection in a adjacent band of lower wavelength. In the case of the "J-band drop out" observed with HST, the object was observed in the 1.60 $\mu$  (H-band)  but was {\it not} seen in the 1.15$\mu$  (Y -band) or 1.25 $\mu$ (J- band).  The absence of  emission in the latter bands  is assumed to occur due to Ly$-\alpha$ absorption by hydrogen clouds in
between the source and us, allowing  for an approximate estimate of the redshift of the object.  More specifically we take as our dropout criterion

\begin{equation}\label{eq:dropout}
\Delta m_{AB} \geq 1.2
\end{equation}

where $\Delta m_{AB}$ is the difference in apparent magnitude between
the two bands of observation, in this case the J and H bands. Observations at longer (near to mid IR) wavelengths are required for photometric determination of objects more distant  than $z=10$, necessitating JWST observations.
\citet{Bouwensetal11} and \citet{Oeschetal11} find  a candidate $z\sim 10$ object in the co-added first and second year observation of the HUDF with the new WFC3/IR camera as a J-band dropout. This object is currently thought to be a galaxy, the
most distant one observed to date,  since the SED is a reasonable match to that of galaxies at $z>9$ and it appears clearly extended \citep{Oeschetal11}. However the absence of spectra and the poor spatial resolution of the image allow us to consider the possibility that this object could instead be a SMDS.  Even though it may be hard to  identify a DS uniquely with HST, the fact that at most one candidate has been found can be used to place bounds on the numbers of dark stars at redshifts up to $z=10$.  In this section we examine 
the observability of DS of various masses in existing HST imaging surveys, and in a later section examine the 
resulting bounds for future surveys with JWST.

\subsection{Comparison of DS stellar output with HST Sensitivity}\label{ComparisonDSHST}

\begin{figure*}
\begin{center}$
\begin{array}{cc}
\includegraphics[scale=0.50]{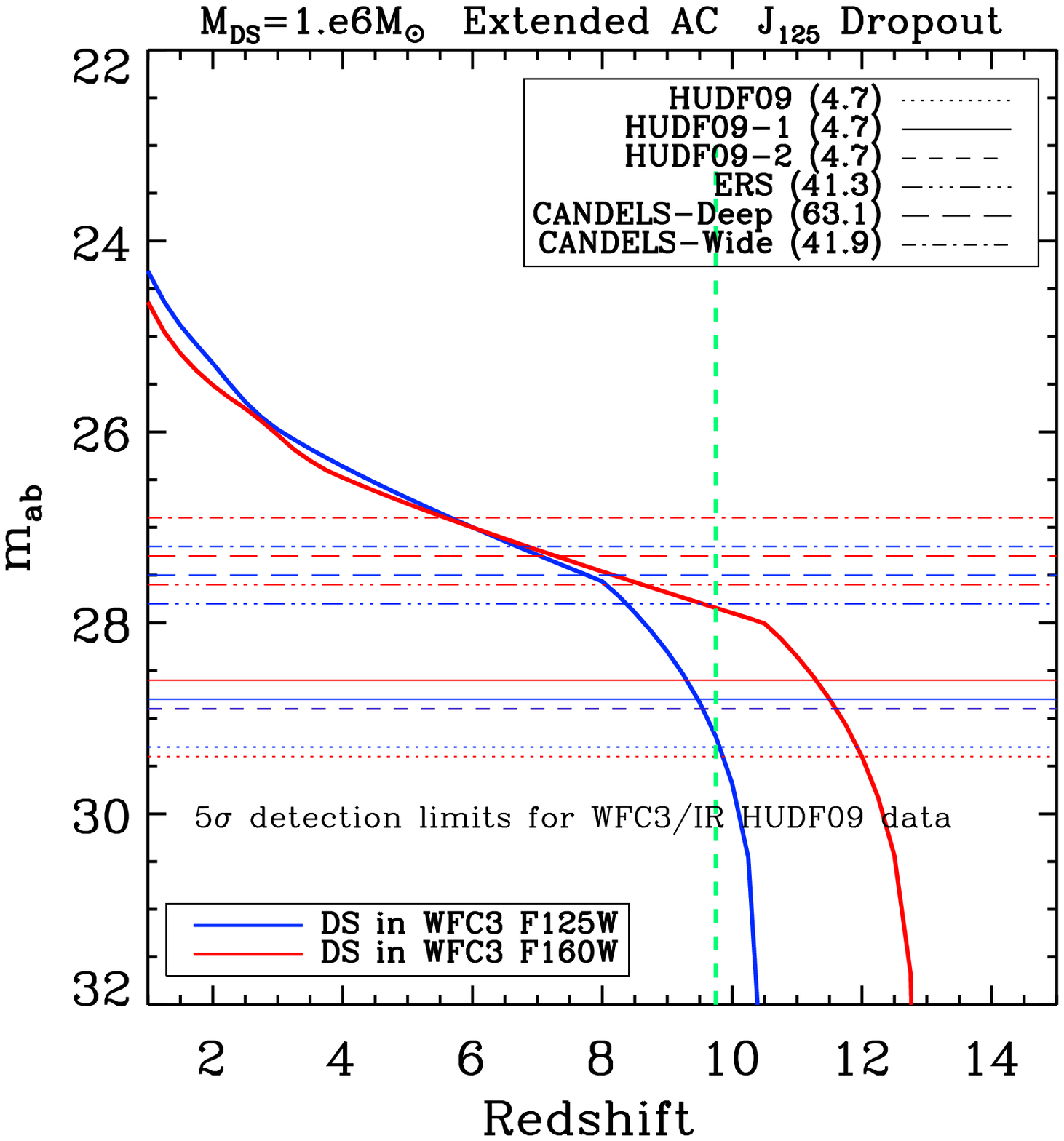}&
\includegraphics[scale=0.50]{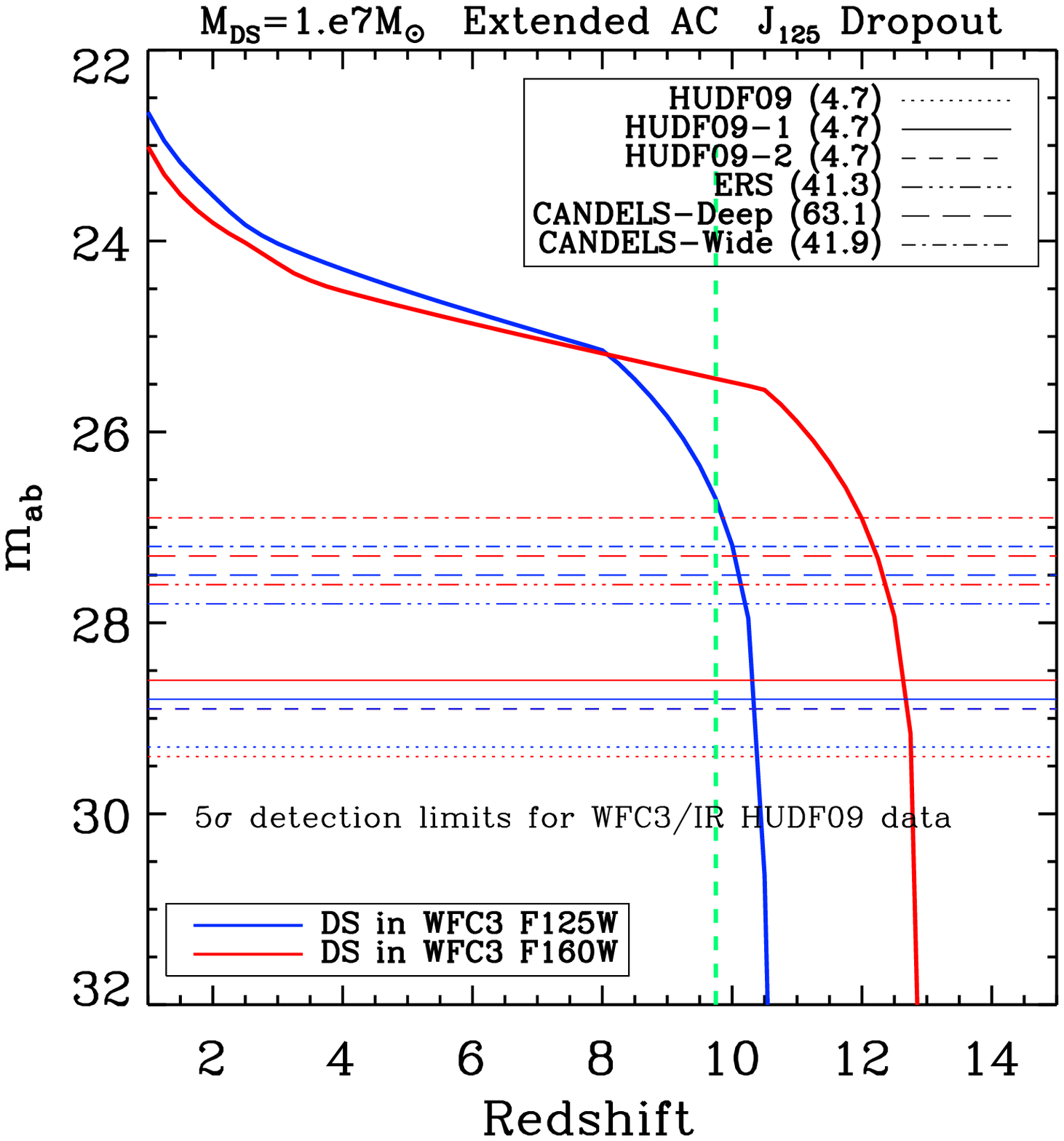}
\end{array}$
\end{center}
\caption{SMDS in HST.  Left (right) panels: Thick curves show apparent
  magnitudes in the H-band [F160W, solid red] and J-band [F125W, blue
  curves]) for the $10^6\Msun$ ($10^7\Msun$) versus the redshift of
  observation for dark star formed via extended adiabatic contraction
  in a $10^7\Msun$ ($10^8\Msun$) halo at redshift of $15$.  Thin horizontal lines indicate  the $5\sigma$ detection limits of the various deep field surveys compiled by \citet{Oeschetal11}, with the areas of the surveys in arcmin$^2$ indicated in the legends. The deepest survey to date is HUDF09 (lowest dotted horizontal lines). The vertical dashed line is placed at the minimum redshift where the J band dropout criterion is satisfied ($z\sim 10$). }
\label{HUDFScansCapOff}
\end{figure*}

\begin{figure*}
    \begin{center}$
      \begin{array}{cc}
        \includegraphics[scale=0.50]{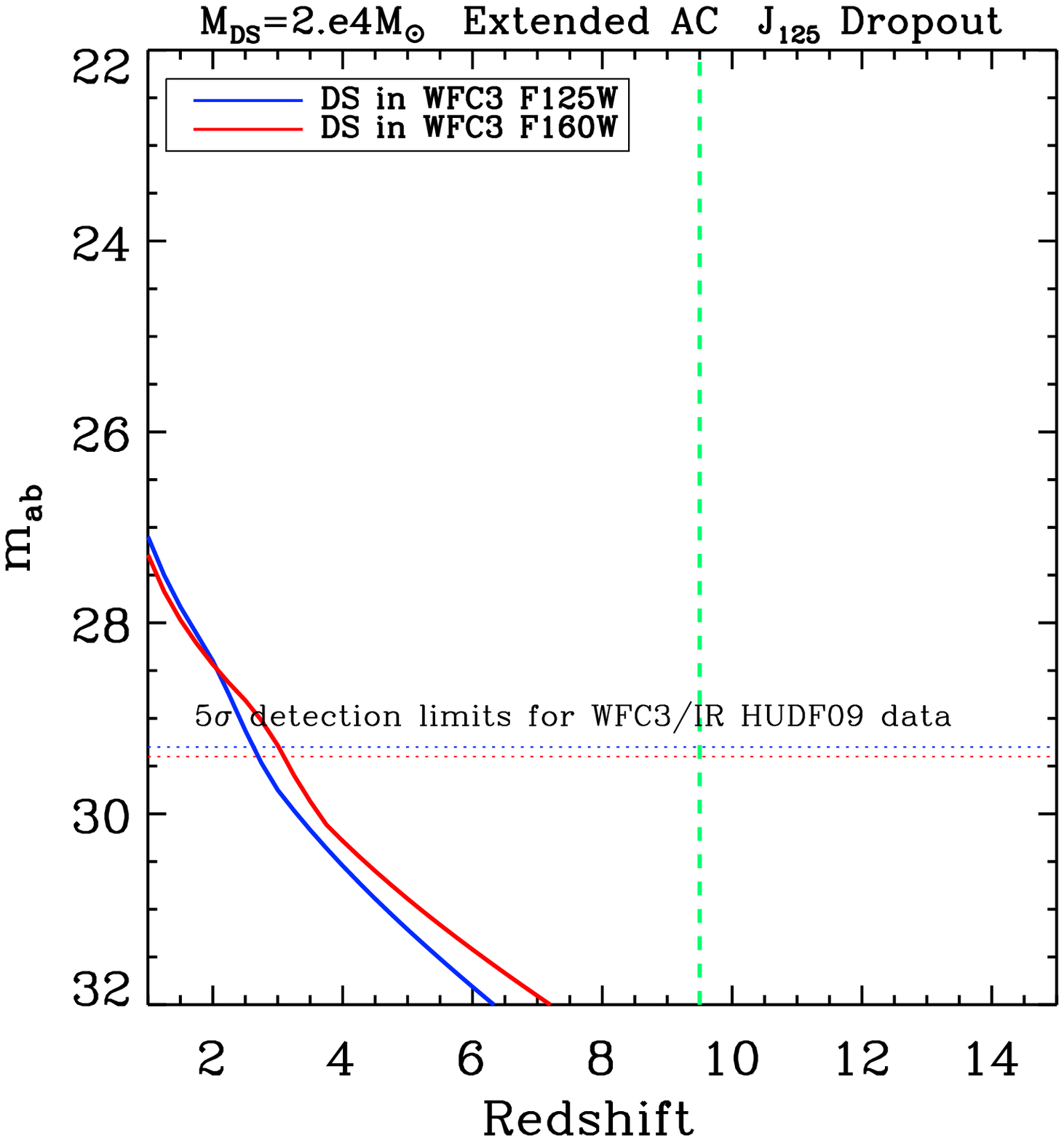}&
        \includegraphics[scale=0.50]{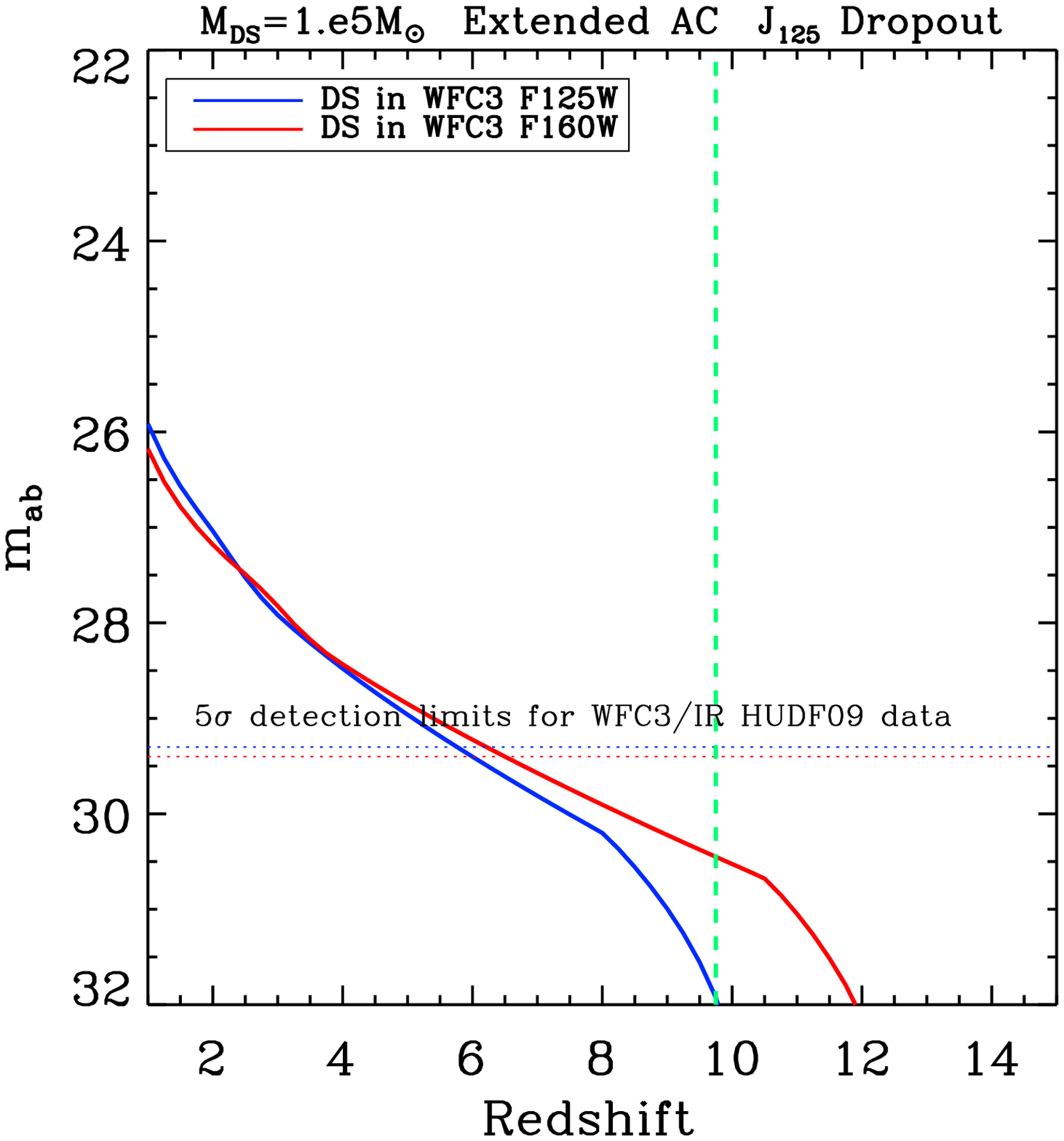}\\
        \includegraphics[scale=0.50]{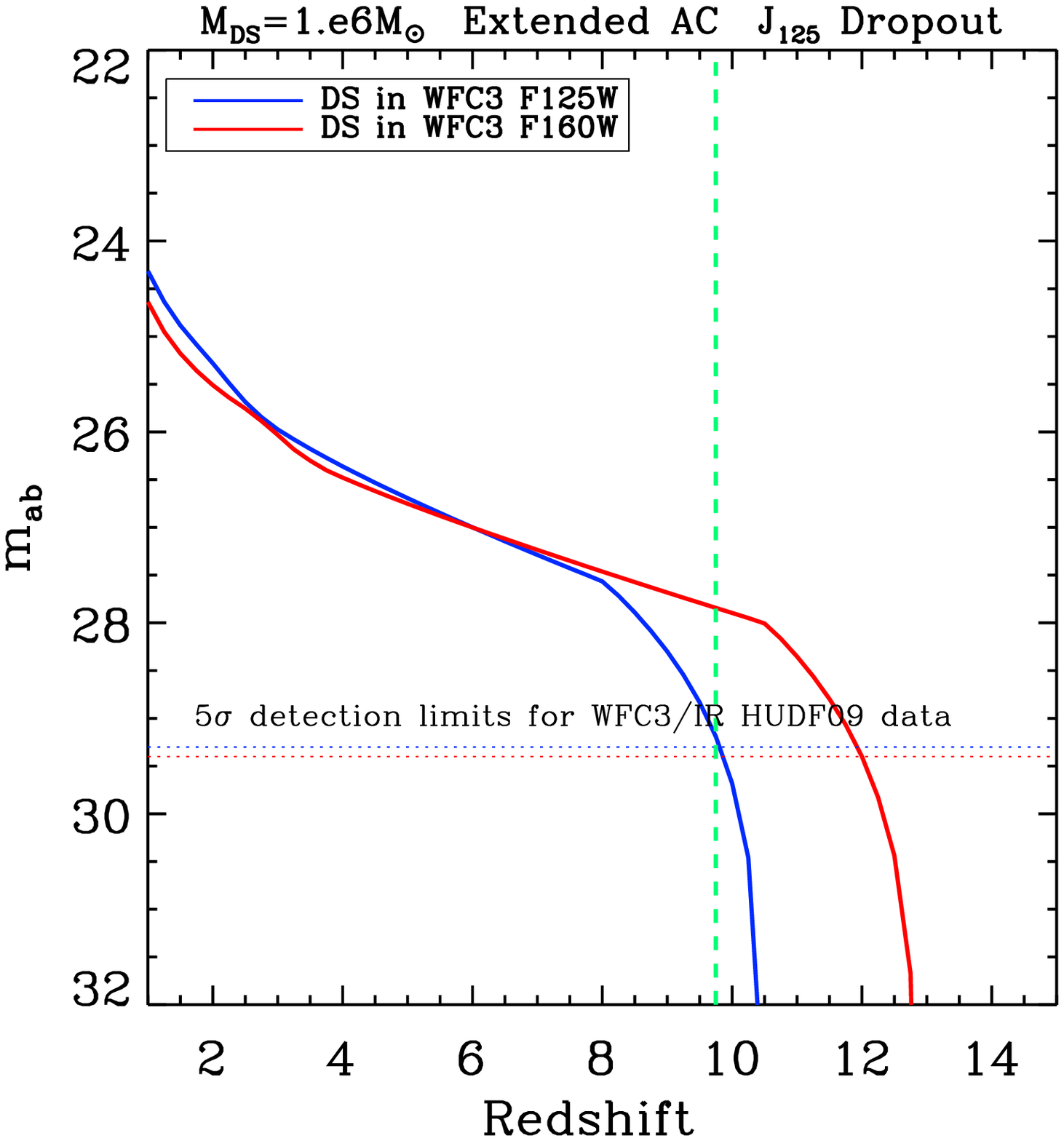}&
        \includegraphics[scale=0.50]{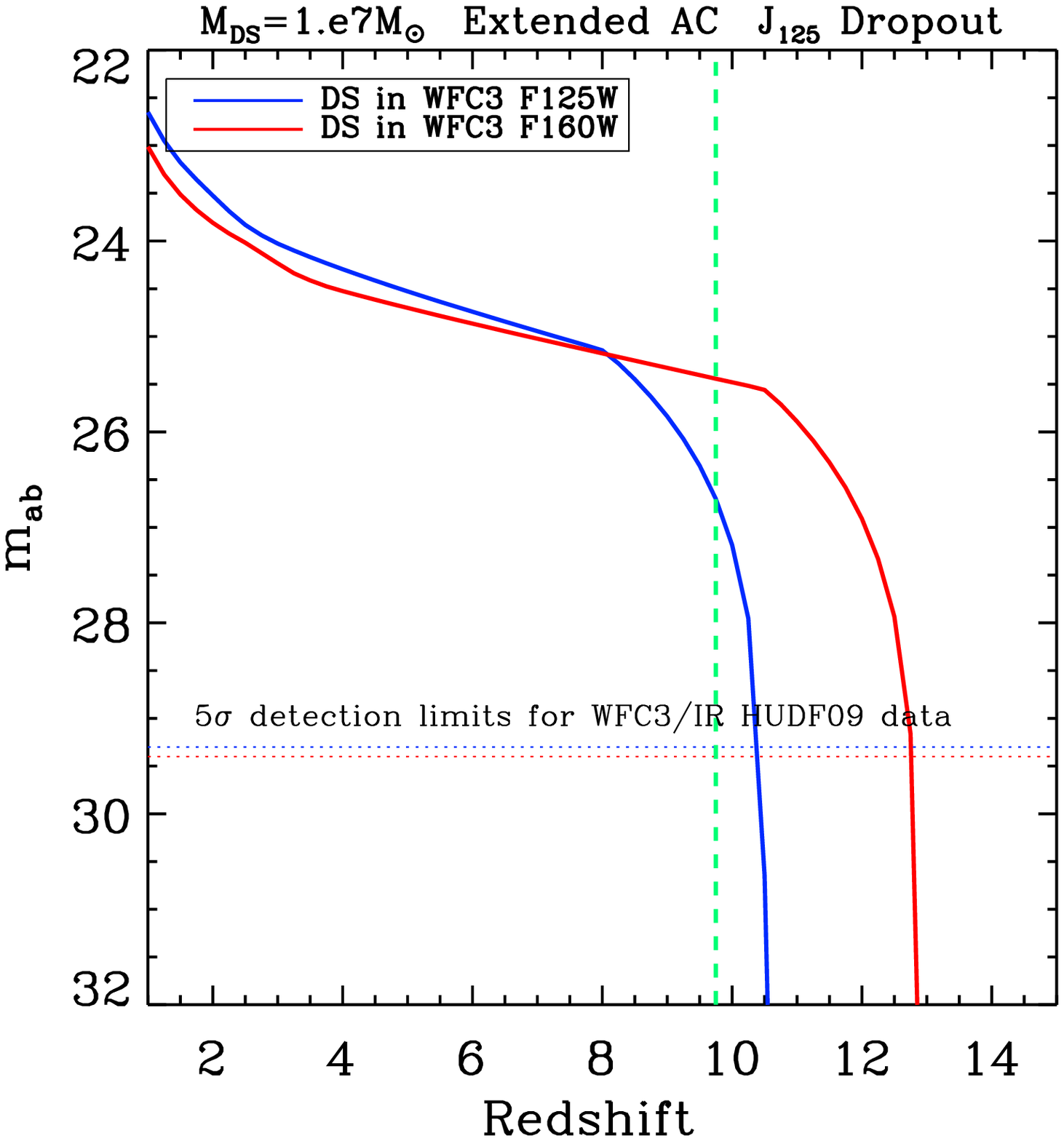}\\
      \end{array}$
    \end{center}
    \caption{SMDS in HST:  \Jband (blue, F125W) and \Hband (red, F160W)) apparent magnitudes $M_{AB}$ for Dark Stars of  mass ranging between $\sim 10^4\Msun-10^7\Msun$ as a function of redshift for WFC3 filters. Here the Dark Stars are considered to be formed via the extended adiabatic contraction mechanism, without any captured DM. The dashed horizontal lines represent the sensitivity limits for the deepest survey available, HUDF09. For the H band the $5\sigma$ depth is $29.4$ whereas for the J band it is $29.3$. The exposure times are $\sim 9.45\times 10^4$ s for the \Jband field and $\sim 1.47\times 10^5$ s for the \Hband field. The green vertical line corresponds to the lowest redshift where the dropout criterion is satisfied. Compared to Figure ~\ref{HUDFScansCapOff} now we explore a wider mass range for the SMDS. Note that SMDS of mass  \tento{5} or lower cannot be observed as J band dropouts with current HST data (another factor of 100 in observing time
would be required) whereas heavier SMDS can be detected.}
\label{HUDFRangeCapOff}
\end{figure*}

\begin{figure*}
\begin{center}$
\begin{array}{cc}
\includegraphics[scale=0.50]{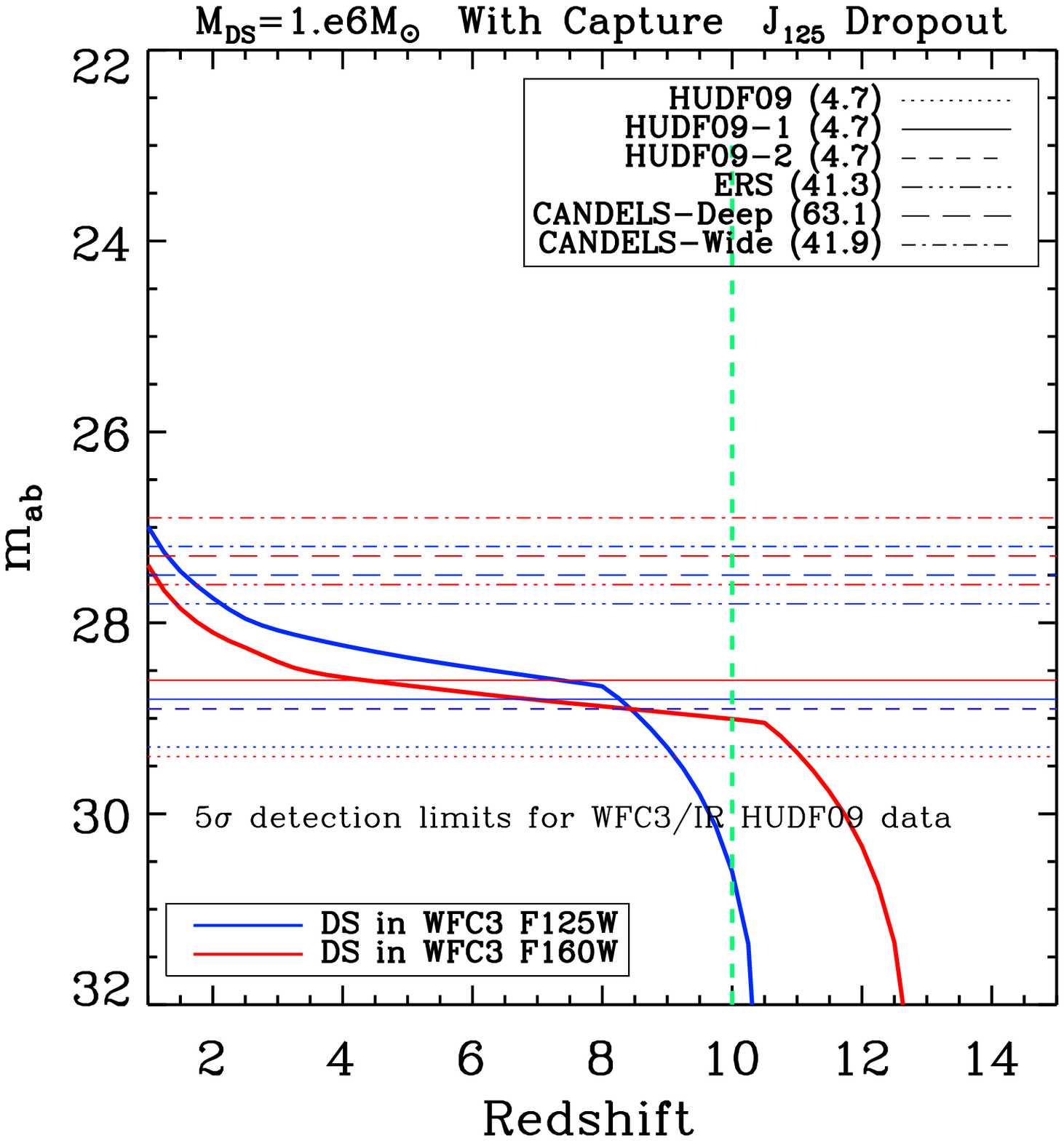}&
\includegraphics[scale=0.50]{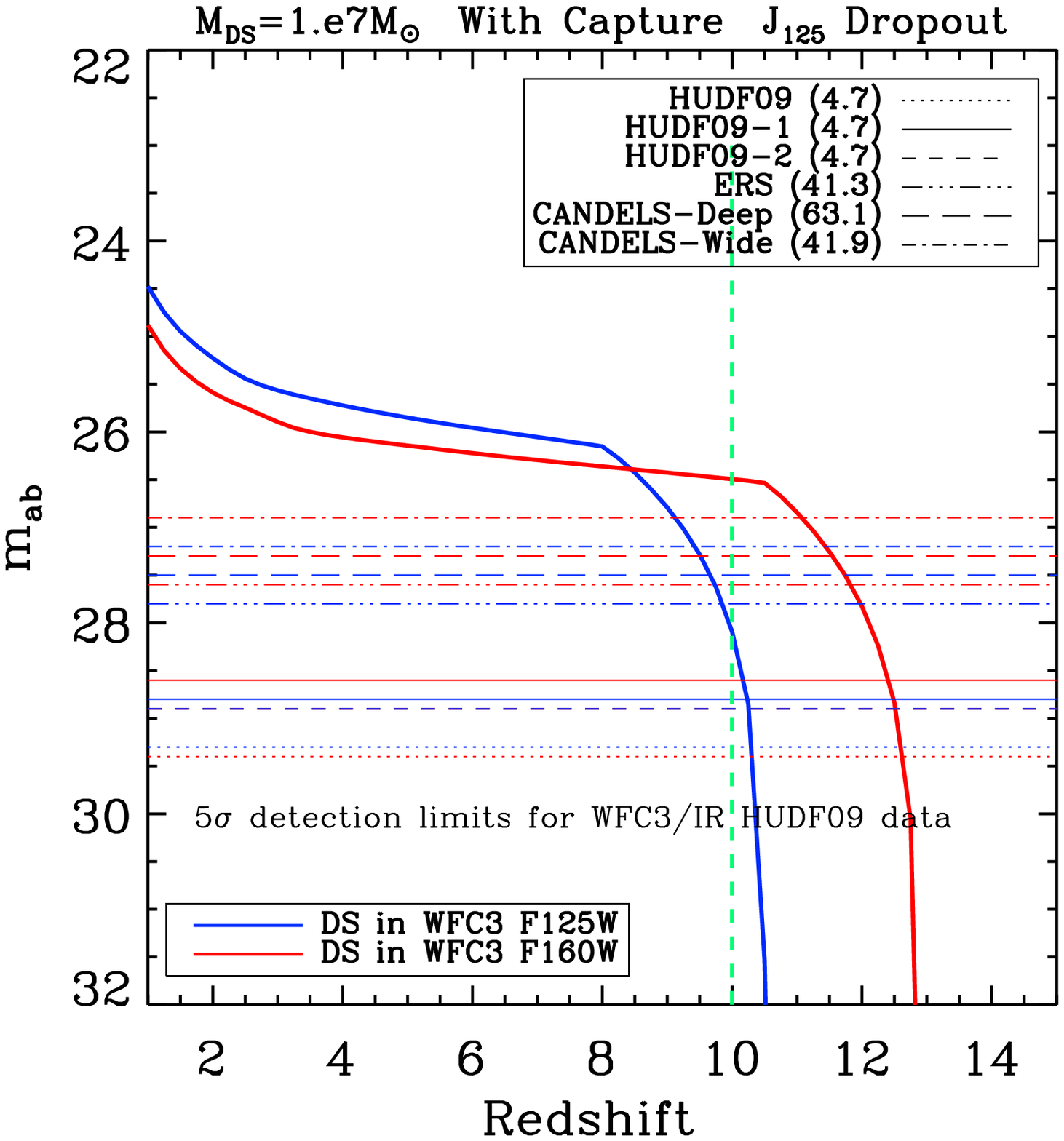}
\end{array}$
\end{center}

\caption{SMDS in HST:  Same as Figure~ \ref{HUDFRangeCapOff} but for $10^6-10^7\msun$ Dark Stars fueled by captured DM.  The dashed horizontal lines represent the sensitivity limits for the deepest survey available, HUDF09.}
\label{HUDFScansCapOn}
\end{figure*}

 Figures~ \ref{HUDFScansCapOff}-\ref{HUDFScansCapOn}  plot the predicted apparent magnitudes of   Dark Stars of $10^4 -10^7 \msun$ at various redshifts  and compare these 
predictions to  sensitivity of various HST surveys  (plotted as thin horizontal lines) in  two HST filters WFC3 F125 (J-band, colored blue) and F160 (H-band, colored red).  
 In these figures, we have assumed that the SMDS  formed 
at z=15 and survived to various redshifts as shown.
In figures~  \ref{HUDFScansCapOff} and \ref{HUDFRangeCapOff}, the Dark Stars are considered to be formed via the extended adiabatic contraction mechanism, without any captured DM;
while in Figure \ref{HUDFScansCapOn}, we consider the case with capture.

  The thick solid curves  show the apparent magnitudes
  $M_{AB}$ for Dark Stars of  various masses  as a function of
  redshift in the  \Jband (F125W,blue) and \Hband (F160W,red).  These solid curves are generated using simulated atmospheres
  spectra from TLUSTY (Fig~\ref{fig:TLUSTYRest}) and redshifting them, 
  ($F_{\nu}(\lambda;z)$), imposing a cutoff at wavelengths lower than the
  Lyman-$\alpha$ if $z\gtrsim 6$, assuming that photons at those
  wavelengths will be absorbed by the neutral hydrogen in the IGM. We use the H and J passbands
  throughput curves ($T^{H,J}(\lambda)$) for the HST WFC3, found at \mbox{\protect{\url{http://www.stsci.edu/~WFC3/UVIS/SystemThroughput/}}}, to compute the observed apparent magnitudes:
              \be\label{eq:mab}
              m_{AB}^{J,H}=-2.5\log\lp\frac{\int d\lambda \lambda
                T^{H,J}(\lambda)F_{\nu}(\lambda;z)}{\int d\lambda \lambda T^{H,J}(\lambda)}\rp+31.4
              \ee

 The  constant $31.4$ is necessary to convert the fluxes to units of
 nJy. $F(\lambda;z)$ is defined by:

\be
F_{\nu}(\lambda;z)=\frac{(1+z)L_{\nu'}(\lambda')}{4\pi D^{2}_{L}(z)},
\ee

where $\lambda$ is the redshifted wavelength,
i.e. $\lambda=(1+z)\lambda'$ and $L_{\nu'}(\lambda')$ is the emitted flux (we
  use TLUSTY to estimate it). The
luminosity distance is labeled by $D_{L}(z)$ and depends on the
chosen cosmology. We define a J-band dropout to be any observation to the right of the green vertical line, corresponding to a  difference in apparent magnitudes of $1.2$ or larger between the  J and H filters as defined in Eq.(\ref{eq:dropout}), (the same criterion as used by \citet{Oeschetal11}).  The location of the green line shows that  J-band drop out technique will also  identify the redshift of any SMDS found in this way to be at $z \sim 10$.

In Figure~ \ref{HUDFScansCapOff} 
 the sensitivity limits from various deep field surveys compiled by \citet{Oeschetal11}: HUDF09, HUDF09-1, HUDF09-2, Early Release Science Data (ERS), CANDELS-Deep and CANDELS-Wide are indicated by different line styles in the legends on the top right of each panel; these
 data are compared to the SMDS case of extended AC (no capture).
 Also shown are the sensitivity limits for various deep field surveys complied in \citet{Oeschetal11}.
In Figure~ \ref{HUDFRangeCapOff} we focus on the most sensitive of these surveys, HUDF09
\footnote{
 HUDF09 has a limiting $5\sigma$ mab of 29.3
    in the J-band for an exposure time
  of 94500 s and 29.4 in the H-band for an exposure time
  of 146711 s for the \Hband band.}.

Similarly, Figure~  \ref{HUDFScansCapOn} 
plots the apparent magnitudes  as a function of redshift for  $10^6 \msun$ (left) and $10^7 \msun$ (right) dark stars which grew via captured DM (rather than via extended AC). 
The SMDS formed with capture are harder to detect: since they are hotter their peak output
is at lower wavelengths (where Ly-$\alpha$ absorption is worse); in addition
their radii are 5-10 times smaller, thus lowering their bolometric luminosities  
 \citep{Freeseetal10}.
In all cases the vertical dashed line is placed at the minimum redshift where the J band dropout criterion is satisfied. 

For SMDS with masses $\leq 10^5 \msun$, the predicted fluxes in both the F125W and F160W filters are too low to be seen in
HST data; the only way around this would be if the object happened to
be gravitationally lensed, as discussed in \citet{Zackrisson2010}. 
The $10^6 \msun$ dark stars can be seen in the F125 (F160) passbands out to redshifts of
9 (11.5) while the $10^7 \msun$  dark star would be detectable out to redshifts of  10.5 ( 13). 
However,  $10^7 \msun$ DSs would be too bright to be compatible with HST data:
they would be several magnitudes brighter than the HST sensitivity, whereas the observed object is just bright enough to be seen. Thus the observed $z=10$ candidate in HST cannot be a $10^7 \msun$ DS.  In addition, if $10^7 \msun$ SMDS formed at higher redshifts, we can place strong bounds on the numbers of them that can survive down to z=10, where they are not found.

We also note that any SMDS that continued to exist to z=6 would have been seen 
 as an $i_{775}$ dropout in HUDF which has a 29.9 $m_{AB}$ detection
 limit for $10\sigma$ detection in the $i_{775}$ passband
 ~\citep{Bouwensetal06}.  Since no candidates exist in the data, this
 makes it clear that SMDS did not survive to z=6. Thus we conclude that it is the $10^6 \msun$ SMDS that serve as the best possible explanation for the J-band dropout at z=10 seen by HST.

\subsection{Using HST observations to constrain the numbers of dark stars}\label{sec:BoundsHST}


We will use  HST data to constrain the fraction
$\mathrm{f_{SMDS}(z_{start})}$ of early DM halos that can host
SMDS.
We focus on SMDS of masses $M_{DS} = 10^6 - 10^7 M_\odot$ since lower mass DS
are not observable in current HST data (unless they are significantly magnified by gravitational sensing or if they form clusters of dark stars 
\citep{Zackrisson2010b}).

Following \citet{Zackrisson2010b, ER:Zackrisson2010b}, we compute the number $N_{obs}$ of DS that  could potentially be observed,
\be\label{maxf}
N_{obs} = \frac{dN}{dz d\theta^2} f_{SMDS}(z=z_{start})\theta^2 f_{surv}f_{\Delta t} 
\ee
and use the fact that  at most one object has been observed with HST at z=10 to obtain bounds on $\mathrm{f_{SMDS}(z_{start})}$, the fraction of DM halos in a given mass range that can host a DS:
\be\label{maxf2}
N_{obs}^{HST} <1
\ee
Here $dN/dz d\theta^2$ is the number of DM halos forming per unit redshift per \arsq in which a given mass DS is hosted (computed from Figure~\ref{FRates1}). We have multiplied by unit redshift interval
 $\Delta z = 1$, since we only consider SMDS formed within a redshift interval equal to one (see the discussion following Eqn. (1)).  Here $\theta^2$ is the total area surveyed in which the SMDS could have been detected, \fsurv is the fraction of DS that survives from the redshift where the DS starts forming, \zstart, until it could be observed as a dropout (at $z \sim 10$ with HST) and \fdt is the fraction of the observational window of  time $\De t$ during which the DS is still alive. Here, $\De t$ is the cosmic time elapsed between the minimum and maximum redshift where the DS could be observed as a dropout.  Please note that those redshifts are different from $\zmin$ and $\zmax$ defined under Equation~ \ref{convert}. For the case of HST, we get $\De t = 6.5 \times 10^7$yr  (the cosmic time between the minimum redshift of $9.5$ and maximum redshift of $10.5$ where the DS could be observed as a J-band dropout computed using Equation \ref{time}).

We estimate the survey area $\theta^2$ in the following way:
For each of the surveys in Figs. \ref{HUDFScansCapOff} and \ref{HUDFScansCapOn}, we have indicated (in parentheses in the plots) the
 area (in arcmin$^2$) observed by the survey.  For DS of a given mass, we can add up the areas
 of all those surveys which are capable of observing DS as J-band dropouts to obtain  a total
 effective area of observability for that DS mass.  In other words, we add the area of all surveys in which the fluxes in the \Hband are still above the sensitivity limits while the fluxes in the \Jband are a least $1.2$ lower in apparent magnitude and below the detection limit of the J-band.
From Figure~ \ref{HUDFScansCapOff} we  estimate $\theta^2= 4.7\times 3$ \arsq as the effective area of the  surveys in which a $10^6\Msun$ SMDS formed via extended AC could have been observed as a J band dropout with HST, since its only for the three deepest surveys, each with an area of 4.7 \arsq, that this SMDS would show up as a dropout.   
For the \tento{6} SMDS  formed via captured DM the detectability is much lower, implying that they could have been observed with HST WFC3 as a J band dropout only in the deepest survey, namely in HUDF09, which has an area of 4.7 \arsq. 

Although the z=10 J-band dropout seen by HST cannot be a $10^7 \msun$ SMDS (as it would be too bright and would show up in both bands), 
still we can apply Eq. (\ref{maxf}) to place an upper bound the numbers of these objects. 
For the $10^7\Msun$ stars formed via extended AC,, this area is increased to $\sim 160$ \arsq, as all surveys compiled could pick this object up as a J band dropout. For the hotter DS fueled by captured DM, we can see from Figure~ \ref{HUDFScansCapOn} that the total area of the surveys in which $10^7 \Msun$ DS could have been detected is $\sim 160$ \arsq 
(similar with the area for the extended AC DS of the same mass). 

 We comment here on the three redshifts of formation we have chosen.  For a conversion between \zform (redshift where the DS reaches its final mass) and \zstart (the redshift where the DS starts accreting baryons) see Table \ref{tb:Rates}. 
 \begin{itemize}
\item Case A: \zform$=10$. Here, we assume that the DS become supermassive only at z=10 and not before.
 We can only constrain the product $f_{SMDS}\times f_{surv}\times f_{\De t}$. The fraction of the observational window during which the DS is alive and can be observed, is $f_{\De t}=\min(\tau-\tau_{min}, \De t)/\De t$, where $\tau_{min}$ is the minimum DS lifetime  that allows the DS to survive to $z=10.5$ where it can be observed as a J-band dropout with HST.  In the case of a \tento{7} SMDS $\tau_{min}\sim 1.15\times10^8$yrs (time elapsed between $z=13$ and $z=10.5$) whereas for the \tento{6} SMDS $\tau_{min}\sim 3.6\times10^7$yrs (cosmic time elapsed between $z=10.7$ and $z=10.5$).
We note that the limits  we place on $\mathrm{f_{SMDS}(z_{start})}$ are only valid at \zstart$\sim 13$ (for the \tento{7} SMDS) and \zstart$\sim 11$ (for the \tento{6} SMDS) as can be seen from Table \ref{tb:Rates}. 
 
 \item Case B:  \zform$=12$. Here we consider the DS to become supermassive at \zform$\sim 12$ and not at later
redshifts. We will assume that the DS could survive until $z\sim 10$ ($f_{surv}=1$) in order to constrain $f_{SMDS}(z_{start})$ using null detection from HST \Jband dropouts. From Table \ref{tb:Rates} we see that the \zstart value for the \tento{7} SMDS in this case is $\sim 16$ and for the \tento{6} SMDS it is $\sim 13$.  In the case of a \tento{7} SMDS $\tau_{min}\sim 2.0\times10^8$yrs whereas for the \tento{6} SMDS $\tau_{min}\sim 1.1\times10^8$yrs.

\item Case C:  \zform$=15$. Here we assume the dark stars become supermassive by \zform$\sim15$. The values for \zstart can be read off from Table \ref{tb:Rates} again. For the \tento{7} SMDS \zstart$\sim 22$ and for the \tento{6} SMDS \zstart$\sim 16$. This case is treated in a similar fashion as case B. For the \tento{7} SMDS, $\tau_{min}=2.9 \times 10^8$yrs (the time elapsed between redshifts 22 and 10.5)  whereas for the \tento{6} SMDS $\tau_{min}=2.0\times 10^8$yrs (the time elapsed between redshifts 16 and 10.5).

\end{itemize}

 From Eqn. (\ref{maxf}), we obtain the following bounds for $10^7
 \msun$ SMDS formed via either extended AC or with capture in each of
 the three cases (A-C):

\be\label{bounds1e7all}
\log f_{smds}(M_{DS}= 10^7\Msun)\leq\left\{
\begin{array}{ll}
-4.5-\log(f_{surv}\times f_{\De t}), &\mathrm{A}\\

-3.4-\log(f_{surv}\times f_{\De t}) , &\mathrm{B}\\

-2.1-\log(f_{surv}\times f_{\De t}) , &\mathrm{C}\\  

\end{array}\right.
\ee

For $10^6 \msun$ SMDS formed via extended AC we get the following limits:

\be\label{bounds1e6all}
\log f_{smds}(M_{DS}= 10^6\Msun)\leq\left\{
\begin{array}{ll}
-4.8-\log(f_{surv}\times f_{\De t}) , &\mathrm{A}\\ 

-4.6-\log(f_{surv}\times f_{\De t}) , &\mathrm{B}\\

-3.8-\log(f_{surv}\times f_{\De t}) , &\mathrm{C}\\  

\end{array}\right.
\ee

\begin{figure*}
\begin{center}$
\begin{array}{cc}
\includegraphics[scale=0.50]{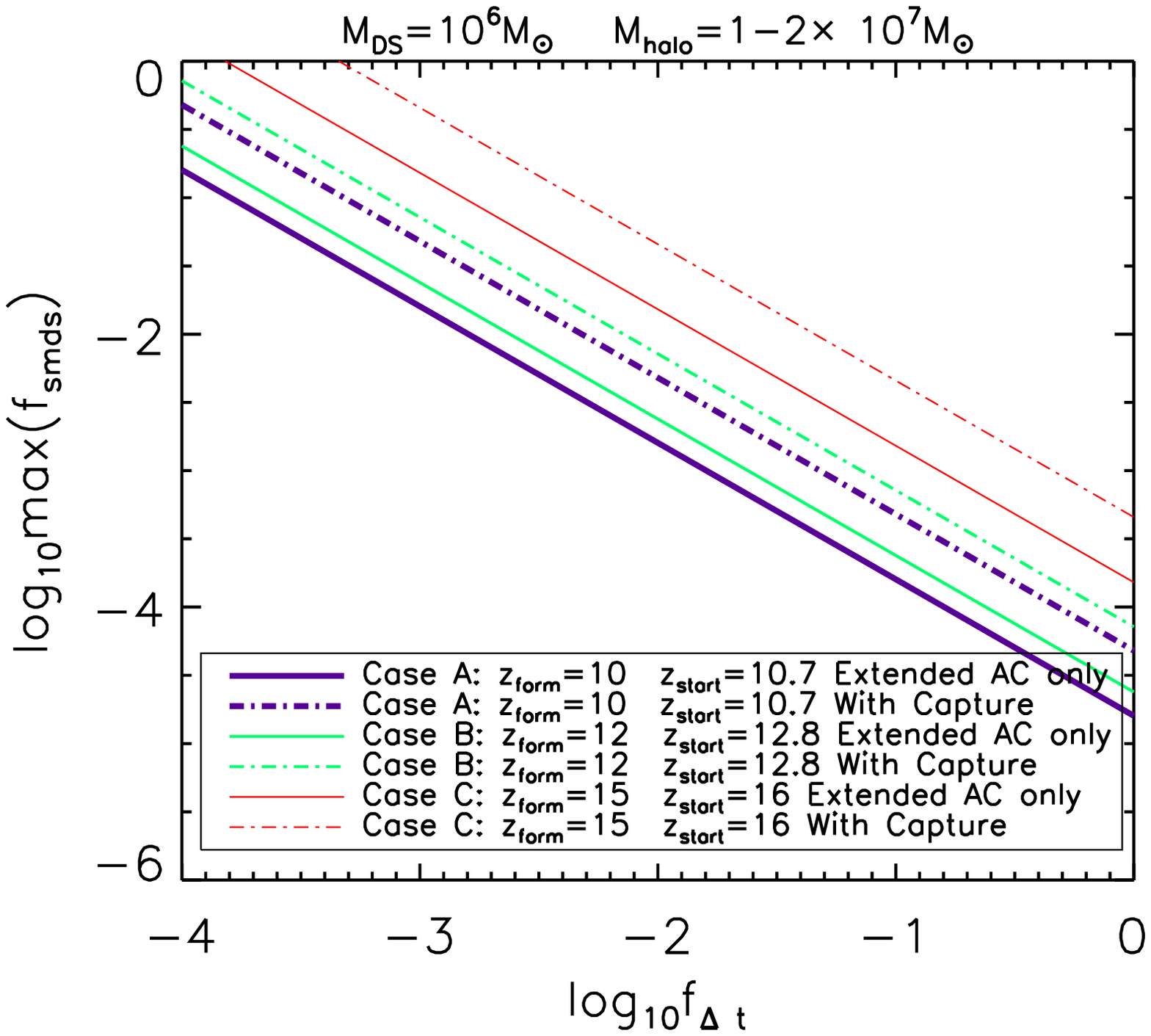} &
\includegraphics[scale=0.50]{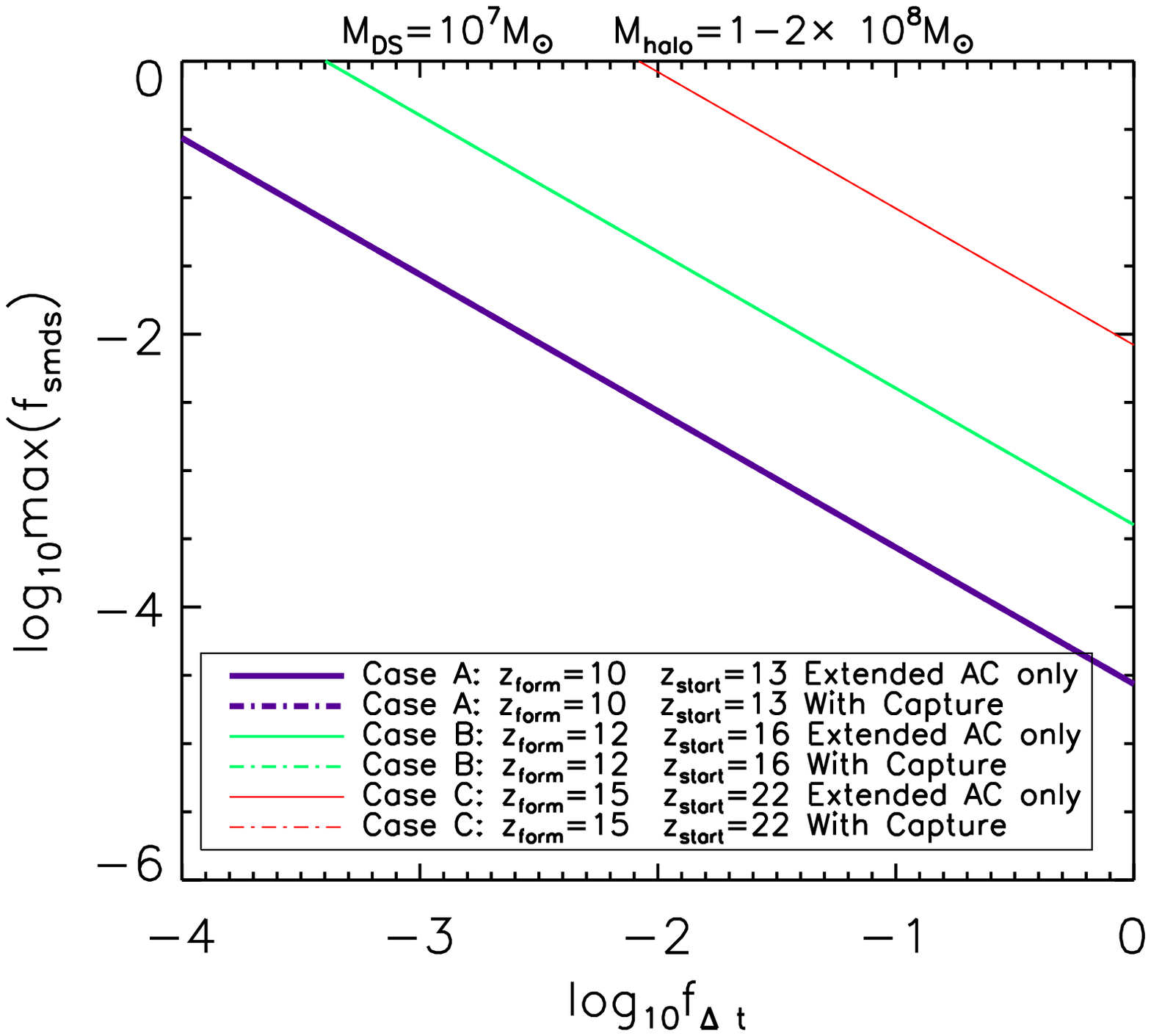}\\
\end{array}$
\end{center}
\caption{Upper bounds from HST on the fraction \fsmds of early halo hosting dark stars of masses as labeled above the plots; values above the lines are excluded.   Different lines correspond to different values of the redshift (\zform$=10, 12,$ and $15$) at which
 the DS attains this mass; see Table ~\ref{tb:Rates}  for the connection between (a) the redshift $z_{start}$ 
 at which the DS came into existence and started to grow and (b) the redshift $z_{form}$ at
 which it reached the supermassive size as labeled.  This plot assumes  \fsurv=1 (i.e. the DS survives long enough to reach the redshift window of observability as a \Jband dropout
 with HST).  However, the DS
 need not survive throughout the entire window; in fact
the horizontal axis in both plots is $\log_{10}f_{\De t}$, for which a value of $0$ corresponds to the DS lifetime being sufficiently large that it survives throughout the redshift window of observability.  Solid lines correspond to DS formation via extended AC (without capture) while dashed lines correspond to DS formation via capture. 
Since DS less massive than $10^6 M_\odot$ are too faint to be detected by HST,
these data do not bound \fsmds for lower mass dark stars.}

\label{FMax}
\end{figure*}

The  values of \zstart that correspond to these values of $z_{form}$ can be found in the last three rows of Table \ref{tb:Rates}. 
The reason that the bounds on the numbers of $10^6\msun$ SMDS are tighter than those on the $10^7 \msun$ SMDS is the following.
In order to reach a larger mass, the DS had to start forming at an earlier redshift and in larger halos; but the numbers of larger halos that
can host DS is smaller at higher redshifts.  Similarly, the bounds in Case A are $\sim 10$ ($\sim 300$) times stronger than the bounds 
in Case C for the \tento{6} (\tento{7}) SMDS. Again the reason for the
very large discrepancy $\sim 300$ is  the fast decrease of the
formation rate of $1-2\times$\tento{8} DM halos at redshifts higher
than $z\sim 15$ as can be seen from Figure~ \ref{FRates1}.  For SMDS lighter than \tento{6} HST cannot be used to place constraints, as those objects are not detectable  with HST as J-Band dropouts. A summary of our bounds  can be found in Figure~\ref{FMax} where we plot the exclusion limits for \fsmds.

\subsection{Other Bounds on Numbers of SMDS}\label{sec:OtherBounds}

Further bounds on the numbers of DS and the halos they form in should result from a  variety
of considerations.  One would be the contribution to reionization.
Work of \citet{Venkatesan:2000} studied stellar reionization with the standard fusion powered first stars (Population III), without any dark stars. 
From comparison with the optical depth to last scattering from early WMAP data, she
bounded the fraction of baryons in halos that can cool and form stars (assuming a Scalo initial 
mass function) to be in the range $f_* \sim 0.01-0.1$.  However, it is not clear how these numbers
would change in the presence of DS and with the updated value for the
optical depth from WPMAP7 \citep{WMAP7}.

More recently, the effects of DS (and the
resultant main sequence stars) on reionization was studied by \citet{Schleicher:2008,Schleicher:2009} and \cite{Scott:2011}.  While DS are fully DM powered, they remain
so puffy and cool that no ionizing photons are produced, and there is no contribution to 
reionization. However, once the DM fuel begins to run out, they contract and heat up as they approach the zero age main sequence (ZAMS) with the onset of fusion, at which point they do produce ionizing photons. For the case of extended AC, and
for DS less massive than 1000$\msun$, \cite{Scott:2011} concluded that the reionization history
of the Universe is unaffected by the DS, compared to  the  case of more standard Pop III stars: the 
DS period of no ionizing photon production is compensated by a short period of high ionizing photon production during approach to the ZAMS.  However, we are not sure what the effect on reionization would be in the case 
of the more massive SMDS.  
On the one hand the more massive stars are hotter and brighter and would emit substantial 
amounts of ionizing photons; on the other hand the more massive the star, the shorter the
lifetime.  

For the case of DS with high capture rates, 
previous studies 
 \cite{Scott:2011} find that reionization is somewhat delayed, decreasing  the integrated 
 optical depth to the surface of last scattering of the CMB.  However, variation of astrophysical
 parameters for the case of standard reionization with standard Pop III stars can produce exactly
 the same effect, so that disentangling these effects will prove difficult.    Nonetheless
 \cite{Scott:2011} do argue that they can rule out the section of parameter space where
 dark stars $\sim 1000 \msun$ with high scattering-induced capture rates tie up more than 90\% of all the first star-forming baryons and live for more than 250 Myr.  Again, their work should be
 extended to the heavier SMDS we study in this paper.

A complicating factor (for both the cases of extended AC and capture) 
is that the SMDS do eventually collapse to BH, and it's not clear how rapidly 
that happens.  If the collapse to BH is rapid, this may cut short the ZAMS phase and
reduce the role SMDS play in reionization.  Second, the SMDS are
likely to have stellar pulsations \citep[][in progress]{Montgomery:2011};
as a consequence it is possible they will lose some mass before reaching the ZAMS.  Third,
even after joining the ZAMS, en route to BH collapse, the SMDS may
blow off some of their mass (\citet{Umeda:2009} suggest 1/2 of their mass).

Alex Heger (private communication) has the following new results
 for early stars (only made of hydrogen and helium) that are nonrotating:
If they are heavier than 153,000 $M_\odot$, no hydrostatic equilibrium solution
exists, i.e. no primordial hydrogen burning star exists. Thus once a fusion
powered star accretes enough mass to get heavier than this, then it
collapses straight to a BH.  For any of our dark stars that are heavier
than this, once they run out of DM, they collapse directly to BH without
contributing at all to reionization.
Rotation might change these results.

 Further, there are implications of DS regarding the fraction of baryons that end up in DSs.  
 Our work assumes that the DS can grow in a DM halo of a given mass until
 almost al the baryons in the halo (assumed to be the baryonic mass fraction in the Universe)
 are accreted onto the DS.  If the total fraction of halos in which such DSs form is too
 large, this implies that most of the baryons in the Universe are trapped inside DS and and it is not clear  how they would contribute any further to galaxy formation.  As mentioned above,
 en route to BH collapse, the SMDS may blow off some of their mass,  reinjecting baryons
 into the surrounding halos and alleviating this problem somewhat.

Further bounds on the numbers of dark stars have been studied in \citet{Sandick:2010}
The remnant black holes from the DS should still exist today, including inside the Milky Way.  They still
have enhanced amounts of DM around them, known as DM spikes.  The DM inside the spikes annihilates
to a variety of final products, with $\gamma$-rays that would be detected by the Fermi Gamma Ray Space Telescope
(FGST).  In \citet{Sandick:2010}, it was noted that most of the 368 point sources 
observed by FGST might in fact be due to DM annihilation in the spikes.  In addition, FGST data were used
to place bounds on the fraction of early haloes hosting DS to avoid overproduction of gamma-rays from annihilation
in the remnant DM spikes. The bounds range from $f_{DS} < 10^{-3} -  1$, depending on the WIMP mass
and annihilation channel.

All of these considerations are beyond the scope of this paper.  For now we take these
arguments to imply that not every early halo can contain a DS.

  \section{Observing Supermassive Dark Stars with JWST}

Dark stars can be detected by upcoming James Webb Space Telescope (JWST). 
Table \ref{tb:JWST_Filters} gives a summary of the sensitivity of the NIRCam and MIRI 
cameras on JWST in various wavelength bands
\footnote{Specifically we show the $10 \sigma$ required
$m_{AB}$ sensitivities for the NIRCam and MIRI wide filters after
$10^4$ seconds exposure derived based on the limiting fluxes published
at \url{http://www.stsci.edu/jwst/instruments/miri/sensitivity/} and
\url{http://www.stsci.edu/jwst/instruments/nircam/sensitivity/table*}.
One can scale the limits to different exposure times, as the limiting
flux $\propto t_{exposure}^{-1/2}$ and converting to $m_{AB}$
magnitudes is just a matter of applying Equation \ref{eq:mab}. For
instance an increase by a factor of $100$ in exposure times would
convert in a gain in the sensitivity limits by $2.5$ AB magnitudes. }.
One can see that the NIRCam is much more sensitive than the MIRI filters, so
that light emitted at wavelengths larger than  5 microns is harder to observe.
 In this section we estimate the number of SMDS that would show up in a typical survey with JWST NIRCam or MIRI cameras, based on the bounds we have just derived in the previous section. 

\begin{table*}
  \begin{center}
    \begin{tabular}{cccccc}
      \hline\hline
            & Filter & $\lambda_{center}$(\microm{} )& $\log_{10}\lambda_{center}$ & $\Delta \lambda$ (\microm{})& $m_{AB}$\\
      \hline 
    NIRCam& F070W    & 0.7      &-0.15                      & 0.175                         & 28.1 \\
          & F090W    & 0.9            &-0.05                   &0.225                          &28.51\\
          & F115W    & 1.15          &0.06                      &0.2875                          &28.72\\
          & F150W    & 1.5            & 0.17                   &0.375                          &28.77\\
          & F200W    & 2.0            &0.30                   &0.5                           &28.75\\
          & F277W    & 2.77           &0.44                  &0.6925                          &28.67\\
          & F356W    & 3.56           &0.55                   &0.89                          &28.55\\
          & F444W    & 4.44           & 0.65                  &1.11                          &27.92\\
          \hline
      MIRI& F560W    & 5.6           &  0.75                 & 1.2                         & 25.65 \\
         & F770W    & 7.7               &  0.89            &2.2                          &25.28\\
         & F1000W    & 10.0           &1.00                   &2.0                          &24.29\\
         & F1130W    & 11.3           &1.05                  &0.7                         &23.32\\
         & F1280W    & 12.8           &1.1                  &2.4                          &23.53\\
         & F1500W    & 15.0           &  1.18               &3.0                          &23.26\\
         & F1800W    & 18.0           &  1.25              &3.0                          &22.31\\
         & F2100W    & 21.0           &1.32               &5.0                          &21.56\\
         & F2550W    & 25.5           &   1.4           &4.0                          &20.28\\

         \hline
    \end{tabular} 

    \caption{$10 \sigma$ sensitivity limits for the JWST wide passband
      filters (fourth column). The $m_{AB}$ limits are derived
      assuming $10^4$ seconds exposure and are based on the limiting
      fluxes published at
      \protect\url{http://www.stsci.edu/jwst/instruments/miri/sensitivity/}
      and
      \protect\url{http://www.stsci.edu/jwst/instruments/nircam/sensitivity/table*}. We
      identify each filter by its name, in the first column. Values in
      the second column correspond to the center wavelength of each
      filter, whereas in the fourth column the values for the passband
      width are given.} 
\label{tb:JWST_Filters}
  \end{center}
\end{table*}

\begin{figure*}
\begin{center}$
\begin{array}{cc}
\includegraphics[scale=0.50]{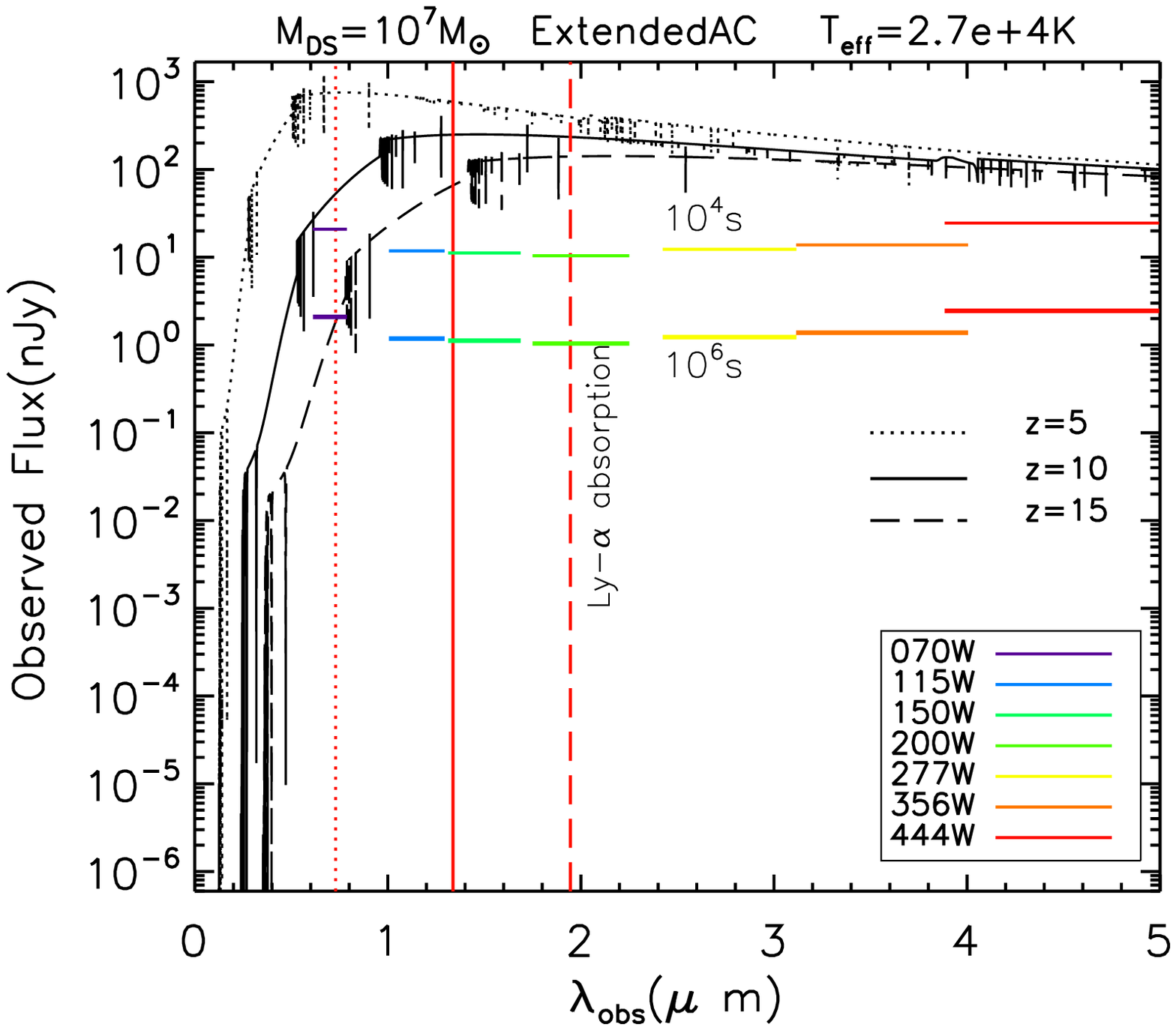}&
\includegraphics[scale=0.50]{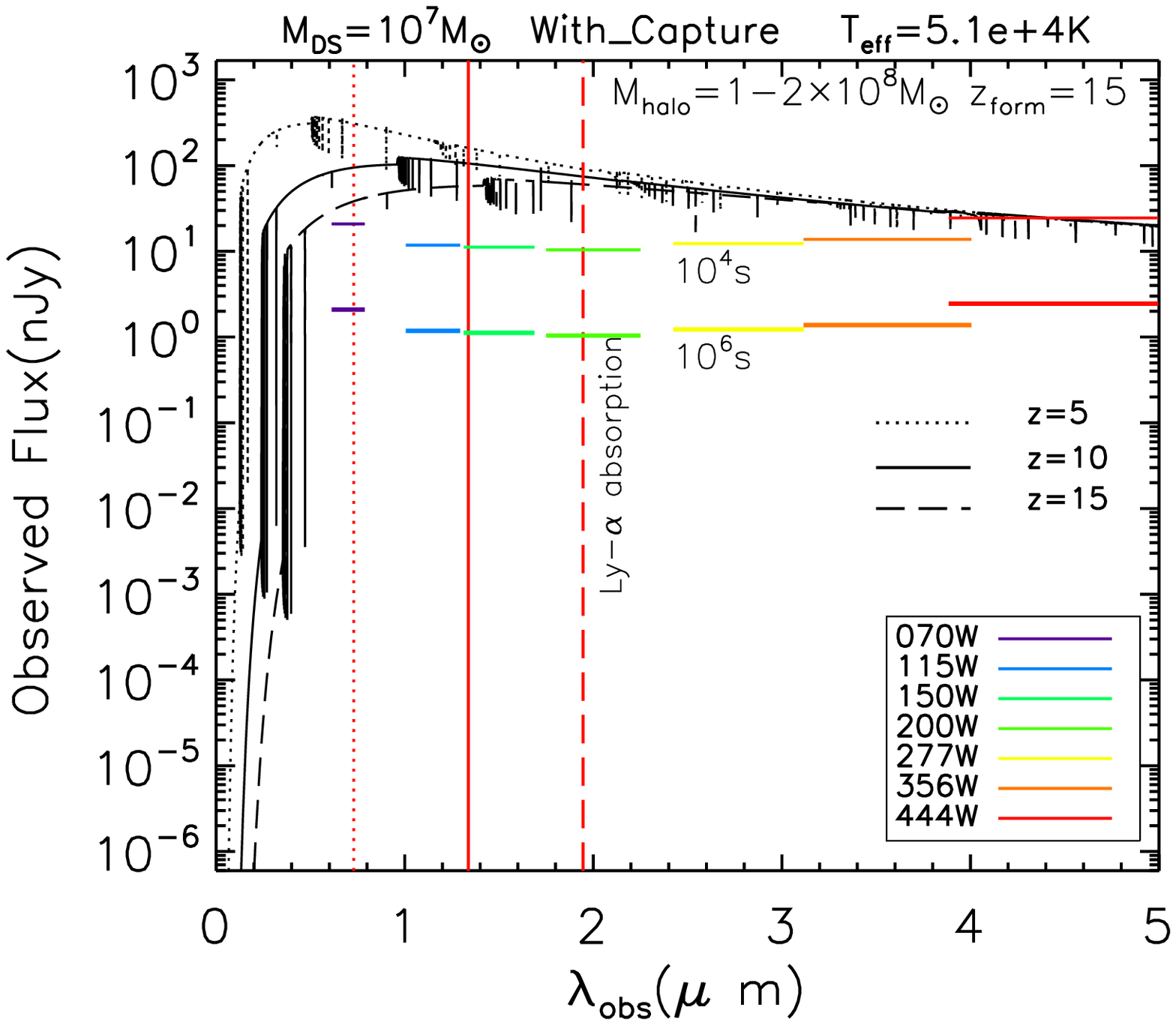}\\
 & \\
\includegraphics[scale=0.50]{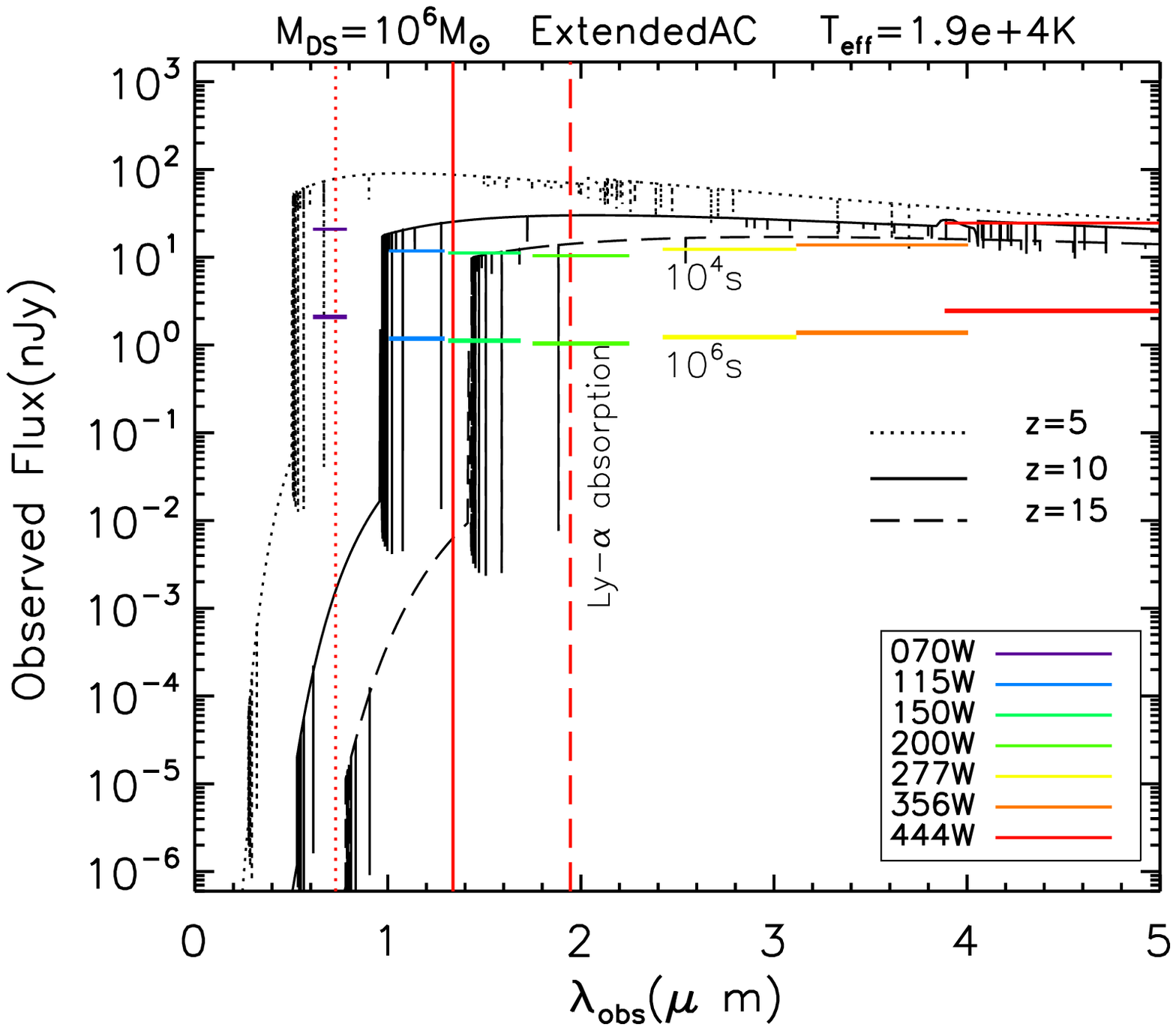}&
\includegraphics[scale=0.50]{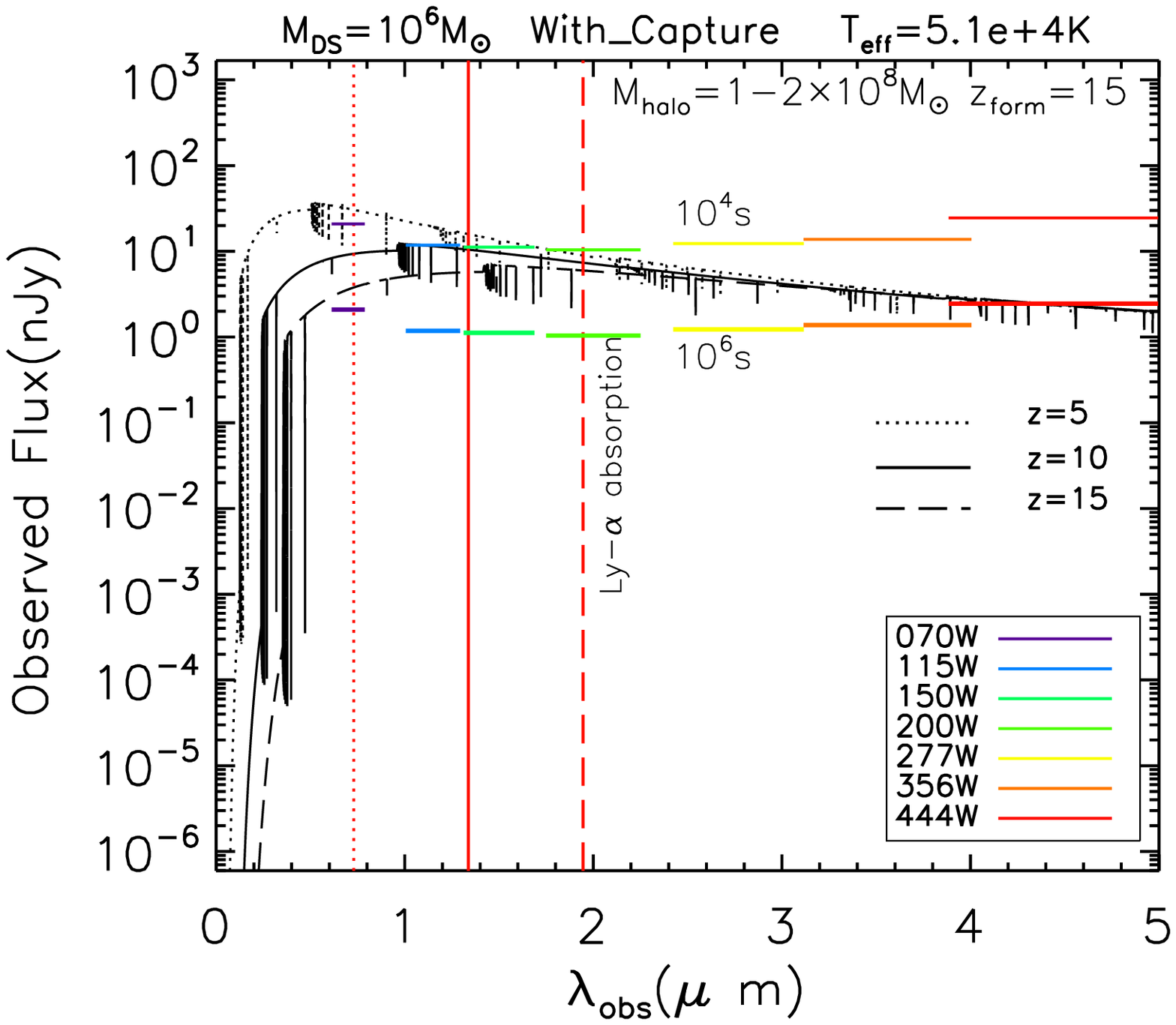}\\
\end{array}$
\end{center}
\caption{Spectra for supermassive dark stars formed at \zform$=15$ compared
with sensitivity of JWST filters. Listed above each panel are the mass of the DS in solar masses, the formation mechanism (extended AC or "with 
capture") and the surface temperature $T_{eff}$.
The fluxes are shown at $z=15$ (dashed line), $10$ (solid line) and $5$ (dotted line) and compared to the detection limits of NirCam wide passband filters. The colored horizontal lines represent the sensitivity limits for the filters as labeled in the legend for exposure times
$10^4$s (upper lines) and  $10^6$s (lower lines).  IGM absorption will decrease the observed fluxes for wavelengths shortward of the vertical red lines, which indicate the Ly-$\alpha$ 
line (1216\AA) redshifted  from the rest-frame of the star.}
\label{TLUSTYRedshifted}
\end{figure*}

\begin{figure*}
\begin{center}$
\begin{array}{cc}
\includegraphics[scale=0.50]{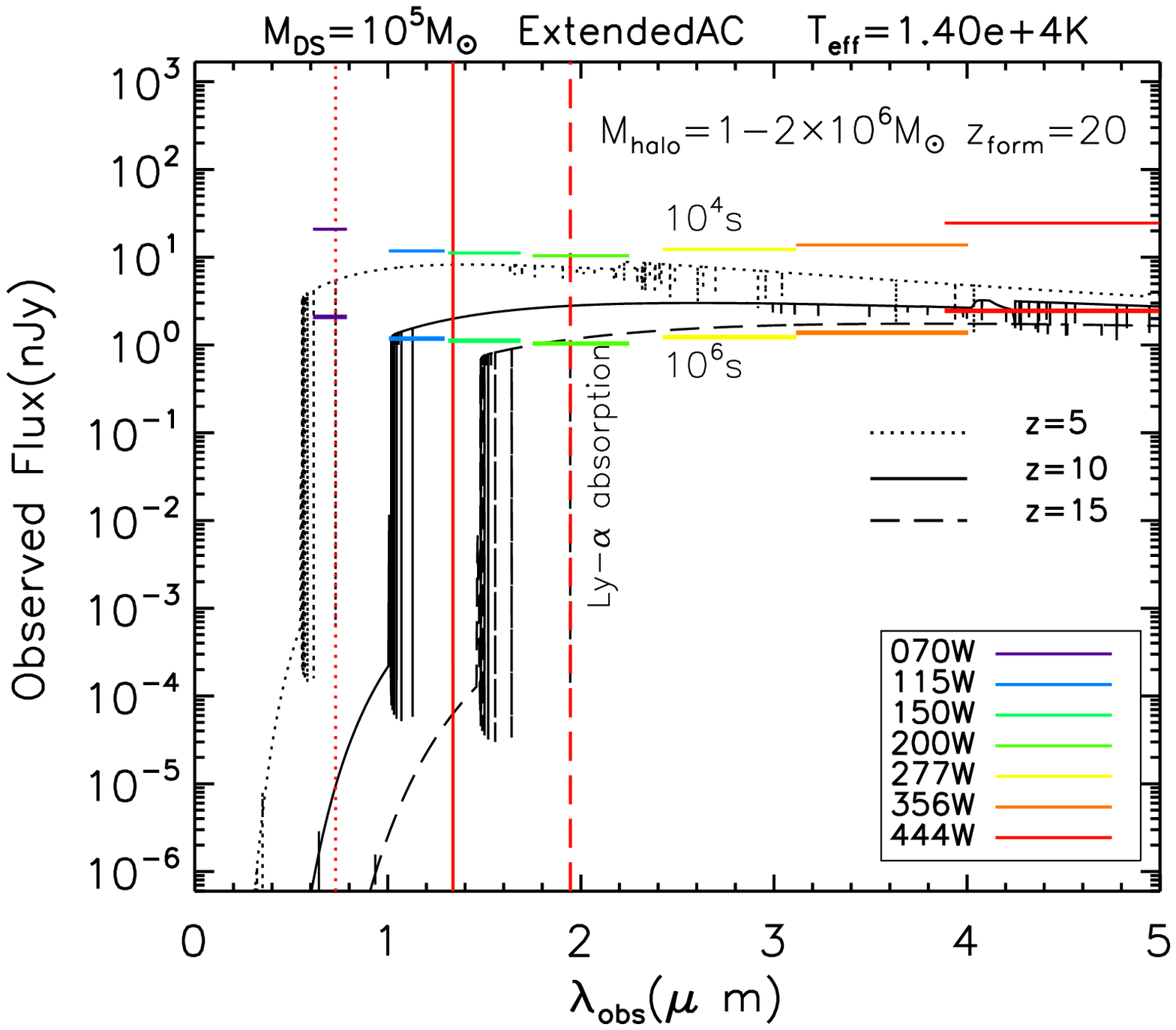}&
\includegraphics[scale=0.50]{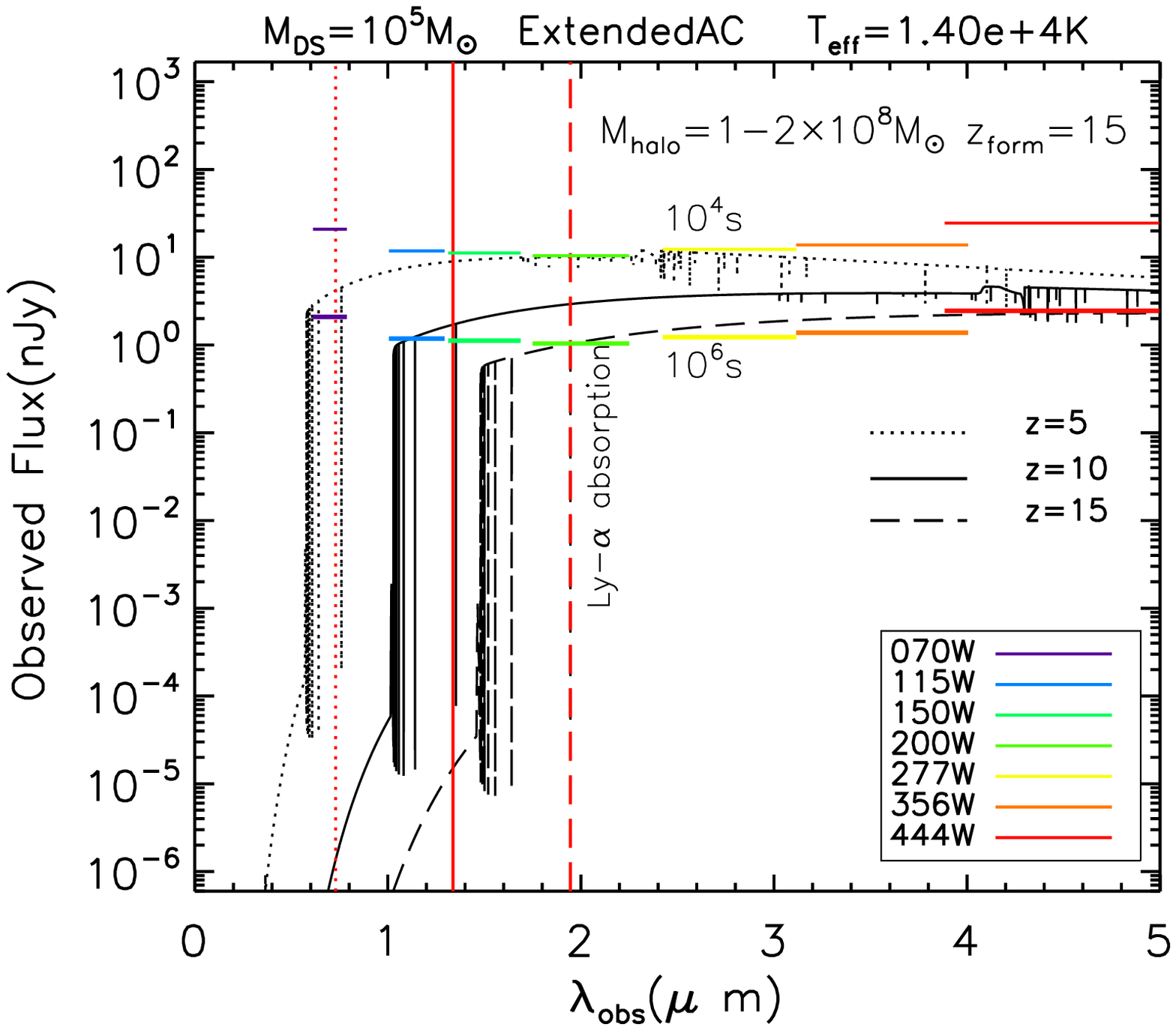}\\
\end{array}$
\end{center}
\caption{Similar to Figure~ \ref{TLUSTYRedshifted}, now for a \tento{5} DS formed either at \zform=20 in a \tento{6} DM halo (left panel) or at \zform=15 in a \tento{8} DM halo (right panel).}
\label{TLUSTYe5}
\end{figure*}

\begin{figure*}
\begin{center}$
\begin{array}{cc}
\includegraphics[scale=0.45]{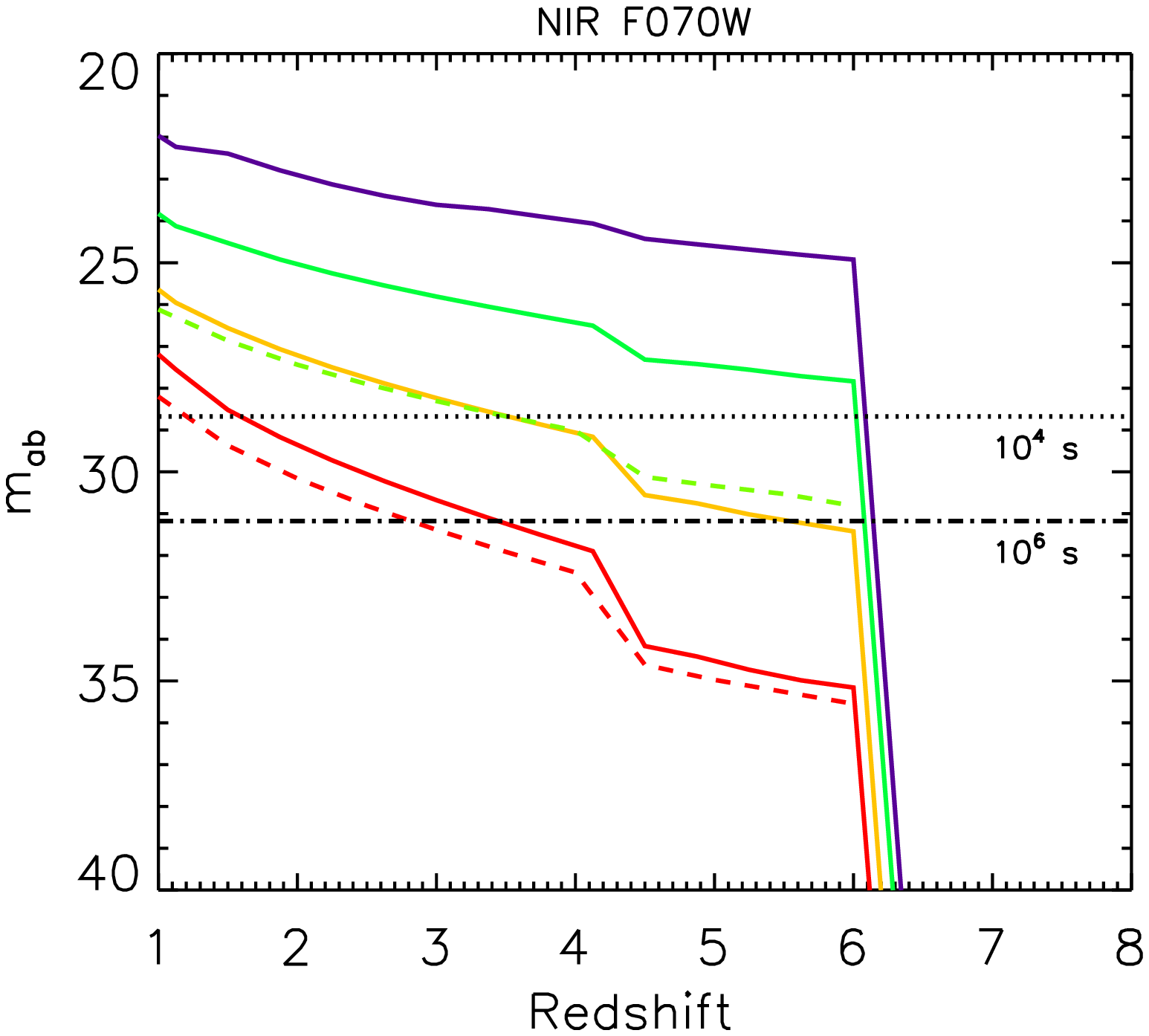}&
\includegraphics[scale=0.45]{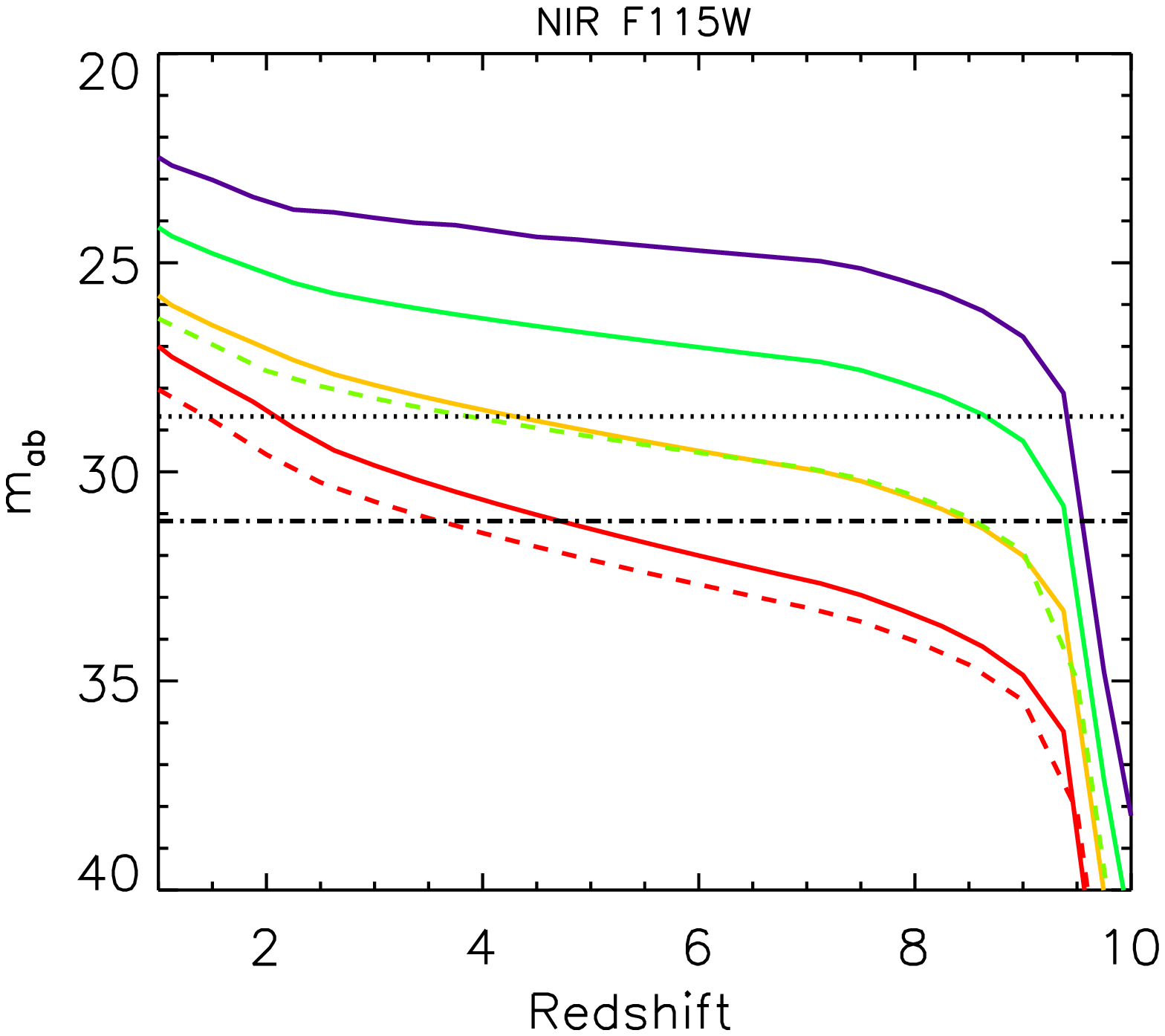}\\
\includegraphics[scale=0.45]{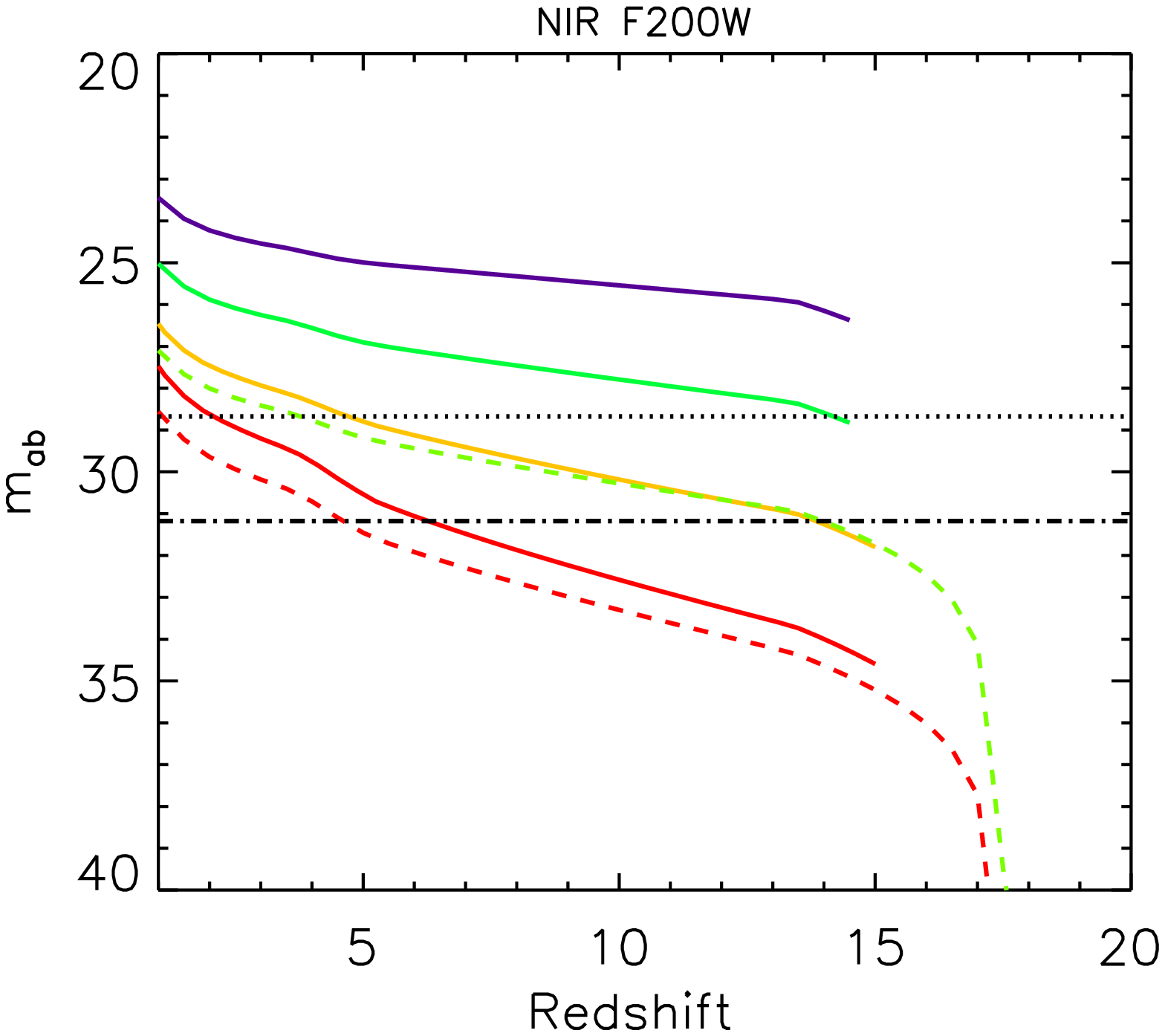}&
\includegraphics[scale=0.45]{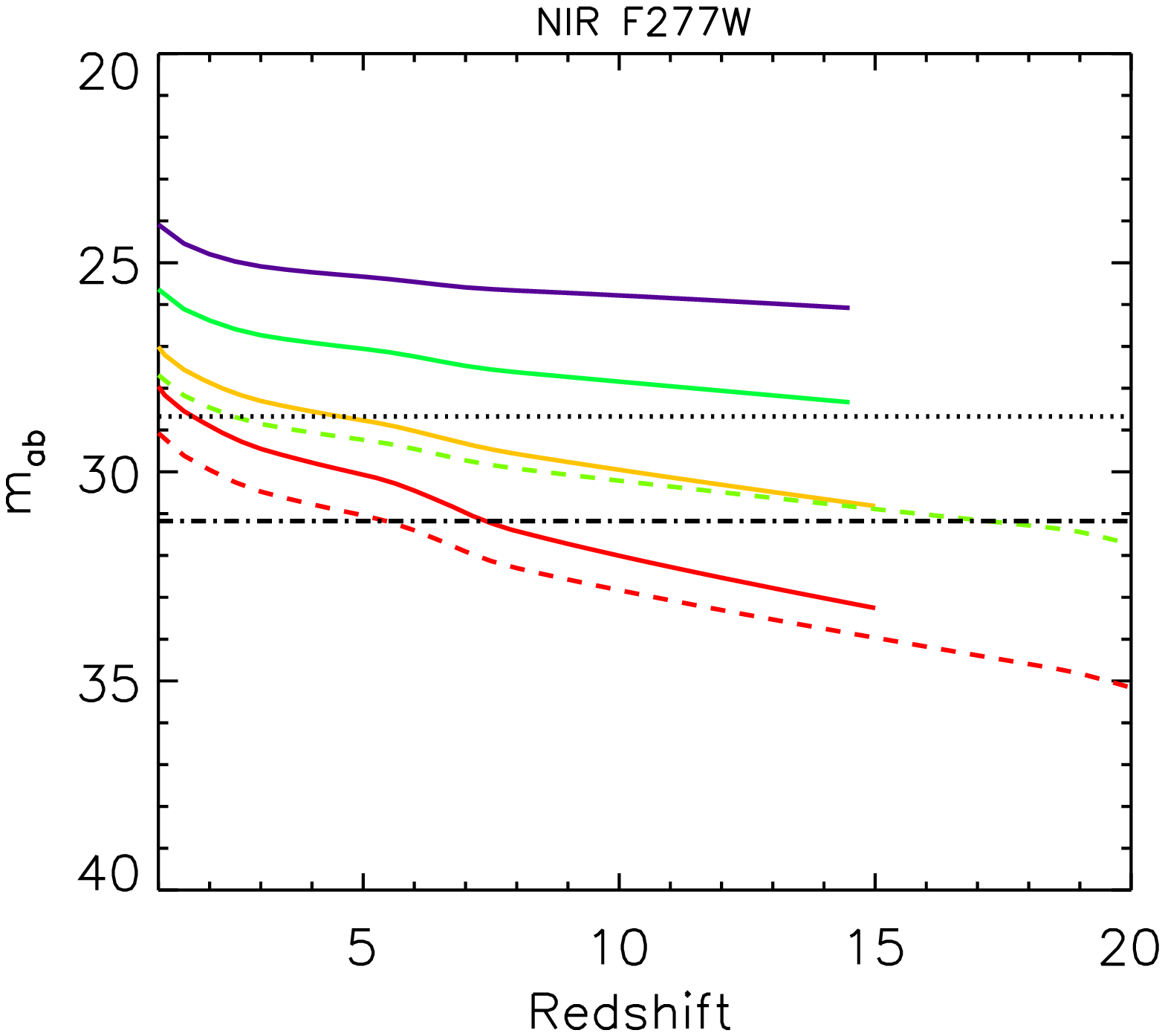}\\
\includegraphics[scale=0.45]{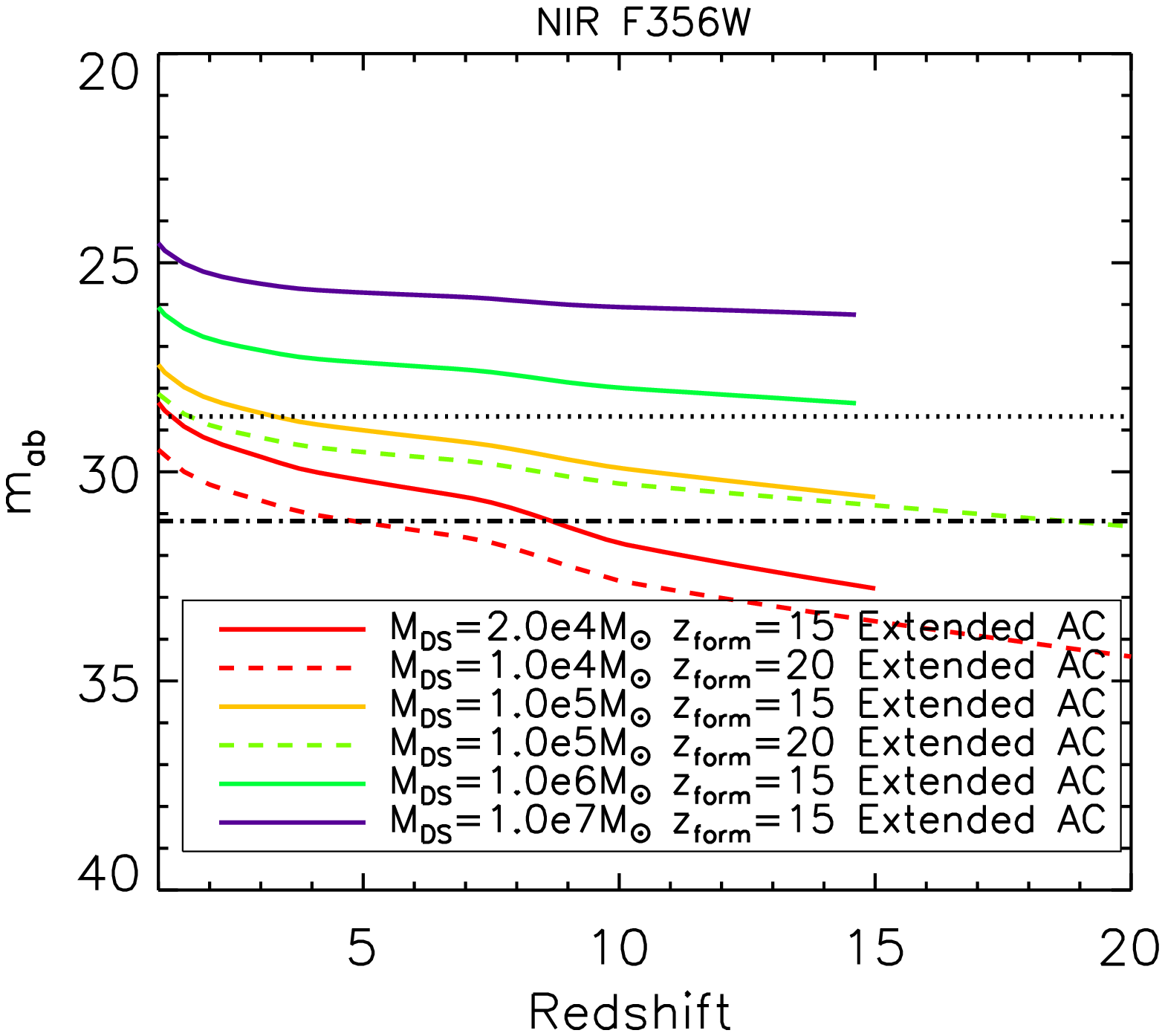}&
\includegraphics[scale=0.45]{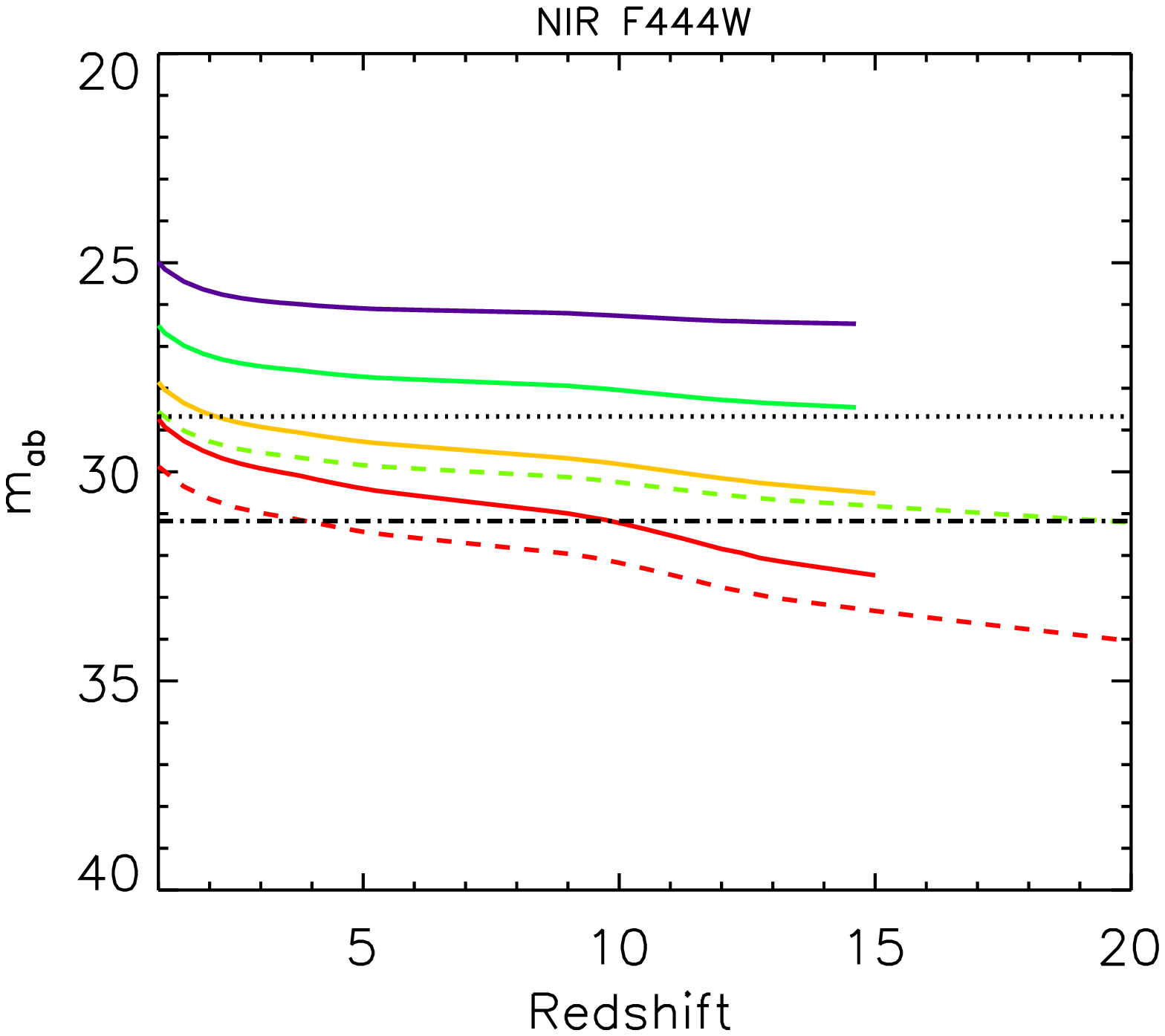}\\
\end{array}$
\end{center}
\caption{Apparent magnitudes as a function of redshift for various SMDS through the NIR camera
  wide passband filters on JWST for the case of formation via extended AC. 
  The number after the letter F and
  before the letter W in the name of each filter corresponds to the
  wavelength in the center of the passband in \microm{0.01} units. The two horizontal lines correspond to
  sensitivity limits for each filter for $10^4$s exposure time (the
  dotted line) and $10^6s$ exposure time (the dash-dotted line). 
  The \zform labeled in the legend is the formation
  redshift when the SMDS reached its corresponding mass. The curves
  corresponding to \zform$=15$ do not extend all the way to z$=20$ because
  at that high redshift the star has not formed yet. The sharp decrease  of the fluxes at various redshifts in the first three panels is due to the Gunn-Peterson trough entering the filters. The higher wavelength filters F277W-F444W would not be affected by the IGM absorption until \mbox{z $\gtrsim 20$}.}
\label{NirCapOff}
\end{figure*}

\begin{figure*}
\begin{center}$
\begin{array}{cc}
\includegraphics[scale=0.50]{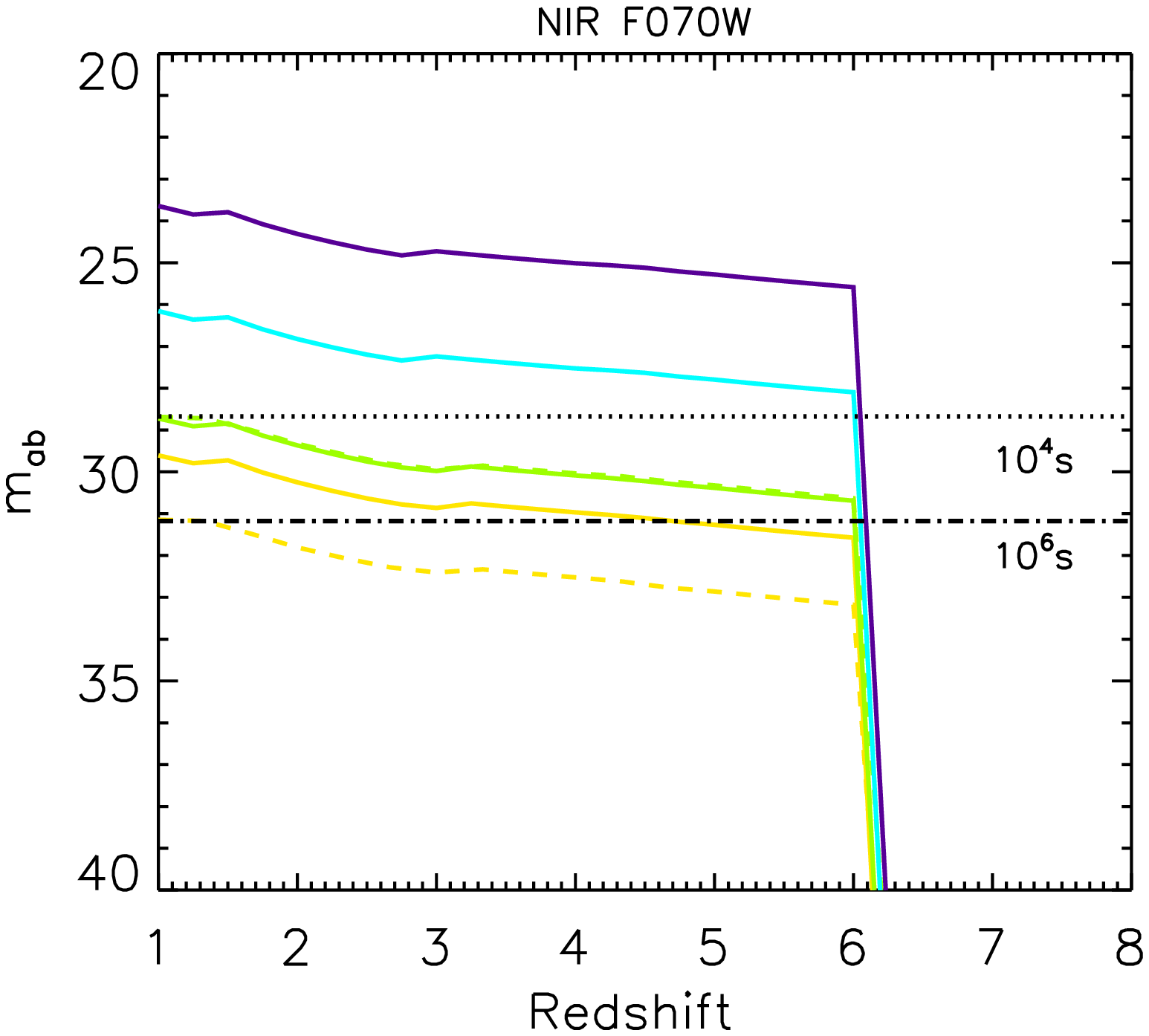}&
\includegraphics[scale=0.50]{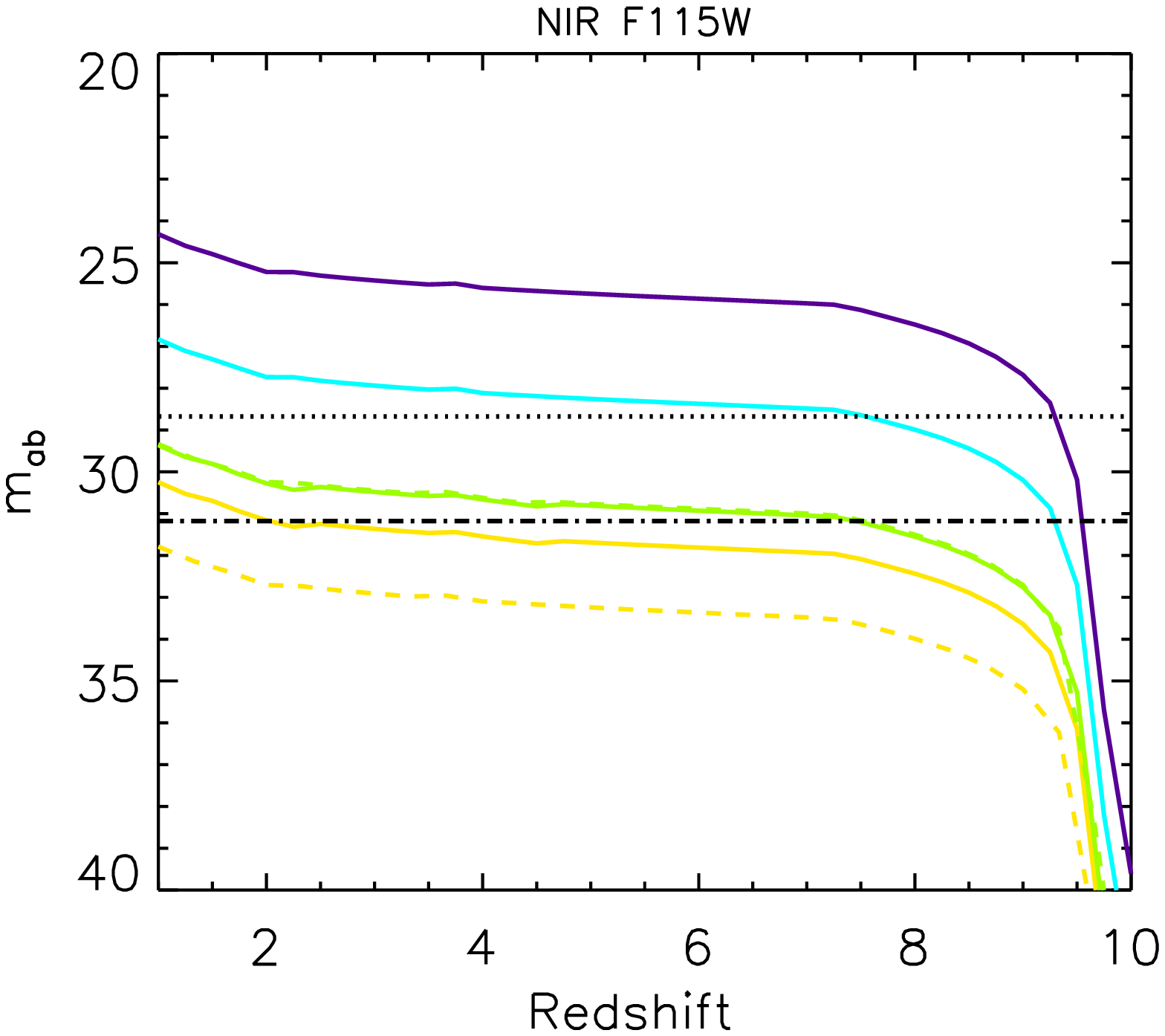}\\
\includegraphics[scale=0.50]{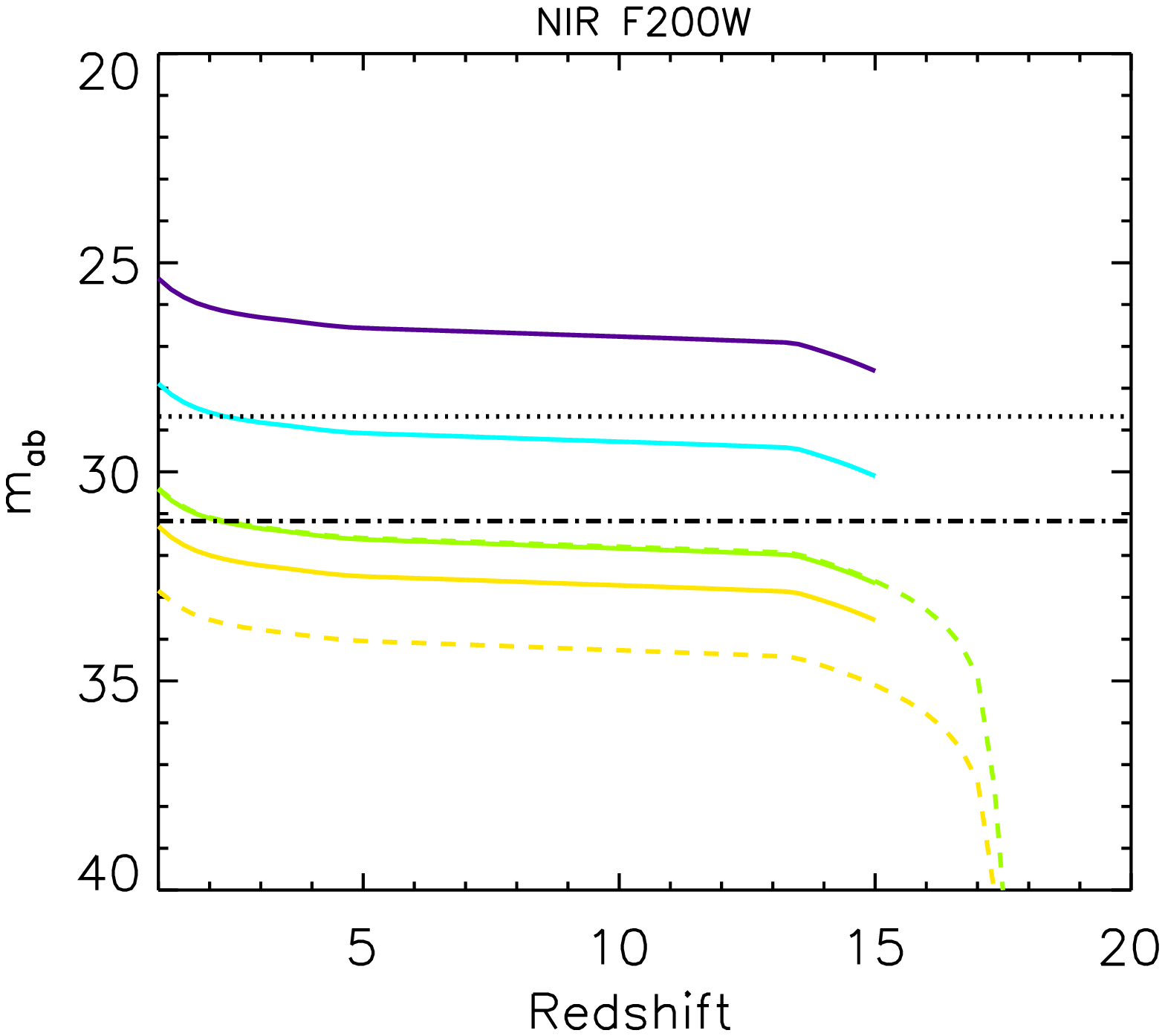}&
\includegraphics[scale=0.50]{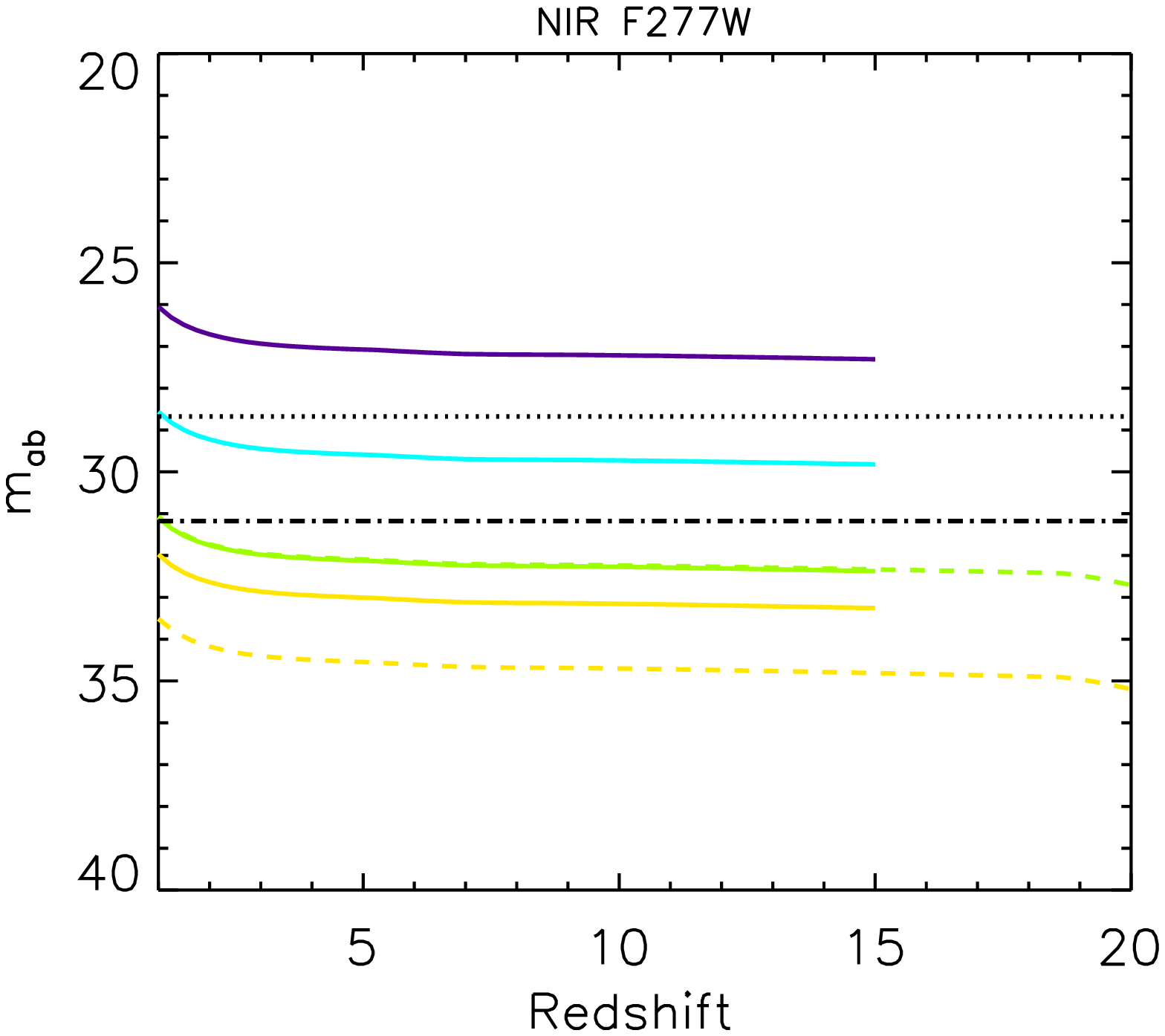}\\
\includegraphics[scale=0.50]{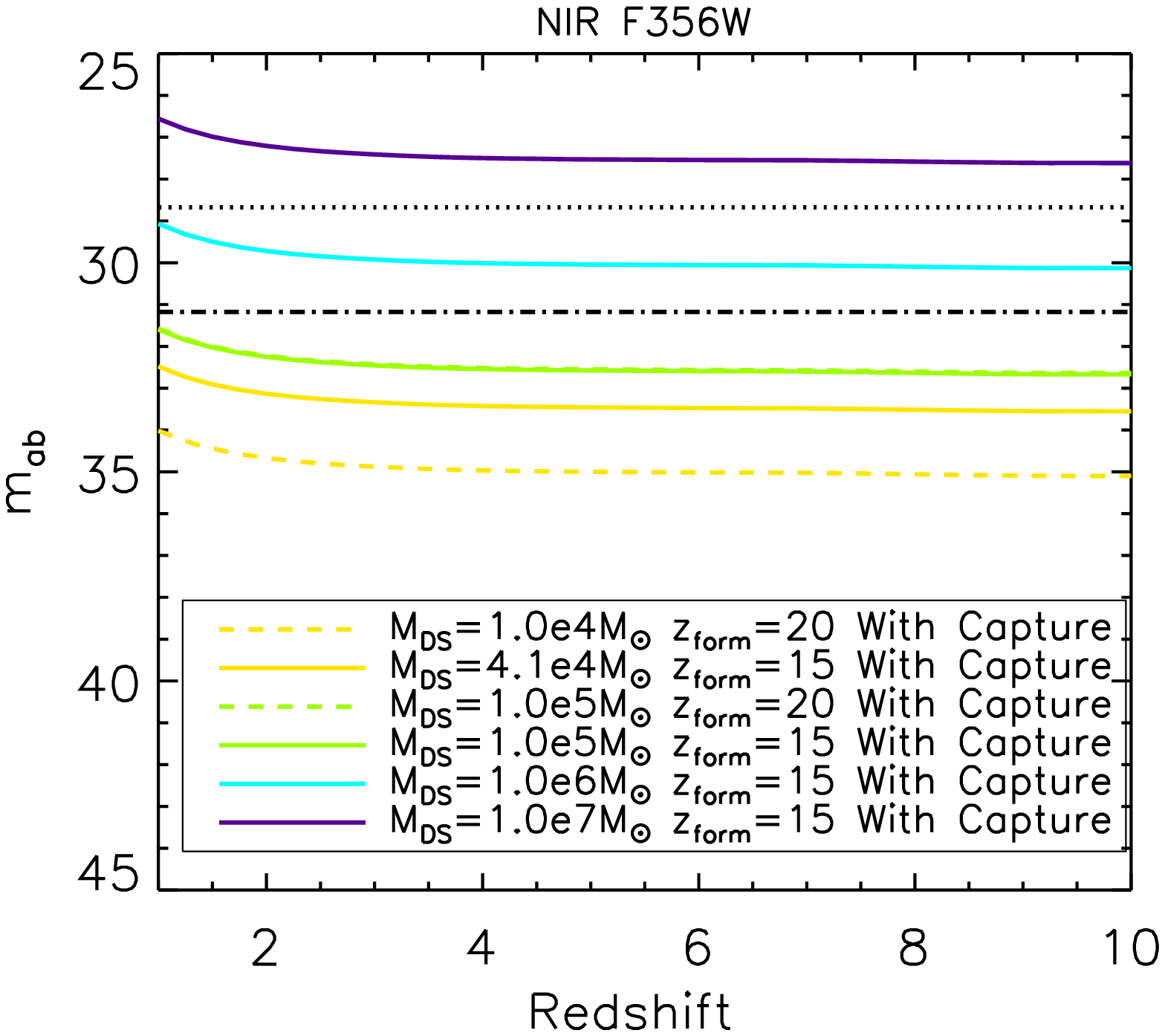}&
\includegraphics[scale=0.50]{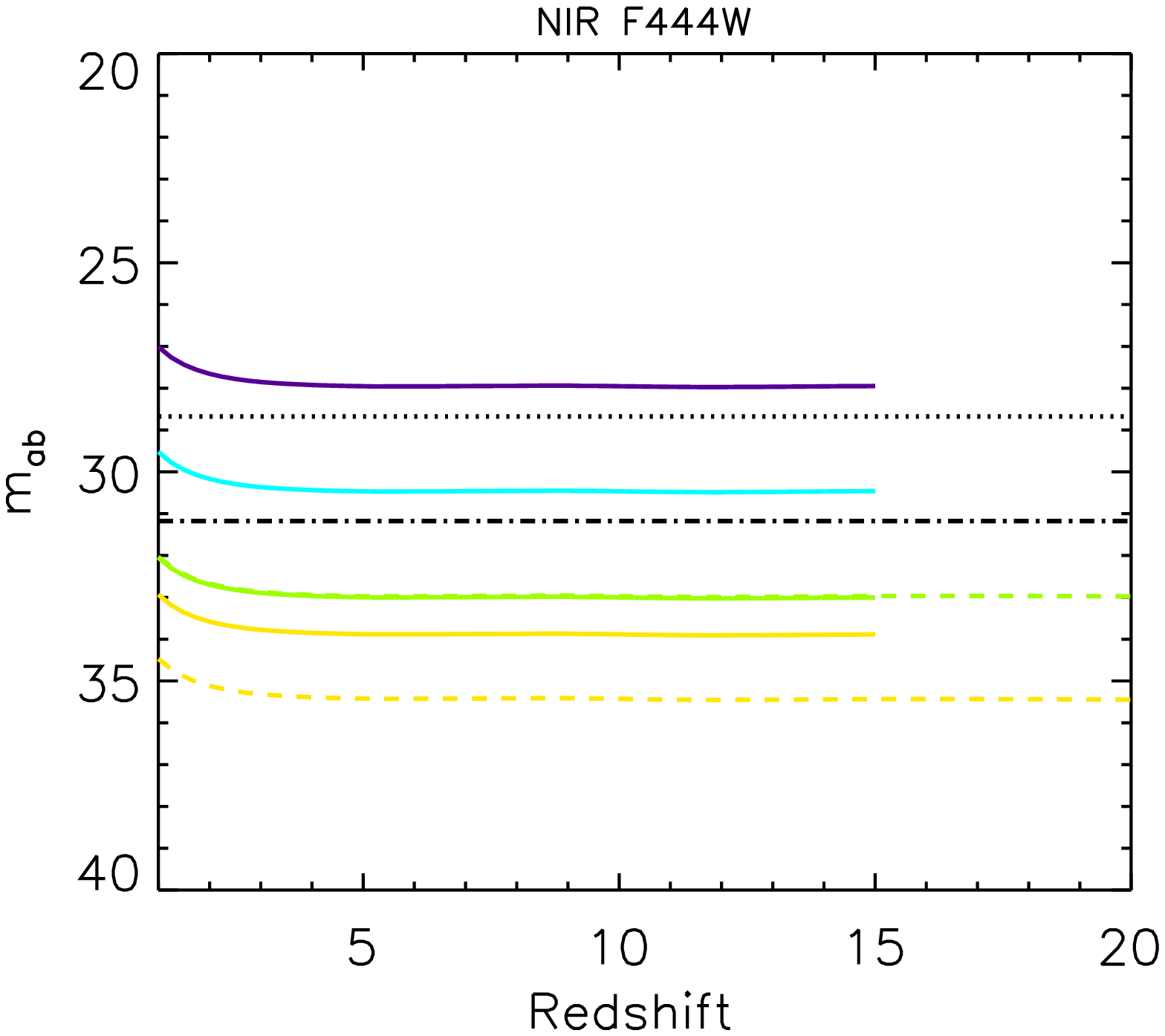}\\
\end{array}$
\end{center}
\caption{Same as Figure \ref{NirCapOff} for SMDS formed ``with capture'' in various JWST bands as labeled.  }
\label{NirCapOn}
\end{figure*}

Figures~ \ref{TLUSTYRedshifted} - \ref{NirCapOn} illustrate the detectability of SMDS with JWST
NIRCam filters.
Figures~ \ref{TLUSTYRedshifted} and \ref{TLUSTYe5} plot the stellar spectra of SMDSs of various masses and formation redshifts as a function of wavelength (for light  emitted at $z=15, 10$ and $5$) and
compare to the sensitivity of JWST filters for $10^4$s and $10^6$s exposure times.
In Figures~ \ref{NirCapOff} and ~ \ref{NirCapOn}, we instead plot the apparent magnitudes
as a function of redshift of emitted light
for various SMDS through the NIR camera  wide passband filters,
with each panel in the figure focusing on a particular JWST broadband filter; in these
two figures the SMDS are formed via extended AC and capture respectively.
 Ly$-\alpha$ absorption  cuts off the photons with wavelengths lower than 1216\AA (in the rest frame); we treat the absorption as being complete. Thus the SMDSs drops below the JWST sensitivity limit at  $z\sim 6$ for the F070W filter and at $z\sim 10$ for the F115W case.

Since the most massive dark stars are the brightest, they are the easiest to detect.
From figures~ \ref{NirCapOff} and \ref{NirCapOn}
one can see that $10^7\Msun$ dark stars, both with and without capture, are individually observable 
in $10^4$ seconds of NIRCam data
even at redshifts as high as $15$ in filters with a passband centered at \microm{2} and higher 
(F200W-F444W filters). 
For the case of a $10^6 \Msun$ SMDS, a longer exposure time of $10^6$s allows
the dark star, both with and without capture, to be  individually observable in all filters from F200W to F444W even at $z\sim 15$.  For $10^5 \Msun$ SMDS, those formed via extended AC are
visible in these filters out to $z \sim 15$ with $10^6$ sec exposure time while those formed with capture are too dim.  Lighter ones $< 10^5 M_\odot$ would not be detectable 
as J-band dropouts but, if they survived to lower redshifts (e.g. z=7) would likely already have been  
seen with HST or other telescopes.
Since the sensitivity of the higher wavelength MIRI filters above 5$\mu$ is worse,
only the $10^7 \msun$ DS are bright enough to be observable in MIRI filters (see the
discussion of Figure \ref{fig:JWST_DS_GAL} in Section \ref{sec:DSvsGal}).

\subsection{Detection at $z\sim 10$ as a \NirJband band dropout with JWST}

\begin{figure*}
\begin{center}$
\begin{array}{cc}
\includegraphics[scale=0.50]{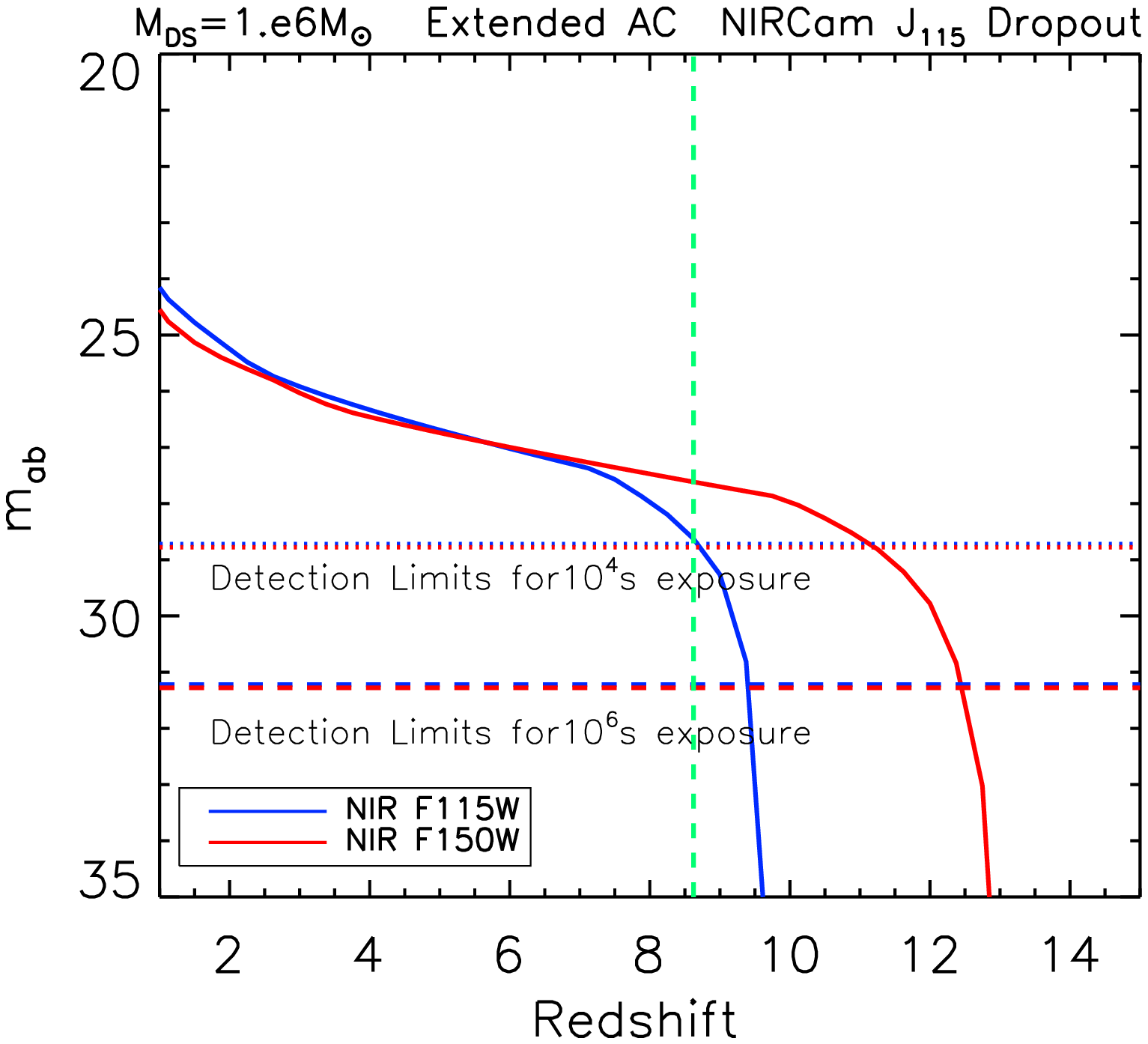} &
\includegraphics[scale=0.50]{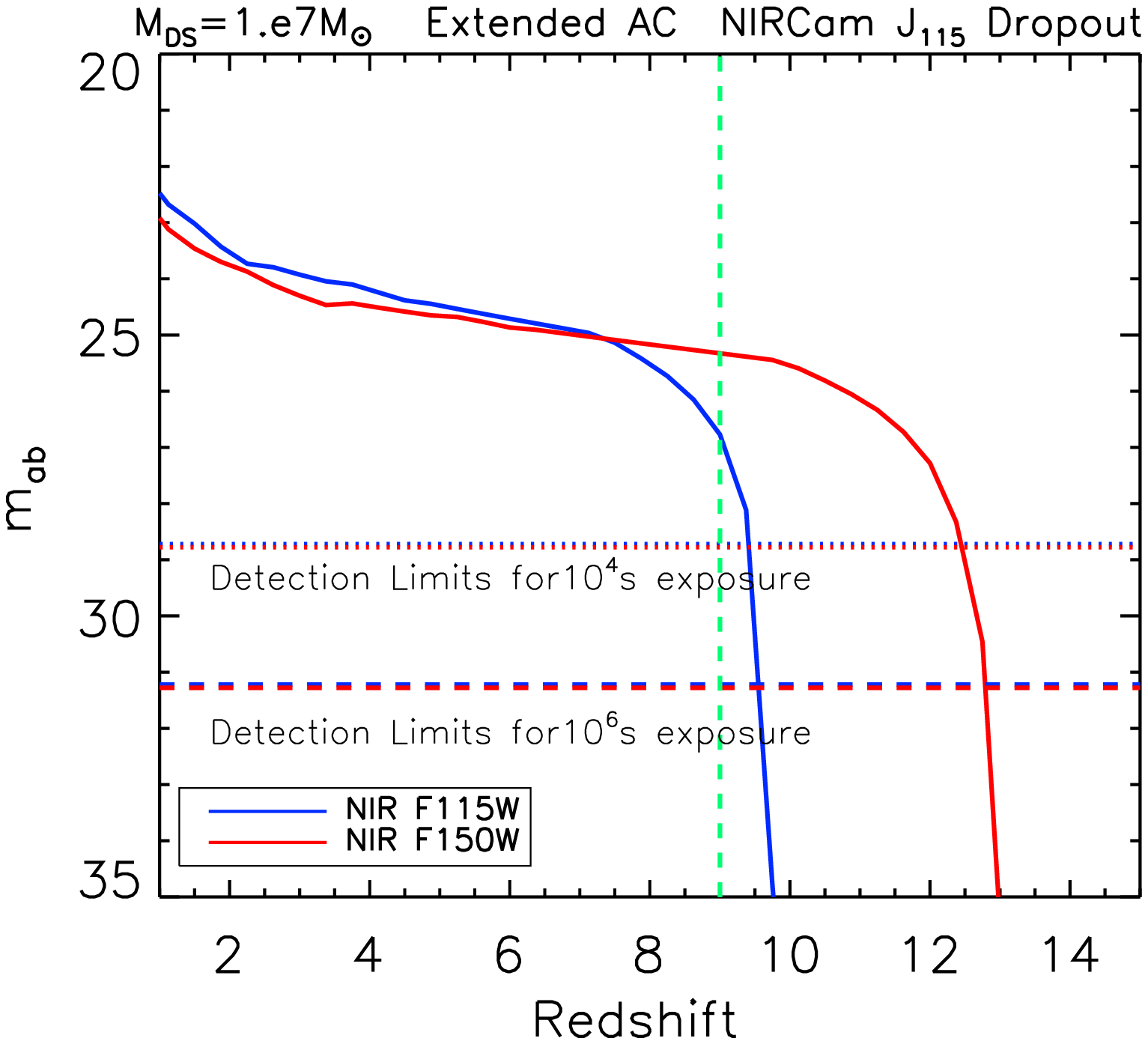}\\
\includegraphics[scale=0.50]{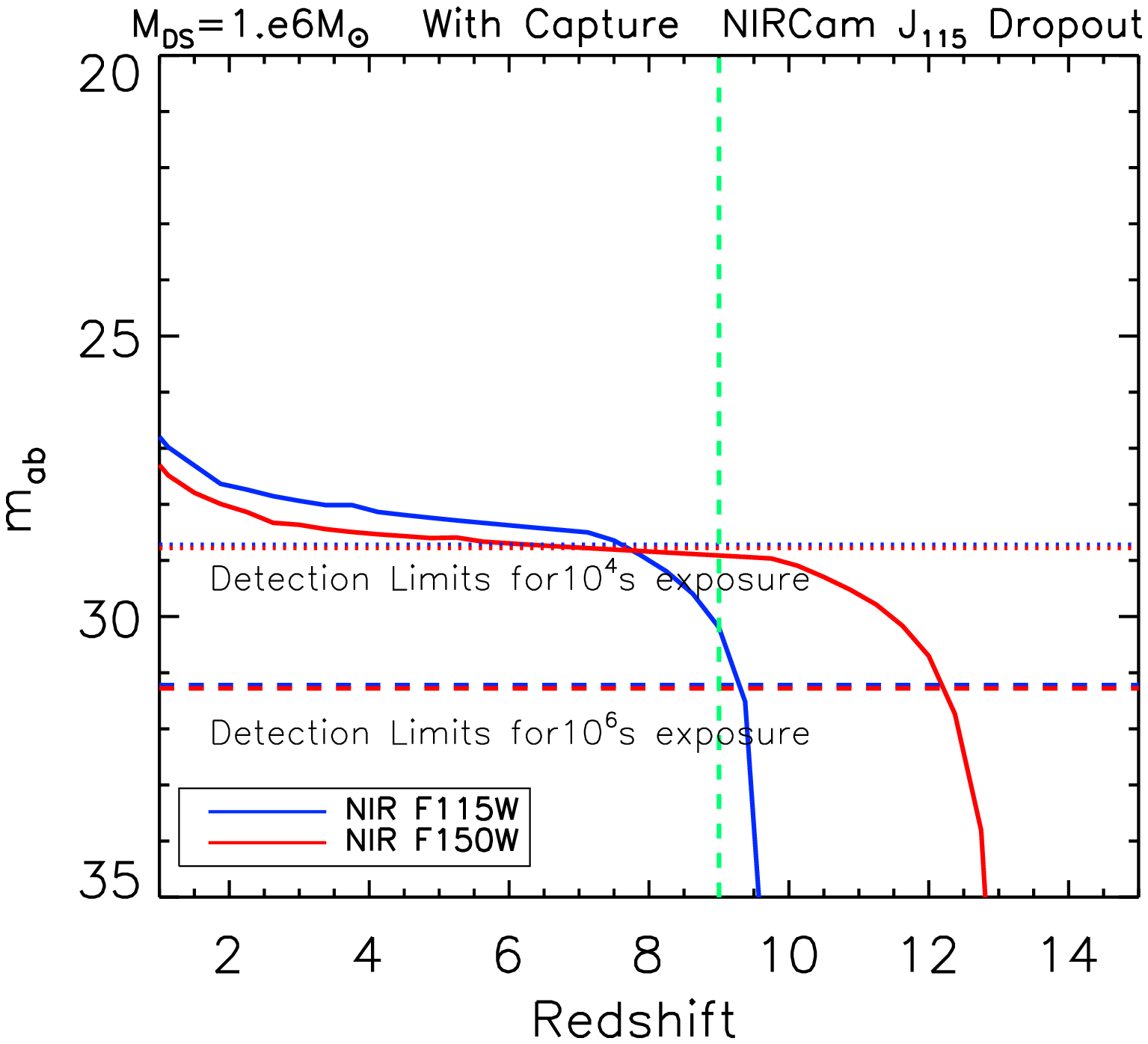} &
\includegraphics[scale=0.50]{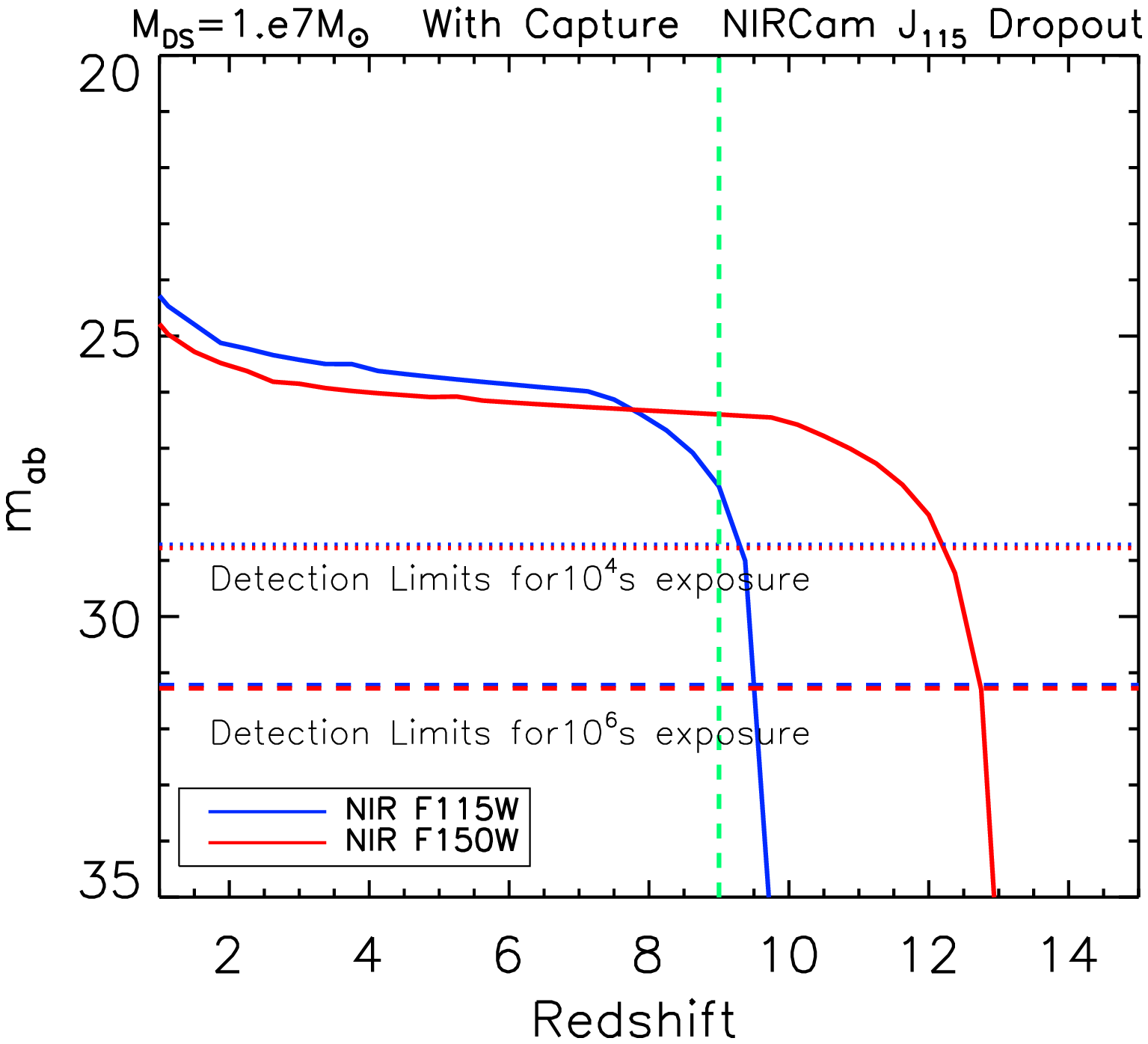}\\
\end{array}$
\end{center}
\caption{SMDS with JWST as \NirJband band dropouts: Apparent magnitudes for various SMDS through the F115W and F150W filters for NirCam.  Top panel: $10^6\Msun$ and $10^7\Msun$ Dark Stars formed without DM capture. Lower panel:  $10^6\Msun$ and $10^7\Msun$ Dark Stars formed "with capture".  The dotted  horizontal lines are obtained from the $10 \sigma$ required sensitivities for $10^4$ seconds exposure data published at \mbox{\protect\url{http://www.stsci.edu/jwst/instruments/nircam/}}; note that the detection limits for the \NirJband and the \NirHband filters differ by only $\sim 0.05$ $m_{AB}$ apparent magnitude and are thus
 essentially indistinguishable. The dashed horizontal lines are obtained assuming $10^6$ seconds exposure time. }
\label{NirJBand}
\end{figure*}

As expected, the dark stars that could have been detected as J-band dropouts with HST
 are also detectable using the same technique with JWST. We again will adopt the same dropout criterion, i.e. $\Delta m_{AB} > 1.2$  in the J and H broadband filters.
 Figure~ \ref{NirJBand} shows the sensitivity of JWST in a
  $10^4$ exposure in the 1.15$\mu$  (J-band) and 1.50$\mu$ (H-band) filters for NIRCam.
  The apparent magnitudes for $10^6$ and $10^7 \Msun$ SMDS with and without capture are
also shown for comparison.   Here, the SMDS form at z=15 and are assumed
to survive to various redshifts as shown.  Comparing Figures~ \ref{HUDFScansCapOff}
and \ref{NirJBand} (see also Table \ref{tb:JWST_Filters}), one can see that
 JWST is about half a magnitude more sensitive than HST to finding SMDS as
J-band dropouts (for $10^5$s exposure time with numbers provided in the literature
as 5$\sigma$ detection in HUDF09 and 10$\sigma$ detection with JWST).

 The three cases of $10^7 \msun$ with or without capture as well as 
 $10^6 \msun$ without capture 
  could be detectable in a JWST survey as J band dropouts in the
  redshift range $9.5-12$ even with the lower $10^4$ second exposure
  times. 
The $10^6\Msun$ Dark Star formed via captured DM (lower left plot) in Figure~ \ref{NirJBand} 
will require  a longer exposure time of $10^6$s (which would correspond to the same exposure time as the 2004 HUDF survey).

In order to predict how many  SMDS would be visible in a JWST deep
field survey we have to assume something about the total field of view
(FOV) of all future JWST surveys in which the stars would be
observable. The FOV of the  NIRCam instrument is $2.2'\times4.4'=9.68$
\arsq (see \mbox{\protect\url{http://www.stsci.edu/jwst/overview/design/}})
This value is likely to be an underestimate. Since HST had multiple surveys with a total of $160$ \arsq,  we will also consider the case of multiple surveys with JWST with a total added area of $\sim 150$ \arsq.  Given the bounds on the numbers of DS from HST from the previous section, 
we find that the number of expected SMDS with JWST
as J-band dropouts is $N\lesssim 1$ and therefore conclude that SMDS are hard to detect with JWST as J-band dropouts. This is expected since HST was already sensitive enough to observe them as J-band dropouts, assuming enough would have survived from their formation redshift until $z\sim 10$. The only improvement  could be made by a larger survey area compared to the one with HST.  For the \tento{6} SMDSs formed with capture, which were detectable only in the $4.7$ \arsq of HUDF09, JWST should be able to provide a larger survey area so that
these objects become more detectable.

\subsection{Detection at $z\sim 12$ as a \NirHband band dropout with JWST}
\label{sec:Hband}

\begin{figure*}
\begin{center}$
\begin{array}{cc}
\includegraphics[scale=0.50]{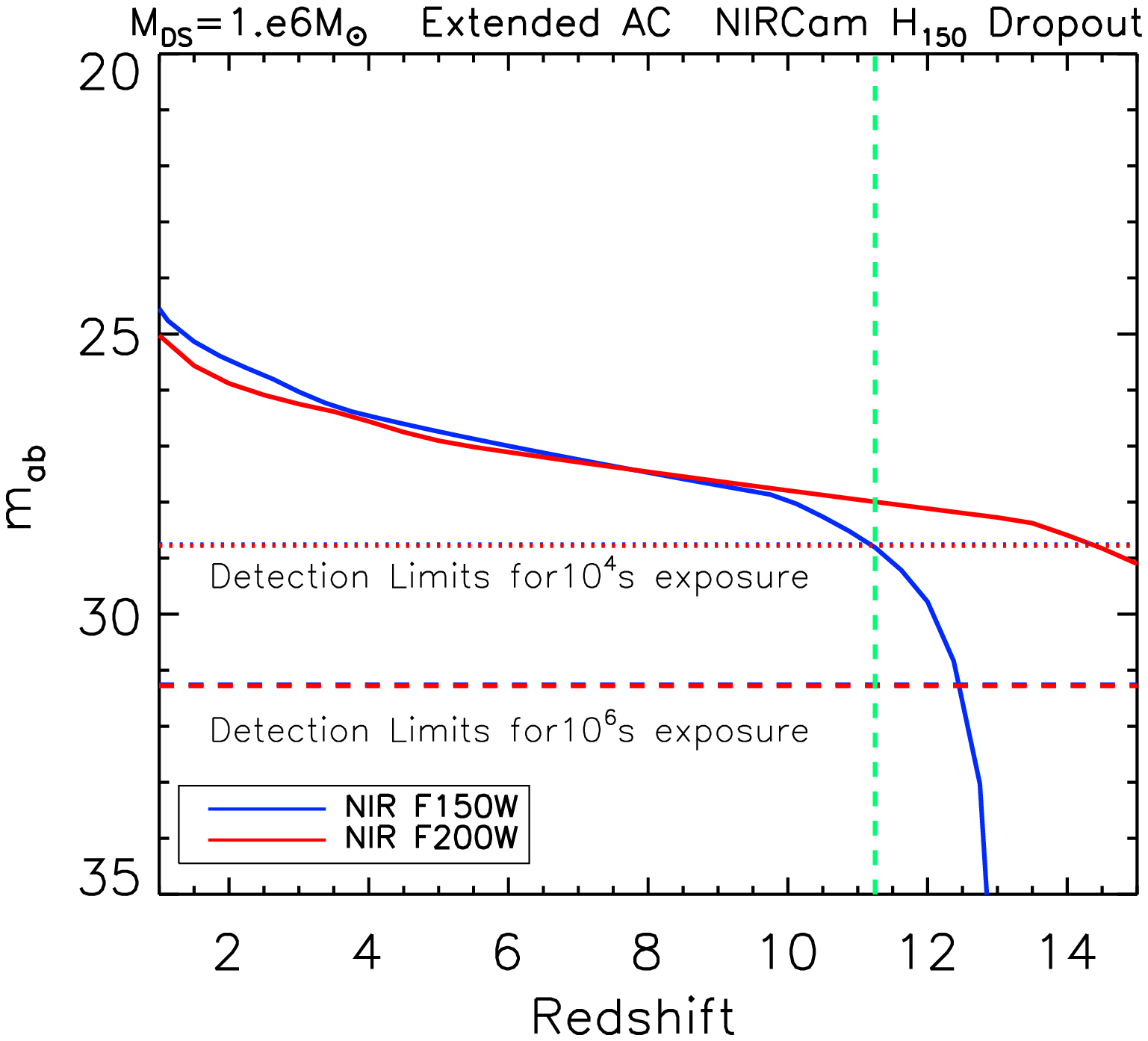} &
\includegraphics[scale=0.50]{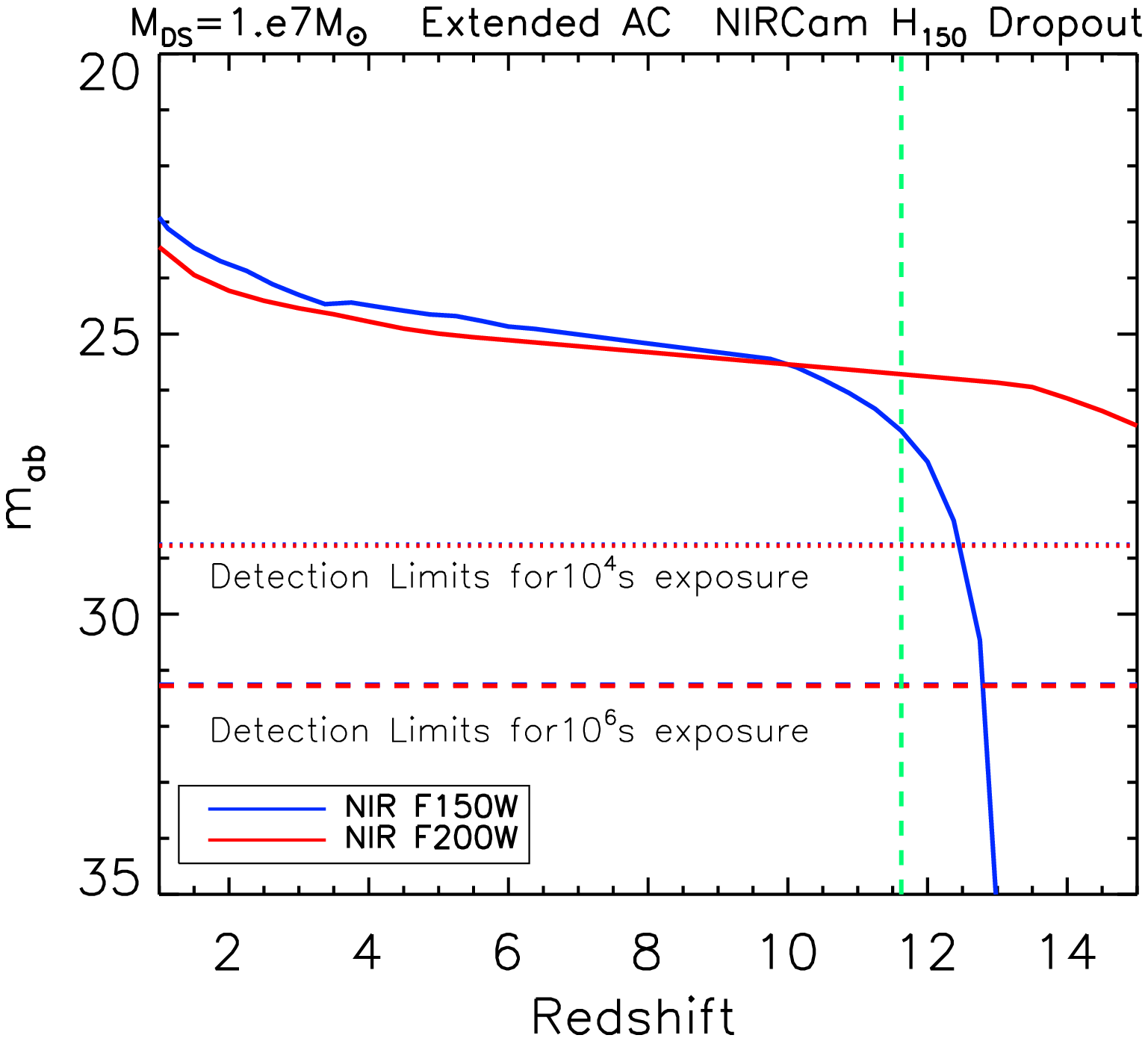}\\
\includegraphics[scale=0.50]{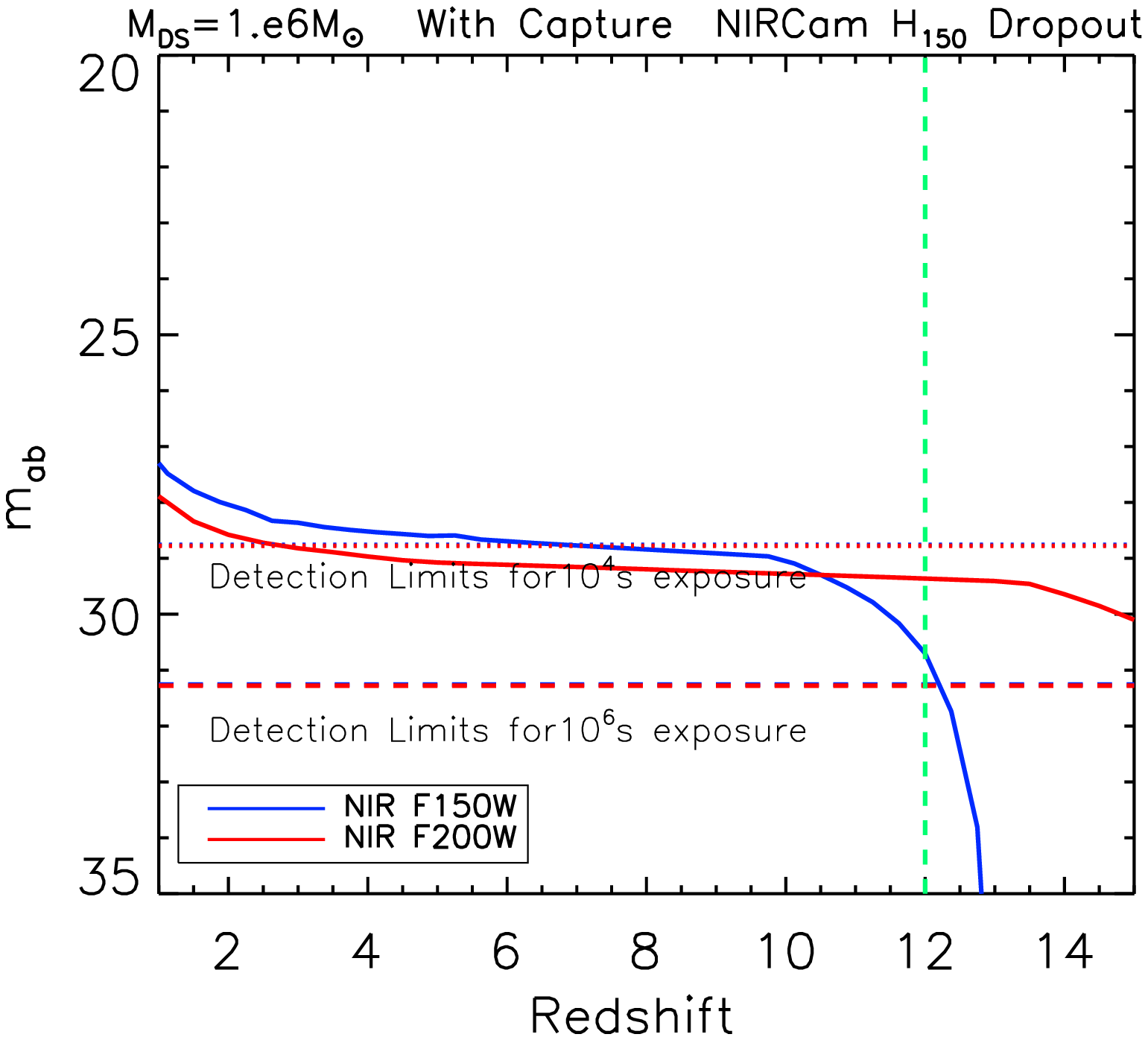} &
\includegraphics[scale=0.50]{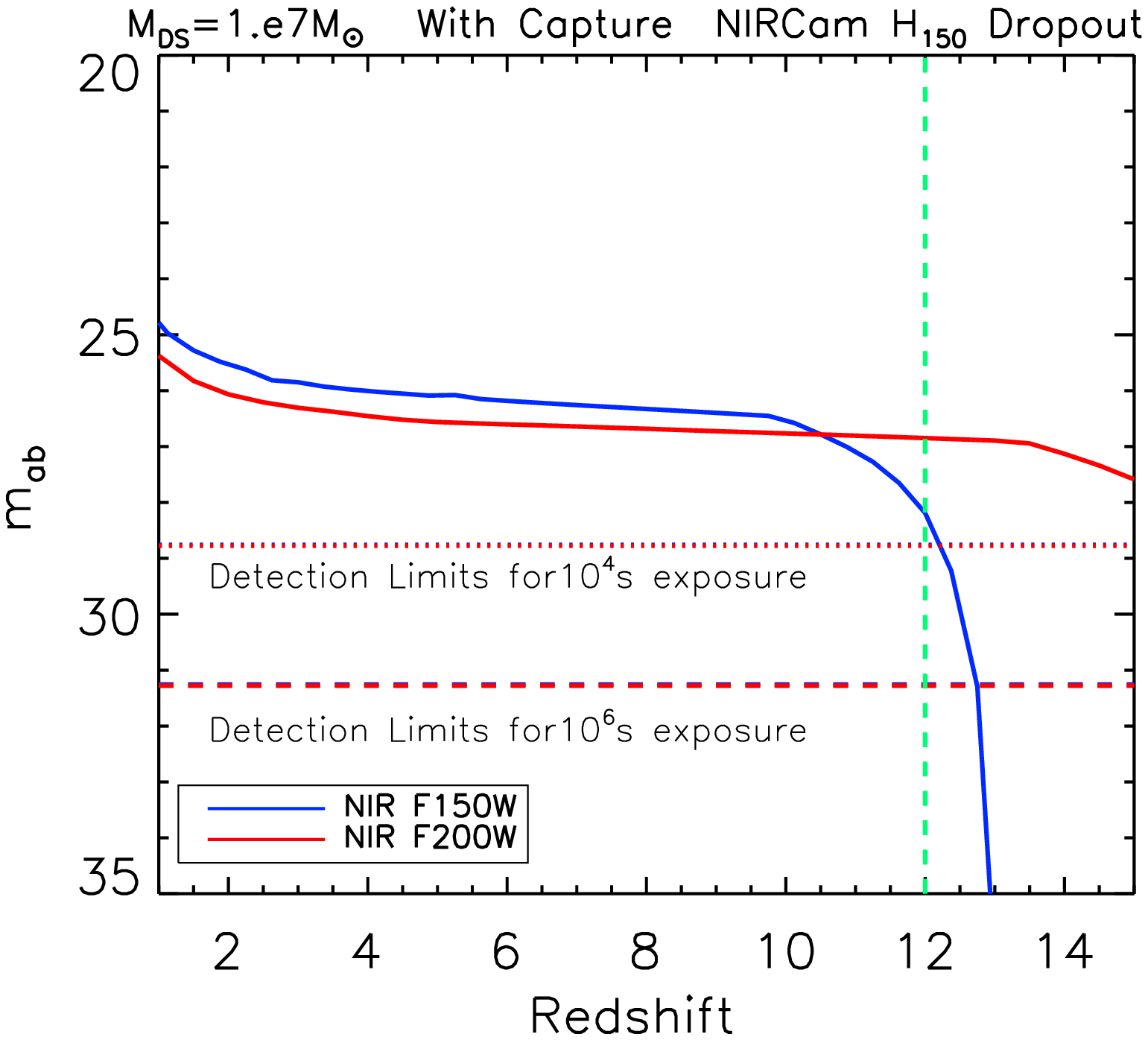}\\
\end{array}$
\end{center}
\caption{SMDS with JWST as \NirHband band dropouts: Apparent magnitudes for SMDS through the F150W and F200W NirCam filters. Those could be used to establish dropout detection criteria in the $12-14$ redshift range.  Top panel: cases of interest ($10^6\Msun$ and $10^7\Msun$)  Dark Stars formed without considering DM capture. Lower panel: $10^6\Msun$ and $10^7\Msun$ Dark Stars formed including DM capture. The vertical green dashed line
indicates the minimum redshift at which the DS will appear as a dropout.}
\label{Nir150200}
\end{figure*}

     Whereas JWST is not particularly better than HST at finding J-band dropouts, it will be
significantly better at finding SMDS as H-band and K-band dropouts at higher wavelengths. 
 In this section
we focus on H-band dropouts, where the object can be seen in the F200 NIRCam filter of JWST but not in the F150 NIRCam filter.  As before in Eqn. (\ref{eq:dropout}), we require
the difference between the broadband fluxes in the \NirHband and \NirKband filters  to be greater than 1.2 AB magnitudes.
We see that the SMDS stellar light seen with JWST's \NirKband filter is essentially unaffected by  
Ly$-\alpha$ absorption until $z\sim 15$, whereas  the IGM absorption will cut off most of the flux in  the \NirHband at $z\gtrsim 11.5$. (see Figure~\ref{Nir150200}).  Hence SDMS can
appear as H-band dropouts.

We will consider the case of SMDS forming at \zform=12, the same as the time of observation.
Figure~ \ref{Nir150200} shows that  the three cases of $10^7 \Msun$
SDMSs with and without capture as well as  $10^6 \Msun$ SMDSs without
capture are all detectable in a JWST survey as \NirHband dropouts in
the redshift range $11.5-12.5$.  DS formed at higher redshifts could be seen all the way out to $z \sim 14, 15$, but likelihood analyses on any objects found as H-band dropouts
with photometry with JWST will probably estimate the redshift at $z \sim 12$.   The  $10^6\Msun$ Dark Star formed via captured DM (lower left plot) 
is too faint to appear as a dropout.    The number of \NirHband dropout events  is given by Eq. (\ref{maxf}) with
 $\Delta z = 1$ and $f_{surv}=1$ since the objects are observed at the same
time they form) and the appropriate survey area $\theta^2$ for JWST must be applied.

Is it reasonable to apply the bounds from HST on the numbers of 
SMDS at z=10 to those at $z=12$?   We will consider three different possibilities, and
summarize all results for the predicted number of H band dropouts with JWST in Table \ref{tb:Hband}.  If we assume that all the SMDS at $z=12$ have the same
properties as those at $z=10$, and that they survive throughout the redshift window observable
by HST, then
the HST bounds are so stringent that JWST will not be able to see many of them.  This
is the case we label "Maximal Bounds".   In particular,  $10^7\msun$ SMDSs would have
been so bright as to be easily seen in HST, and the resultant stringent bounds imply that
only $N_{obs}\sim 1$  DS  would be found even with multiple surveys with 150
\arsq FOV.   For  \tento{6} SMDSs the bounds from HST are slightly weaker because
the objects are not as bright, so that 10 (32) of these might be found per 150 \arsq field
for DS that grew via extended AC (with capture).
Since the ones with capture are fainter and harder to see, (counterintuitively) the weaker HST
bounds imply that more of them might be found with JWST.

However, it is very likely that there are more SDMS at z=12 (the JWST window) than at z=10
(the HST window).  For one thing, the host halo
formation in this mass range peaks at $z\sim12$ (see Fig~3).
Moreover  at lower redshifts ($z\sim 10$) the DM halos that could host those SMDSs are much more likely to merge to form even larger halos. 
In addition, after the first SDMS die (before z=10), they turn into
fusion powered stars that produce ionizing photons, which disrupt the formation of DS at lower redshifts. 
Indeed the strong halo clustering at high redshift would cause the possible
formation sites to be preferentially close to or inside the HII regions
during reionization, potentially leading to strong suppression of star formation; 
due to this mechanism \citet{Iliev:2006sw} found  a suppression of \tento{8}-\tento{9} 
halos by an order of magnitude due to Jeans mass filtering in the ionized and heated
H~II regions.  

We will thus recalculate the number of DSs detectable with JWST using
weakened bounds from HST.  We will take \fsmds \fdt $f_{surv} = 1.5
\times 10^{-2}$ as our "Intermediate Bounds" case. This case could imply
that not all minihaloes host DS, or that not all DS survive throughout the  z$\sim$10 HST observability window.
In this Intermediate Bounds case, hundreds or thousands of SMDS are potentially observable.

For comparison, in the Table we list as a third case the full number of DM haloes that could in principle host DS. If all of these
contained DS one would expect up to $\sim 450,000$ DS with JWST. However, as discussed in Section \ref{sec:OtherBounds}
this would be extremely unreasonable as there would be no baryons outside of DS left for galaxy formation.
Our results for the detectability of SMDS as H-band dropouts
with JWST are summarized in Table \ref{tb:Hband}.

\begin{table*}
 \begin{center}

 \begin{tabular}{ccccc}

\multicolumn{5}{c}{ Upper limits on  numbers of SMDS detectable with JWST as \NirHband dropout}\\
 \hline\hline

$M_{DS} (\msun)$ &  Formation Scenario &       Bounds from HST                & $N_{obs}^{FOV}$ &  $N_{obs}^{multi}$    \\

 \hline
$10^6$                    & Extended AC         &     Maximal Bounds           &   $\lesssim 1$       & 10       \\
$10^6$                            & With Capture         &   Maximal Bounds     &   2                 &  32        \\
$10^7$                       & Any                          &   Maximal Bounds    &   $\lesssim 1$       & $\sim1 $\\
 \hline
$10^6$                       & Extended AC         & Intermediate           &   45                   & 709       \\
 $10^6$                       & With Capture         &   Intermediate &137                 &  2128    \\ 
$10^7$                          & Any                          & Intermediate        &   4                   & 64         \\
\hline
$10^6$                         & Extended AC         &     Number of DM halos     &   28700                   & 444750       \\
$10^6$                      & With Capture         &     Number of DM halos       &   28700                 &  444750       \\
$10^7$                       & Any                       &       Number of DM halos      &   155                   & 2400         \\

\hline
 \end{tabular} 

   \caption{Upper limits on the number of SMDS detections as \NirHband dropouts with JWST.
      In first three rows  (labeled "Maximal Bounds") we assume that all the DS live to below z=10 where they
      would be observable by HST, and we apply the bounds
    on the numbers of DS \fsmds from HST data in Section \ref{sec:BoundsHST}.
     The middle three rows (labeled "Intermediate") relax those bounds by assuming that only $\sim 10^{-2}$ of the
     possible DS forming in z=12 haloes make it through the HST observability window.
       For comparison we also tabulate in the
     last three rows the total number of potential DM host halos in
     each case. We also split the number of observations in two
   categories, $N_{obs}^{FOV}$ and $N_{obs}^{multi}$. The first
   assumes a sliver with the area equal to the FOV of the instrument
   (9.68 \arsq), whereas in the second we assume multiple surveys with
   a total area of 150 \arsq. Note that for the case of the \tento{7}
   SMDS the predictions are insensitive to the formation mechanism. }
   \label{tb:Hband} 
 \end{center}
 \end{table*}

\subsection{Detection at $z\sim 15$ as a \NirKband dropout with JWST}     

\begin{figure*}
\begin{center}$
\begin{array}{cc}
\includegraphics[scale=0.50]{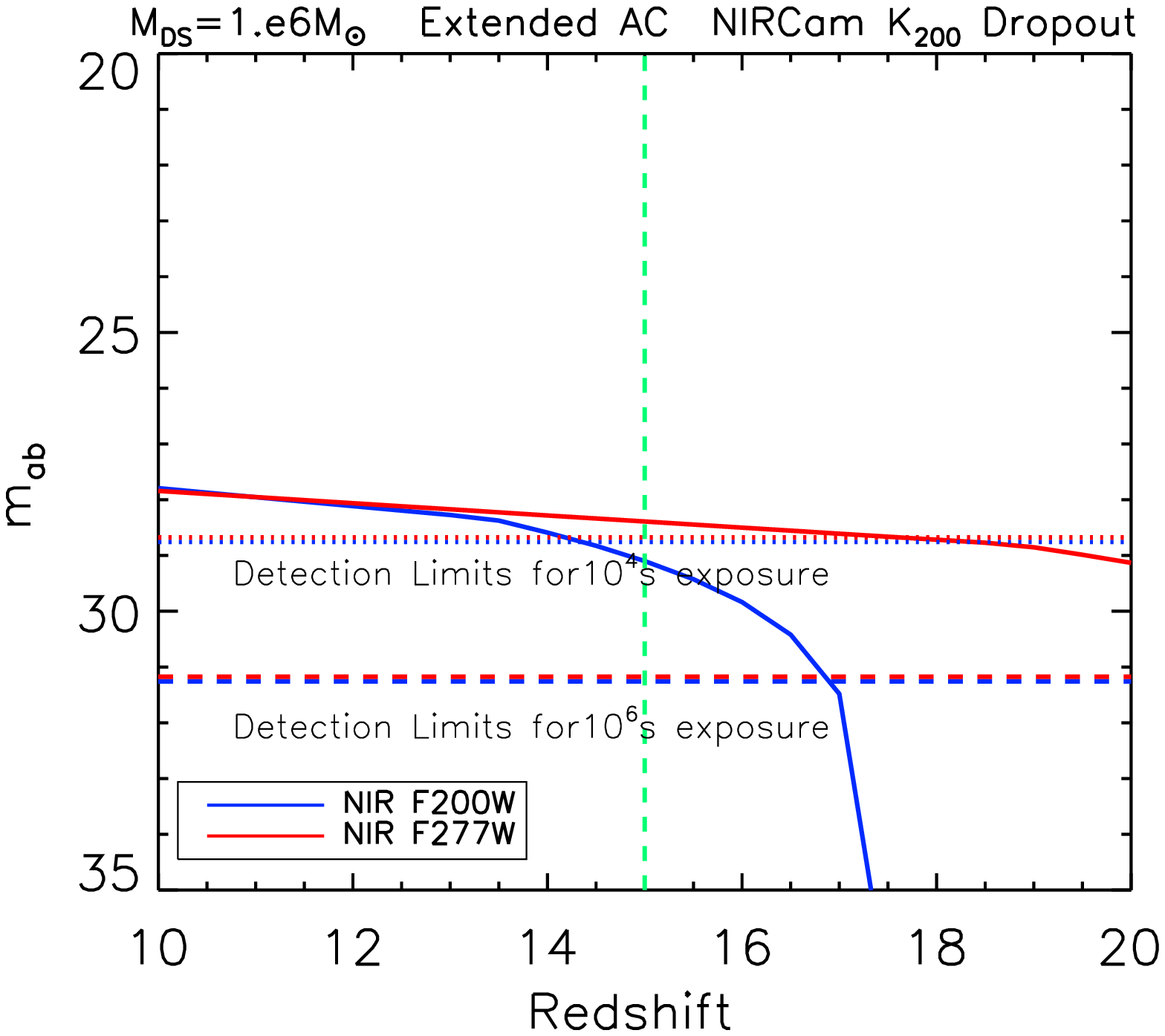} &
\includegraphics[scale=0.50]{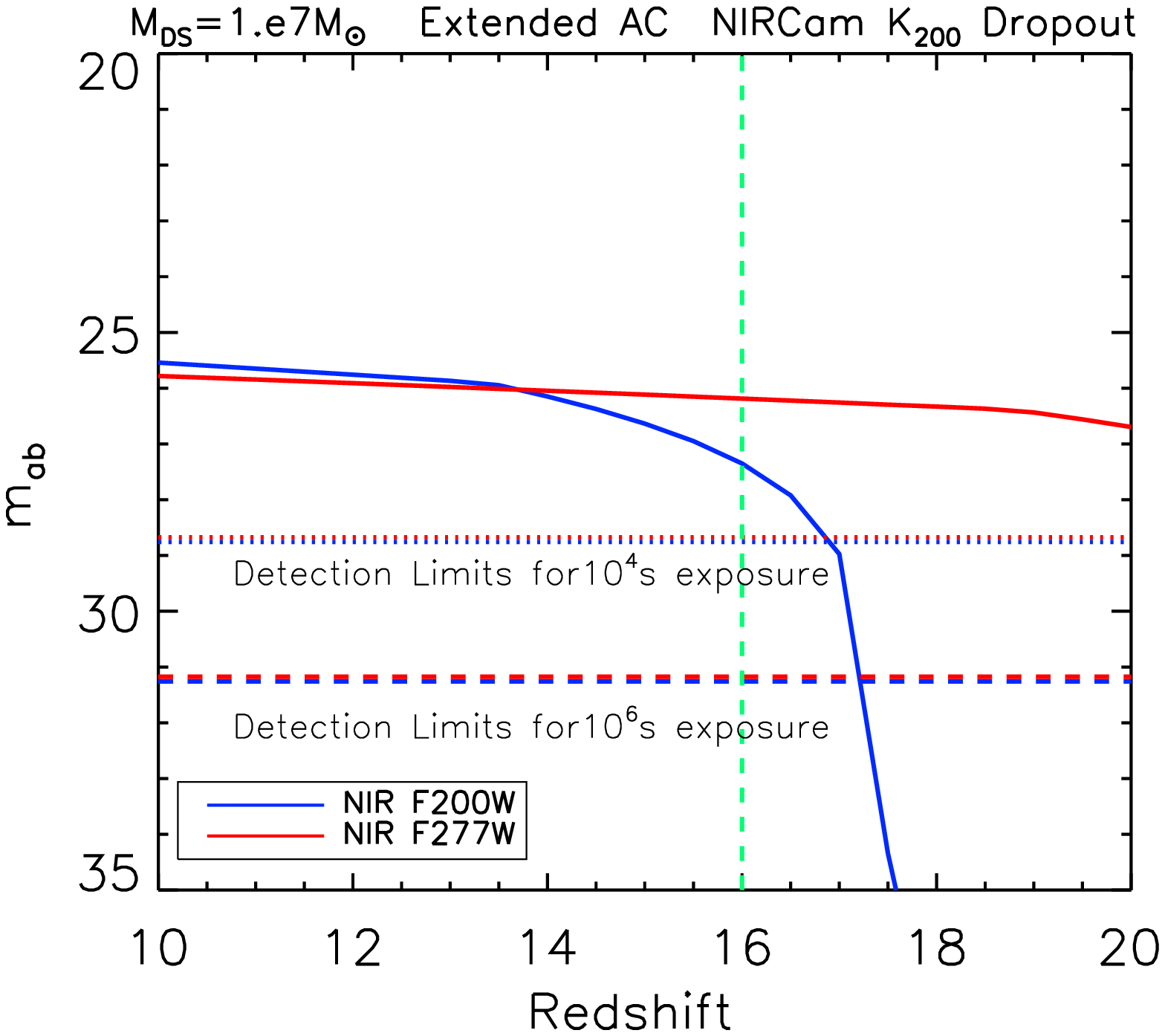}
\end{array}$
\end{center}
\caption{SMDS with JWST as \NirKband band dropouts: Apparent magnitudes for SMDS formed without DM capture through the F200W and F277W NirCam filters. Left panel: for the $10^6\Msun$ dark star. Right panel: for the $10^7\Msun$ dark star. The vertical green dashed line
indicates the minimum redshift at which the DS will appear as a dropout.}
\label{Nir200277}
\end{figure*}

DS at  $z\gtrsim 14$ can be detected as \NirKband band dropouts using the F200 and F277 NirCam filters in 
JWST, as shown in Figure~ \ref{Nir200277} for $10^6$ and $10^7 \Msun$ SMDS formed via
extended AC (no capture) at \zform=20. 
To qualify as a \NirKband dropout the difference in
magnitudes between the F277W and F200W filters must be greater than 1.2.  
As for the case of H-band dropouts above, we use HST data to bound the number of possible K-band dropouts,
under three different assumptions:  (i) Maximal Bounds where every DS survives through the HST observability window
at z$\sim$10; (ii) Intermediate Bounds with $\sim 10^{-2}$ of the possible DS surviving that long, and (iii) for comparison
simply counting every possible hosting halo.   Our results for predicted numbers  of SMDS observable 
as K-band dropouts with JWST are summarized in
 in Table \ref{tb:Kband}

The $10^6\Msun$ DS  could be observed in the redshift range $z\sim
15-17$ as a \NirKband dropout for $10^4$ seconds exposure.   For the case of Maximal Bounds from HST, we predict at most
  $N_{obs}^{multi}\sim1$.   For the Intermediate Bounds case, 
the possible number of detections is increased to roughly
 5 for the case of a 9.68 \arsq FOV or to 75 for the 150 \arsq
 case.  The (unreasonable) case where every possible halo hosts a DS 
 shows the maximal number of $10^6 M_\odot$ SMDS observable as K-band dropouts to be $\sim 70,000$.
 In the case of the $10^7\Msun$ star,
it would appear as a \NirKband dropout in the $16-20$ redshift
range. However due to the sharp drop in the formation rate of DM halos
in the $1-2\times$\tento{8} at such high redshift the number of
dropout events we predict in this case is at most $\sim 1$ (other than for the unreasonable case where every single possible halo
hosts a SMDS). The results for the detectability of SMDS as K-band dropouts
with JWST are summarized in Table \ref{tb:Kband}.

\begin{table*}
 \begin{center}

 \begin{tabular}{ccccc}

\multicolumn{5}{c}{Upper limits on numbers of SMDS detectable with JWST as \NirKband dropout}\\
 \hline\hline
 $M_{DS} (\msun)$                   &  Formation Scenario & Bounds from HST& $N_{obs}^{FOV}$ &  $N_{obs}^{multi}$    \\

 \hline
 $10^6$                                  & Extended AC                &Maximal Bounds  &   $\lesssim 1$       & $\sim$ 1       \\
 $10^7$                                  & Any                              & Maximal Bounds  &   $\ll 1$       & $\ll 1 $\\

 \hline
 $10^6$                                   & Extended AC               & Intermediate         &   5                   & 75       \\
 $10^7$                                   & Any                             &  Intermediate    &$\ll 1$                  & $\lesssim 1$  \\

\hline

 $10^6$                   & Extended AC         &      Number of DM halos    &4511        &  69900  \\
$10^7$                    & Any                        &      Number of DM halos &       8               &      116    \\
\hline 
\end{tabular} 

   \caption{The number of SMDS detections as \NirKband dropouts with JWST.   Cases are the same as above in Table \ref{tb:Hband}.}
\label{tb:Kband} 
 \end{center}
 \end{table*}

\section{SMDS vs Pop~III galaxies with JWST}\label{sec:DSvsGal}

A key question in the discovery of dark stars with JWST will be the ability to differentiate these objects from  other sources at high redshifts.  
Assuming that a population of potential $z>10$ candidates is identified by the drop out techniques
 described in previous sections, the most significant contaminant population at these redshifts is 
likely to be galaxies dominated by Pop III stars.  In this section we focus on ways to differentiate 
between SMDSs and galaxies containing Pop III stars.
\citet{Zackrisson2010}  showed that DSs in the mass range $< 10^3
\msun$ could be easily distinguished from galaxies in the redshift
range $z=0-15$ (including galaxies containing Pop III stars),
supernovae, AGN, Milky Way halo stars as well as Milky Way brown
dwarfs by their extremely red colors in color-color plots. The DS
considered there have all $T_{eff}\lesssim 9000K$, which leads to an
decrease of the ratio (B) of the fluxes to the left and right of the
Balmer jump located at \microm{0.365} with temperature \citep[see
Sec. 8.3 of][]{Rutten:2003}. The significant Balmer jump in the case
of DS with mass $\lesssim$\tento{3} will lead to very red
$m_{365}-m_{444}$ colors at z=10, offering a distinct signature, as
pointed out in \citet{Zackrisson2010}. Here we study instead the much heavier SMDSs
with $M_{DS}>10^5 M_\odot$.  These heavier stars are intrinsically
much brighter and thus easier to find as dropouts.  However, they are
also hotter than $10000K$, leading to values of the ratio B to
increase with temperature, as explained in Sec. 8.3 of
\citet{Rutten:2003}.  For the SMDS we consider here the Balmer jump is
insignificant, therefore its much more difficult distinguishing them from potential
interlopers based on the technique proposed in \citet{Zackrisson2010}
for the smaller $\sim$\tento{3} DS.  In this section we begin a discussion of 
this issue, restricting our studies to what can be learned from JWST directly.  Future studies
will be required, in which we investigate also the possible role of spectroscopy with TMT and GMT,
and other upcoming observatories in differentiating Pop III galaxies from dark stars.

The earliest Population III stars  (in the absence of dark matter 
heating) are expected to have masses in the range 10-100 $M_\odot$ - too faint to be 
seen as individual objects with JWST \citep{Oh:1999, Oh:2001,Gardner:2006,Rydberg:2011ck}.  However a galaxy containing $10^5-10^7 \msun$ of Pop III stars might  indeed be detectable. \citet{Zackrisson:2011} presented a comprehensive study of the integrated spectra signatures of Pop~III stars in the wide filters of JWST. Their main findings are that Pop~III galaxies could be detectable to redshifts as high as 20 if the stellar population mass is $\sim 10^7 \Msun$ (or in the case of $10^5\Msun$ stellar population mass up to redshifts of 10). A similar study by  \citet{Pawlik:2011}, who examined nebular emission lines from
early galaxies, came to the same conclusion:  thousands of these may be found with JWST.
 Moreover, \citet{Inoue:2011} and  \citet{Zackrisson:2011} have
proposed selection criteria using two of the  filters of JWST:   \citet{Inoue:2011}
argued for using two NIRCAM filters and  \citet{Zackrisson:2011} argued for adding imaging
in two MIRI filters
to more cleanly  differentiate between Pop~III galaxies and Pop II or Pop I galaxies at z$\sim7-8$.  
Indeed these authors found that galaxies containing Pop III stars at high redshift
are typically brighter in most JWST filters than later generations of stars; thus galaxies with
Pop III stars would be the most likely source of confusion in identifying dark stars.

Using the Yggdrasil\footnote{We highly recommend watching the movie
Thor to understand this name.} model grids \citep[][see \url{http://ttt.astro.su.se/~ez/}]{Zackrisson:2011}   we compare signatures in the NIRCam passbands of Pop~III galaxies at z$\sim10-15$ with those of SMDS. All the nomenclature  used here for Pop~III galaxies follows \citet{Zackrisson:2011}: we consider three different Initial Mass Functions (IMF) for Pop~III galaxies:
\begin{itemize}
\item{Pop~III.1: A zero-metallicity population with an extremely top
    heavy IMF and a Single Stellar Population (SSP) from
    \citet{Schaerer:2002qf} with a power-law IMF ($dN/dM \propto M^{-\alpha}$). The population has stellar masses in the
    range $50-500 \Msun$ and a Salpeter slope $\alpha=2.35$ for the
    entire mas range.}
\item{Pop~III.2: A zero-metallicity population with a moderately
    top-heavy IMF. A SSP from
    \citet{Raiter:2010dq} is used. This model has a log-normal IMF
    with characteristic mass $M_c = 10 \Msun$ and dispersion
    $\sigma=1\Msun$. The wings
    of the mass function extend from $1$ to $500\Msun$.}
\item{Pop~III, Kroupa IMF: In view of recent simulations
    \citep[e.g.][]{Greif:2010rr} the mass of Pop~III stars might be
    lower than previously predicted. Therefore in  this case a normal
    \citet{Kroupa:2001} IMF,  usually describing  Pop II/I galaxies, is
    used. The stellar masses range in the $0.1-100\Msun$ and the SSP is a rescaled version of the one used in \citet{Schaerer:2002qf}}
\end{itemize}

Following  \citet{Zackrisson:2011} we  further subdivide the models into two types,
based on the amount of nebular emission.  The first galaxies are expected to have significant
ionized gas surrounding them. Depending on how compact the HII region is, the escape fraction
for ionizing radiation from the galaxy into the IGM can vary anywhere from 0-1.  Hence we  consider the two extreme possibilities: 
\begin{itemize}

\item{Type A galaxies: If the gas covering fraction $f_{cov}=1$, then there is maximal nebular contribution to the SED
 and no escape of Lyman continuum photons.}
\item{Type C galaxies: If $f_{cov}=0$,  there is no nebular contribution to the SEDs and instead
stellar light dominates the SED. 
We will not consider here the intermediate case of Type B galaxies.}
\end{itemize}

 \citet{Zackrisson:2011} argue that  the nebular emission typically 
 dominates the spectrum of young Pop~III galaxies at z$\sim10-15$; e.g. at z=10 nebular 
 emission dominates for galaxies younger than 10~Myr. All young or
 star forming galaxies are expected to have significant contribution to the
 SEDs from nebular emission, and this effect is increased with lower
 metallicity or a more pronounced top heavy IMF. 
  Hence we will predominantly focus on Case A of maximal nebular 
 emission from the early galaxies.
 
\begin{figure*}
\begin{center}$
\begin{array}{cc}
\includegraphics[scale=0.50]{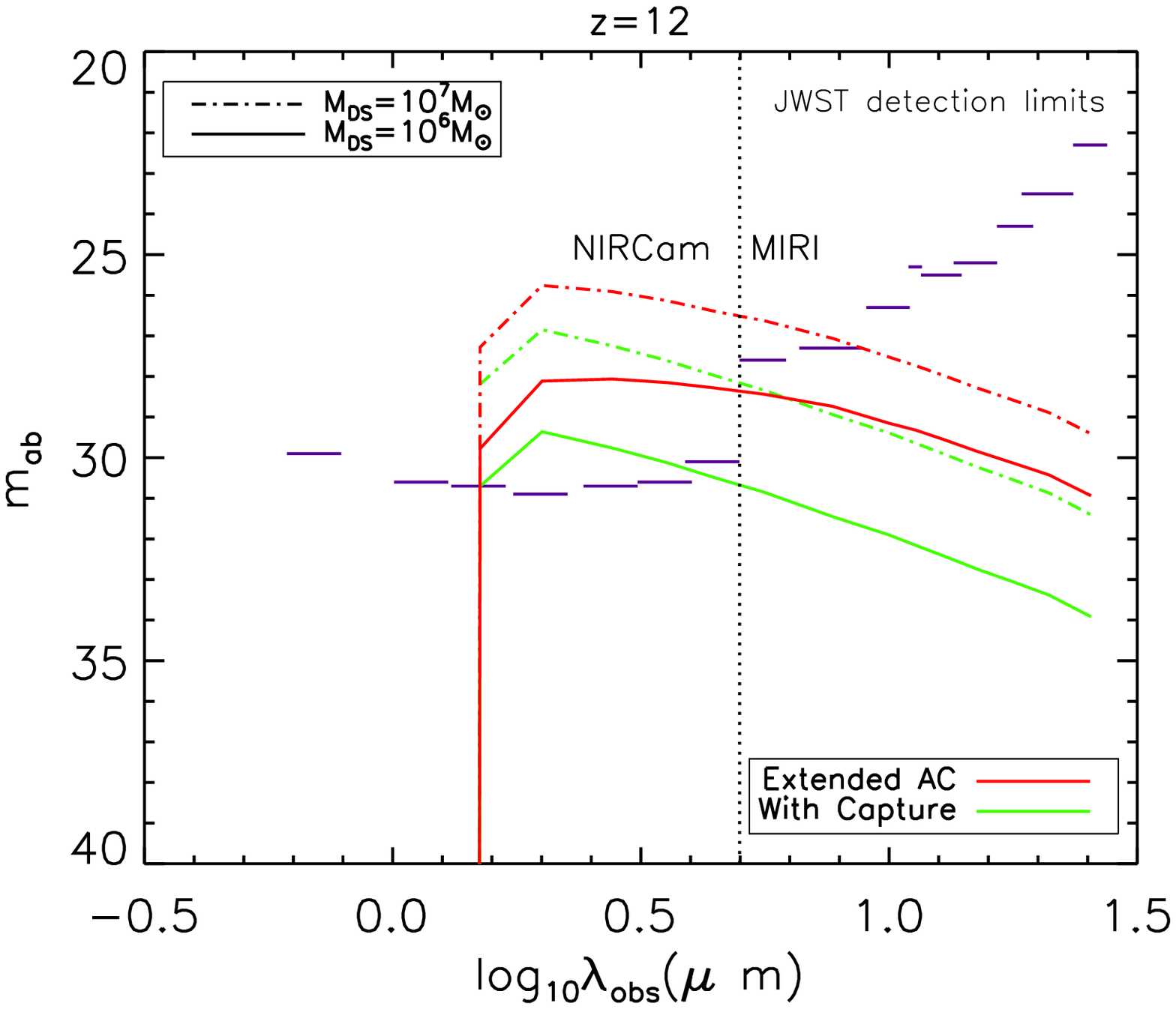}&
\includegraphics[scale=0.50]{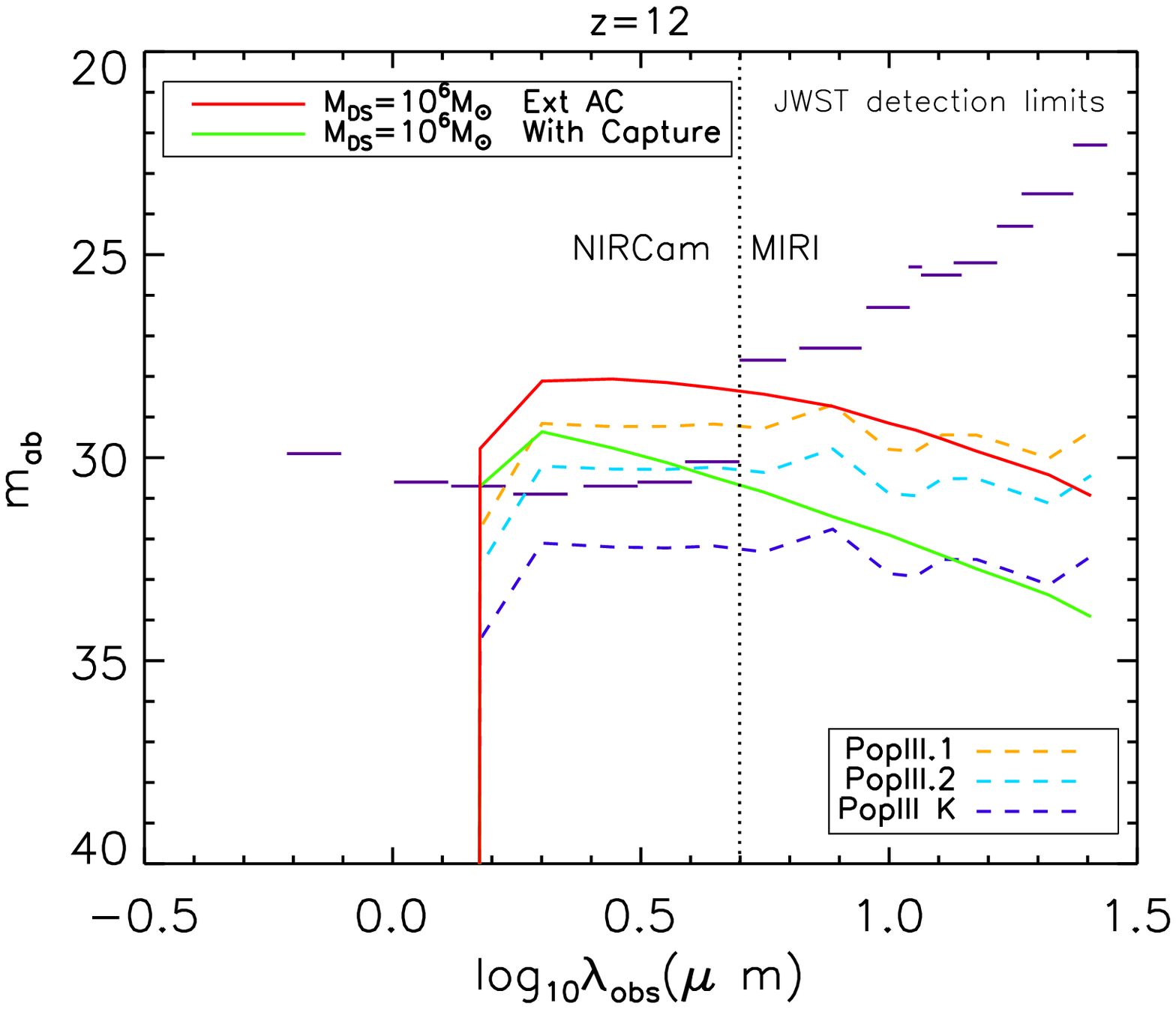} \\

\end{array}$
\end{center}
\caption{JWST detection limits and apparent magnitudes of SMDS and
  Pop~III galaxies. Each filter's sensitivity limit for JWST NIRCam
  and MIRI wide filters for 100 hours of exposure time is plotted as a
  segment at the corresponding wavelength. Left panel: Apparent
  magnitudes for SMDS in the \tento{6}(solid
  lines)-\tento{7}(dash-dotted lines) mass range for both
  extended AC (red lines) and Capture (green lines) formation mechanisms.  Right panel: Apparent
  magnitudes of SMDS of \tento{6} (solid lines) and Pop~III
  instantaneous burst galaxies of 1 Myr age (dashed lines) with the
  same stellar population mass as the SMDS. }
\label{fig:JWST_DS_GAL}
\end{figure*}

In Figure~ \ref{fig:JWST_DS_GAL} we plot the  SEDs (in apparent magnitudes) of
SMDS and Pop~III galaxies at z=12 as a function 
wavelength. Our interest is in their detectability with the NIRCam and MIRI cameras on JWST. The  vertical dotted line demarcates the wavelength ranges covered by the two instruments, and the
dark blue horizontal segments  represent band widths and the
sensitivity limits of individual filters assuming a 100 hour exposure
\footnote{For NIRCam we did not plot the F090W filter, since the
  throughput profile is not yet available.}.  
  In the left panel we plot the apparent magnitudes for $10^6$ and $10^7 M_\odot$ DS formed via 
both Extended AC and Capture mechanisms. We have previously discussed (see
Figures \ref{NirCapOff} and \ref{NirCapOn}) that
both $10^6$ and $10^7 M_\odot$ DS are bright enough to be observed by the NIRCam filters.
On the other hand, in the less sensitive MIRI filters, 
 $10^7 M_\odot$ DS can be seen  in the lowest two wavelength filters (F560W and F770W) but $10^6 M_\odot$ DS are too faint to be observed.
 
In the right panel we compare the observed SEDs of $10^6 M_\odot$ SMDS (solid curves) with Pop III galaxies (dashed curves).
For the galaxies we assumed an instantaneous starburst (at $t=0$) and used the results from the Yddgrasil code at 1 Myr after the burst. 
 The light from the galaxies is assumed to be dominated by nebular emission
(Type A) \citep{Zackrisson:2011} for galaxies younger than 10 Myr.
We have taken the stellar mass of the galaxy to be the same value
$10^6 M_\odot$ as the DS mass.  SMDS are brighter than the galaxies in all filters
in which the objects are potentially visible.
The sharp cut off  in flux at  $\log_{10} \lambda_{obs} \sim 0.02$ is due to Ly-$\alpha$ absorption.
For a stellar population mass of \tento{6}, Pop III.1 galaxies are detectable as a \NirHband dropouts  in a deep field survey with an exposure of 100 hours; Pop~III.2  are still just above the sensitivity limits; but Pop~III galaxies with a Kroupa IMF are  not detectable as \NirHband dropouts.

Let us imagine that an object has been detected as a photometric dropout at some redshift,
say an H-band dropout at $z=12$.  Our goal is to determine the nature of this object, i.e., to 
differentiate SMDS from first galaxies with JWST.  One approach would be to exploit
the emission lines in galaxies that are not shared by the DS.  \citet{Pawlik:2011}
have shown that there would be several major signatures in the spectra for Pop III galaxies with significant nebular emission (our
  Case A), including the HeII line at \microm{0.1640} and  H$\alpha$ emission.  
  They found that JWST spectrometers (NIRSpec and MIRI) are indeed sensitive enough to
  detect these emission lines, thereby potentially finding
 up to  tens of thousands of star-bursting
galaxies with redshifts z $\geq$10  in its field of view of
$\sim 10 {\rm arcmin}^2$.  They also found that the He1640
recombination line is only detectable in significant numbers for the case of zero-metallicity starbursts with top-
heavy IMF.  They noted that their
 estimates are consistent with
previous estimates of JWST starburst counts \citep[e.g][]{Haiman:1998,Oh:1999}.
A third possibility would be to detect the continuum limit of the Balmer series at 0.3646
$\micro$m in the rest frame.  

In short, if followup spectroscopy is done on an object found as
as dropout with JWST,  the detection of a HeII 1640 emission line 
or an H$\alpha$ emission line would most likely indicate that the object is a 
 Pop III galaxy with nebular emission rather than a SMDS (later stellar populations
 e.g. Pop II would also be missing these emission lines, but would not be as bright
 as either Pop III galaxies or SMDSs).  We do, however,
note one caveat:
if there is any supernova  (SN) explosion that can result
from the end of SMDS evolution, there might be another way to
 make He II 1640 radiation.  When the  SN remnant shock reaches
 the radiatively cooling stage of its evolution (i.e. when
 postshock gas cools radiatively faster than it does by adiabatic
 expansion), the shock becomes a "radiative shock", and that usually
 means that gas cools from a postshock temperature above a million
 degrees, down to $10^4$K or below, and He II line emission will
 also occur.    The shocks that do this need not only be SN
explosion shocks, but could also be halo virialization shocks,
 for halos large enough to have virial T high enough to ionize
 He II to He III. On the other hand, as discussed above, Heger
 (personal communication Oct 2011)
 finds that (in the absence of rotation), fusion powered stars more massive than 153,000$M_\odot$~ collapse directly to supermassive black hole seeds rather than blowing up as SN.

While the detection of emission lines most likely indicates that the object is a Pop III galaxy
rather than a SMDS, the lack of emission lines leaves both possibilities still open. One might
therefore ask how well the underlying continuum spectrum can be
determined with JWST. The UV continuum slopes for
galaxies in the redshift range $2-8$ have been analyzed using HST data in
the literature \citep[e.g][]{Bouwens:2009,Bouwens2010Blue,Dunlop:2011,Finkelstein:2011,McLure:2011}.
The value of $\beta$ can be determined by converting photometric colors in relevant filters 
 (as in \citet{Bouwens2010Blue} or
\citet{Dunlop:2011} for HST), but as noted in \citet{McLure:2011} the
photometric errors can be quite
large, of ${\cal{O}}(1)$. Based on our initial estimates it will be
difficult to disentangle the SMDS from PopIII galaxies based on UV
continuum slopes calculated from AB magnitude colors in NIRCam or via
spectroscopy with NIRSpec.   A
detailed study of how well this separation can be done based on UV spectra is the subject of future work.

\begin{figure*}
\begin{center}$
\begin{array}{cc}
\includegraphics[scale=0.50]{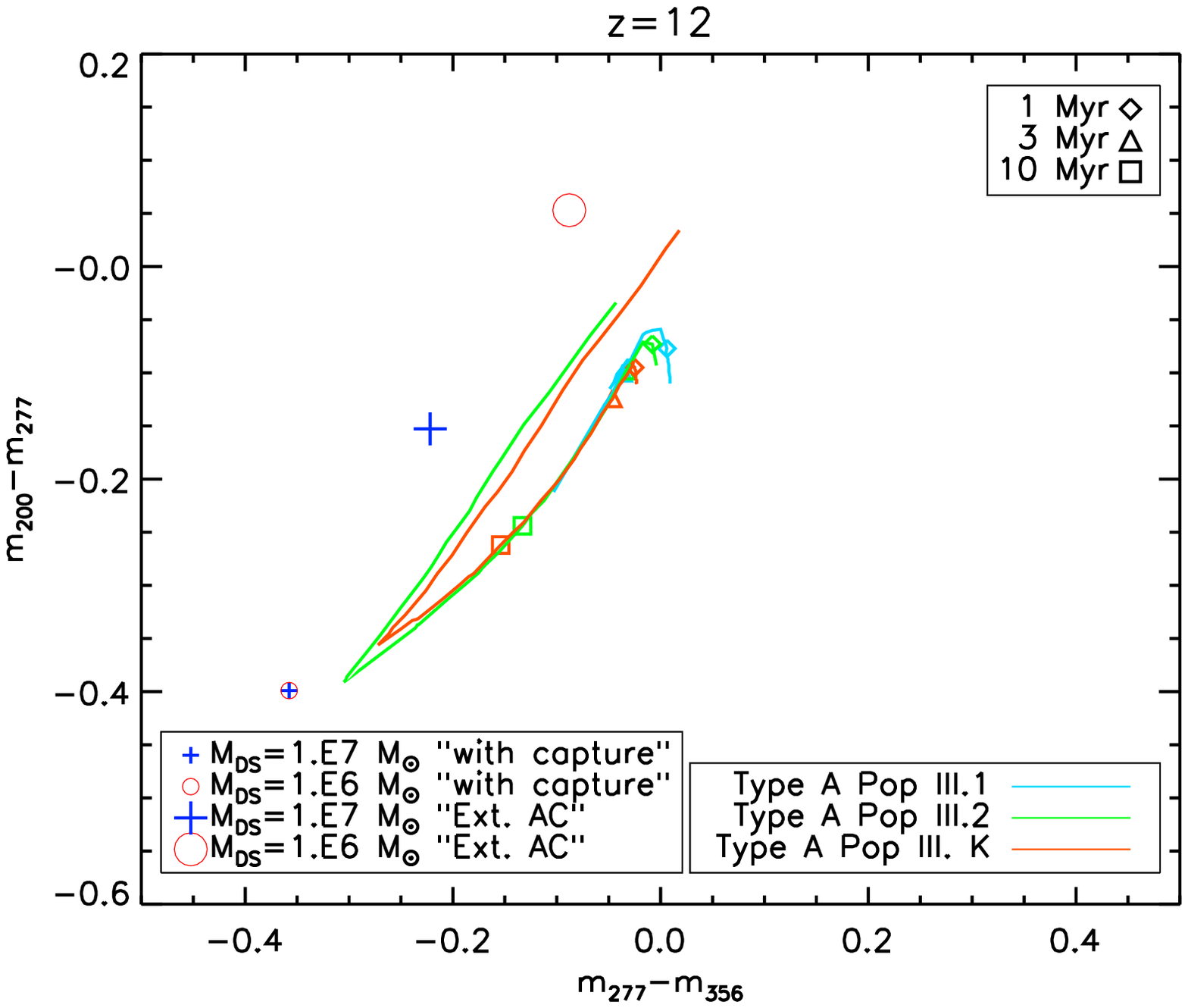} &
\includegraphics[scale=0.50]{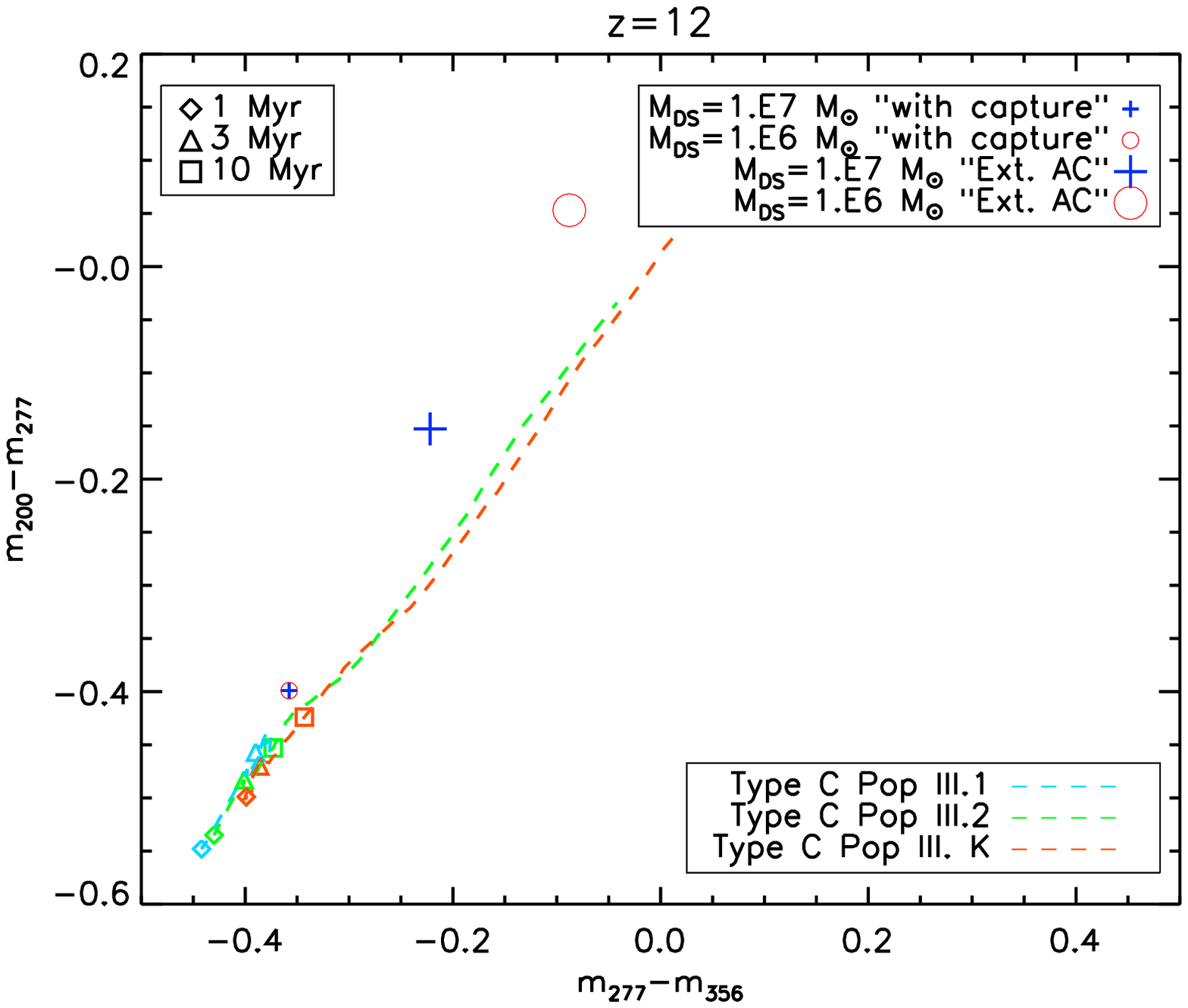} \\
\includegraphics[scale=0.50]{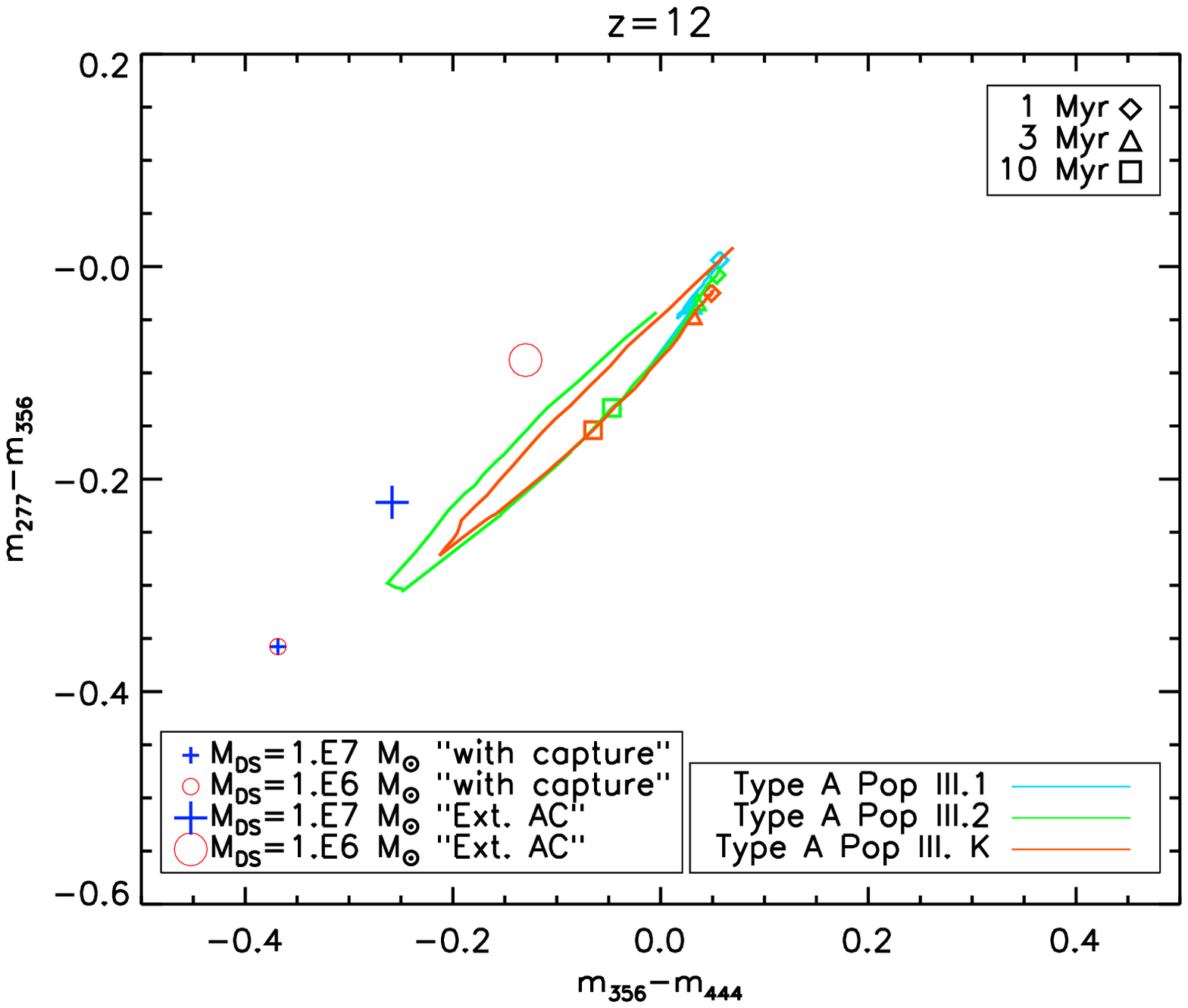}&
\includegraphics[scale=0.50]{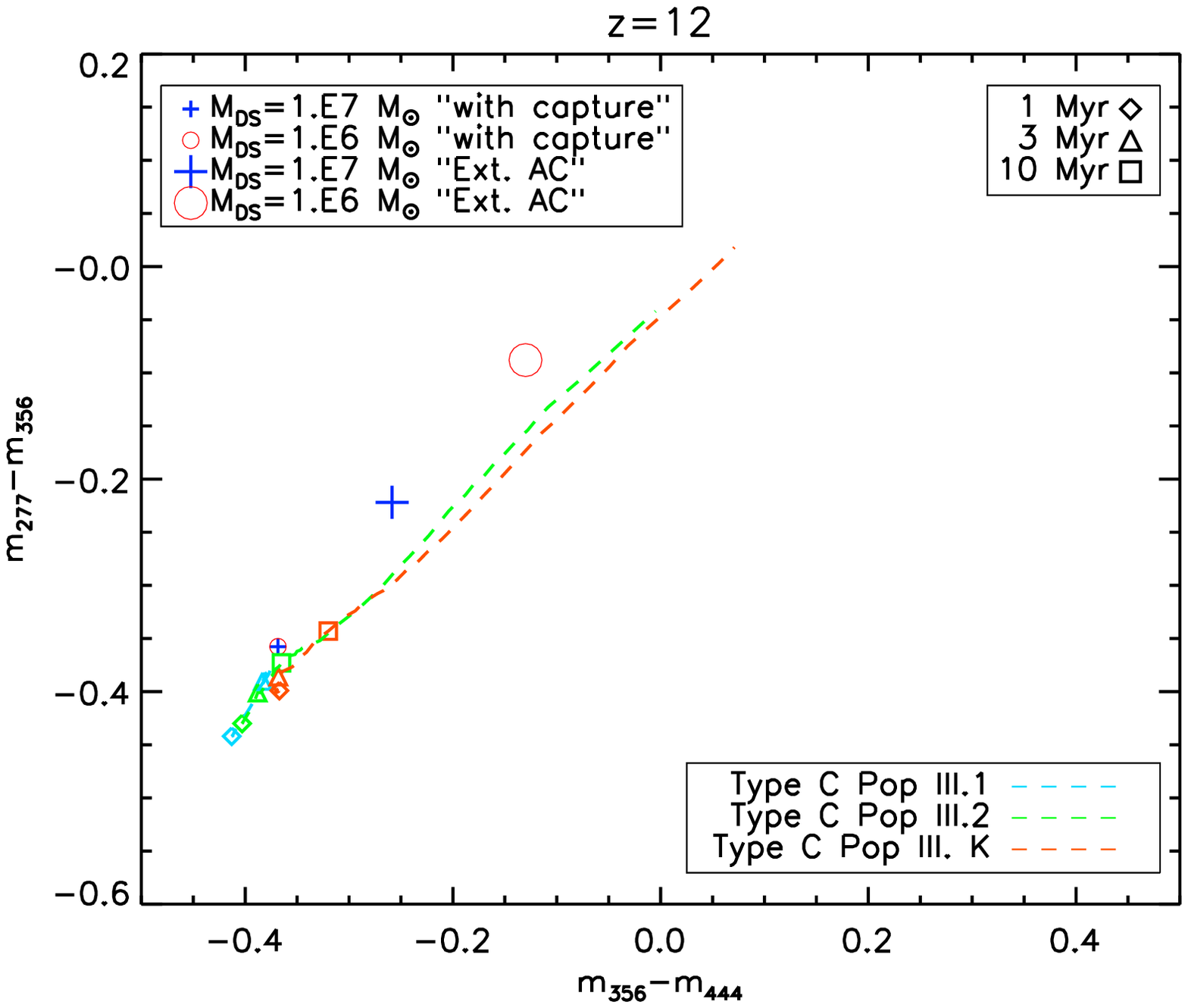}
\end{array}$
\end{center}
\caption{Signatures of SMDS and instantaneous burst Pop~III galaxies in $m_{277}-m_{356}$ vs $m_{200}-m_{277}$ (top row) and  $m_{356}-m_{444}$ vs $m_{277}-m_{356}$ (bottom row) color diagrams. The left column corresponds to Type A Pop~III galaxies (maximal nebular emission) and the right column to Type C Pop~III galaxies (no nebular emission).  The solid lines are evolutionary tracks for Pop~III galaxies obtained using the Yggdrasil model grids introduced in \citet{Zackrisson:2011}. The points along the evolutionary tracks single out the evolution at three different ages of the galaxies. \tento{6} (\tento{7}) SMDS are represented by circle (cross) symbols in the diagrams. For the extended AC case we chose a larger size symbol compared to the SMDS formed ``with capture''.}
\label{fig:DiffColors}
\end{figure*}

The  HeII line in Pop~III type A galaxies due to nebular emission  at \microm{0.1640} would fall within the F200W filter of JWST for sources at redshifts 9.7 $<z<$ 12.7.
The strength of the line is modeled in the Yggdrassil code
    for all PopIII galaxies we have considered.
Since this line is pronounced in Pop III galaxies but not in DS, one could examine
 the difference in the two magnitudes $m_{150}-m_{200}$, which would be
significantly more negative (i.e bluer  $m_{150}-m_{200}$ colors) for DS than
 for the galaxies.  
One should be able to see this effect for objects at $z<12$.  At
higher redshifts, however, 
the Gunn-Peterson cuts off significantly the fluxes in the F150W filter, so that it would be impossible
 to distinguish an increase of F200W flux (due to HeII in Pop III galaxies) from a decrease of F150W flux (due
 to Ly-$\alpha$ absorption). 

Another approach to distinguish between different types of objects is their location in 
color/color plots.
 Previously  \citet{Inoue:2011} and  \citet{Zackrisson:2011}  used color/color plots to
 distinguish between different types of galaxies:  ones with Pop III.1 stars vs. ones containing a later population of stars.  In Figure~ \ref{fig:DiffColors} we try out the possibility of differentiating
DS vs. galaxies, based on their locations in color/color plots. 
   In the left  and right panels we study  Pop~III 
 instantaneous burst galaxies of Type A (maximal nebular emission) and Type C (no nebular emission) respectively.
  We focus here on objects at z=12 as this is the most favorable redshift to look for SMDSs.
  The empty circle (cross) symbols correspond to magnitudes for SMDS of \tento{6} (\tento{7}). The solid lines represent the evolutionary tracks of Pop~III galaxies obtained using the Yggdrasil model grids, with points marking three different ages (diamonds for 1Myr, triangles for 3 Myr, and squares for 10 Myr). We note that, due to the similar temperatures, SMDS formed ``with capture'' of either \tento{6} or \tento{7} occupy the same spot on the diagrams. 
  
 Pop~IIIA galaxies with lifetimes less than 10 Myr will exhibit redder colors than SMDS
 in the $m_{356}-m_{444}$ (see lower left panel) due to the increased fluxes in the F444W filter due to the Balmer emission lines from the galaxies.
One might therefore hope to distinguish between SMDS and Pop~III type A galaxies at $z\sim 12$ would be to look for red colors in  $m_{356}-m_{444}$.   Indeed for the case in Figure~ \ref{fig:DiffColors}, this technique would work:   only the Pop III.1 galaxies are bright enough
to compete with SMDS (see Figure~ \ref{fig:JWST_DS_GAL}), yet these are in a distinctly different location on the
color/color plot from the SMDS.
However, in the figure we have taken the stellar
     mass of the galaxies to be $10^6 M_\odot$, while this number could be an order of
     magnitude higher, which would drive the Pop III lines in the Figure closer to the SMDSs.
     As pointed out before, the error bars in the magnitudes
     for $10\sigma$ detections are $\sim 0.15$, comparable to the differences in magnitudes
     between galaxies and SMDS.  
     In addition, in the  figure we have  taken a specific star formation rate (instantaneous burst). For comparison  we have also tried a constant SFR and found that the results do not change much
     (the colors become slightly bluer). Differentiating between SMDS and Pop III galaxies (of
     uncertain stellar mass and metallicity) with
     such color-color plots will be difficult. 

Differentiating between SMDS and galaxies containing Pop III galaxies is an important issue.
 Using JWST, the best bet is
to look for emission lines of He 1640 or H$\alpha$.  If these are found the object is not likely
to be a SMDS.  On the other hand, if these are not found, then differentiating between SMDS and Pop III.1 galaxies
may be difficult with JWST.  Further studies with other instruments,
specifically ground based spectrometers, may prove to be helpful.

\section{Summary and Conclusions}

The first phase of stellar evolution may have been dark stars, powered by dark matter annihilation.  These
form inside early $10^6-10^8 M_\odot$ halos at z=10-50.  Initially DS are puffy objects with masses of  $1-10 M_\odot$
and radii $\sim$10 A.U.  As long as they are DM powered,  their surface temperatures  ($\sim 10^4$K) remain cool
enough to allow continued growth via accretion of baryons until they become supermassive $M_{SMDS} \sim 10^6, 10^7 M_\odot$.
The requisite DM fuel can be acquired in two ways: (i) extended adiabatic contraction due to DM particles on chaotic or box
orbits in triaxial haloes and (ii) capture of DM particles via elastic scattering off nuclei in the star.
In this paper we have studied the detectablility of Supermassive Dark Stars formed via both mechanisms with upcoming
JWST observations.

In order to determine their observational characteristics, 
we obtained the spectra of SMDSs with the TLUSTY stellar atmospheres code (Figure~ \ref{fig:TLUSTYRest}).
We used N-Body simulations for structure formation at high 
redshifts \citep{Ilievetal10}  to obtain estimates for the numbers of  DM halos capable of hosting
SDMS (Figure~ \ref{FRates1} and Table ~\ref{tb:Rates} ).  
Then we used HST observations to set limits on their detectability.

Both $10^6$ and $10^7 M_\odot$
SMDS could be seen in HST data and would be detected as
J-band dropouts.  Since \citet{Bouwensetal11} report only one plausible
z$\sim$10 object in the data, we used the fact that at most one
observable DS at this redshift can exist to obtain bounds on
the possible numbers of DS in Eqns. \ref{bounds1e7all} and \ref{bounds1e6all}.

SMDSs are bright enough to be seen in all the
wavelength bands of the NIRCam on JWST, while detection is more difficult in the less sensitive higher wavelength
MIRI camera.
We showed that SMDSs could be seen as J-band,
H-band, or K-band dropouts, which would identify them as
z$\sim$10, 12, and 14 objects respectively.  

The strong point of JWST will be its sensitivity to longer wavelengths than HST,
corresponding
to light from higher redshifts where SMDSs may be found.
While JWST is not particularly better than HST at finding J-band dropouts, 
it will be significantly better at finding SMDS as H-band
and K-band dropouts.  

We can summarize our predictions for the numbers of SMDS seen as H-band dropouts with JWST as:
\be\label{Hbandevents6}
N_{obs}= 4.4 \times 10^5  f_{smds}  f_{\De t}  (\theta / 150 {\rm arcmin})^2 
\,\,\,\,\,\,\,\,\,\,\,\, (M_{DS} = 10^6 M_\odot) 
\ee
\be\label{Hbandevents7}
N_{obs}= 2.4 \times 10^3  f_{smds}  f_{\De t}  (\theta /150 {\rm arcmin})^2
\,\,\,\,\,\,\,\,\,\,\,\, (M_{DS} = 10^7 M_\odot) 
\ee 
where  we have scaled the results to 150 arcmin$^2$ survey area, which would require multiple 
surveys by JWST.

Similarly,  our predictions for the numbers of SMDS seen as K-band dropouts are:
\be\label{Kbandevents6}
N_{obs}= 7 \times 10^4  f_{smds}  f_{\De t}  (\theta / 150 {\rm arcmin})^2 
\,\,\,\,\,\,\,(M_{DS} = 10^6 M_\odot, {\rm  AC})
\ee
\be\label{Kbandevents7}
N_{obs}= 120  f_{smds}  f_{\De t}  (\theta /150 {\rm arcmin})^2
\,\,\,\,\,\,\, (M_{DS} = 10^7 M_\odot) .
\ee 
$10^6 M_\odot$ SMDS formed via Capture are not detectable.

Although these numbers are quite large, as we have emphasized throughout it is quite
likely that \fsmds $f_{\De t} << 1$.   If the DS survives to $z \sim 10$, HST observations bound this product. 
Our final predictions for numbers of SMDS that could be detected by
JWST are found in Tables \ref{tb:Hband} and \ref{tb:Kband}.

Differentiating between SMDS and galaxies containing Pop III galaxies is an important issue.
 Using JWST, the best bet is
to look for emission lines of He 1640 or H$\alpha$.  If these are found the object is not likely
to be a SMDS.  On the other hand, if these are not found, then JWST will have trouble differentiating
between SMDS and early galaxies.  Thus  further estimates are required
using instruments such as Giant Magellan Telescope,
TMT, LSST, and others. 

As argued by Heger (Heger, private communication) in the absence of a dark star phase, the
characteristic mass for
 big BHs at birth is 153,000 $M_\odot$ (i.e., once a fusion powered star accretes this
much mass it can no longer sustain hydrostatic equilibrium and collapses directly to a BH.
With a dark star phase, the DS could instead grow to a larger mass while DM powered, and
then collapse directly to a BH; thus in this case the BH could be born with larger masses.
 Future observations of large BHs
might thus be able to differentiate  someday between formation mechanisms via dark stars
or fusion powered stars.

{\it SMDS mass as a function of halo mass:}
Although we have assumed in this paper that DS grow to the point where they
consume most of the baryons in the haloes that host them,  one can examine how 
the results would change if we were
to stop the growth at a smaller fraction of the total baryonic content.
For the case of "maximal bounds" we can show that the resulting predictions for JWST remain identical.
For example,  the case we considered in the paper of
 $10^6\msun$ SMDS that grew inside $\sim 10^7 \msun$ haloes, can be compared
instead to the case of $10^6\msun$ SMDS that grew inside $\sim 10^8 \msun$ haloes.
For the case of "maximal bounds", which assumes that HST bounds at z=10 apply directly to 
SMDS at z=12 (i.e. that the SMDS at z=12 survive all the way to z=10),
we find that our results are completely unchanged.   
The  number of $10^8 M_\odot$ haloes is smaller than the number of $10^7 M_\odot$ haloes
at both redshifts z=10 (so the HST bounds are weaker) and at z=12 (where the JWST observations are made).
Thus the two effects cancel exactly.  One can see this cancellation in the following way.
The numbers of SMDS observable in either HST or JWST are given by the same equation,
Eq.({\ref{maxf}). The two factors
$\frac{dN}{dz d\theta^2}$ and  $f_{SMDS}(z=z_{start})$ change depending on the hosting
halo mass, but their product remains the same since it is set by HST bounds in Eq.(\ref{maxf2}).
Thus  the numbers with JWST  are unchanged regardless of halo size.

The current decade is a time of great excitement in the physics community
regarding the possibility of detection of the dark matter particle.  Three approaches
are being pursued in the hunt for Weakly Interacting Massive Particles:
direct detection (including DAMA, CDMS, XENON, COGENT, CRESST, ZEPLIN,
TEXONO, COUPP, and many others worldwide), indirect detection (including
PAMELA, FERMI, ICECUBE), and colliders (LHC).  
Many of these experiments have indeed found hints of a signal, though confirmation
in more than one type of detector of the same particle remains a goal.
Dark Stars offer a fourth
possibllity for the discovery of WIMPs, or of learning about their properties.  
If WIMPs are indeed discovered, then it is very reasonable to expect to find
Dark Stars in the sky that are WIMP powered.  It is even possible that the WIMPs have
the property that they will be seen first by JWST before any other experiment.
In either case the prospect of finding a new type of star in the next premier NASA 
mission is greatly exciting.

\bigskip
\noindent {\it Acknowledgements:} 
 We are extremely grateful to Pat Scott for providing spectra for Dark Stars with the 
 TLUSTY code and to E. Zackrisson for making the Yggdrasil model grids publicly available. This research is  supported in part by Department of Energy (DOE) grant   number DE-FG02-95ER40899 and  by the Michigan Center
for Theoretical Physics.   KF thanks the Texas Cosmology Center (TCC) where she was a Distinguished Visiting Professor. TCC is supported by the College of Natural Sciences and the Department of Astronomy at the University of Texas at Austin and the McDonald Observatory. KF  also thanks the  Aspen Center for Physics for hospitality during her visit. MV acknowledges support from NSF grant  number AST-0908346. ITI acknowledges support from the Southeast 
Physics Network (SEPNet) and the Science
and Technology Facilities Council grant number ST/I000976/1.The authors acknowledge the Texas Advanced Computing Center (TACC) at The University of Texas at Austin for providing HPC resources that have contributed to the research results reported within this paper. URL: http://www.tacc.utexas.edu\\

\bibliographystyle{mn}
\bibliography{RefsCos}

\end{document}